\documentclass[preprint,aps,prb,nofootinbib,longbibliography]{revtex4-1}
\usepackage{amssymb}
\usepackage{setspace}
\usepackage{amsmath}
\usepackage{amsfonts}
\usepackage{hyperref}
\usepackage{graphicx}
\usepackage{float}
\usepackage{subfigure}
\usepackage{color}

\DeclareMathOperator{\OMi}{\kappa-(\text{ET})_2\text{Cu}_2(\text{CN})_3}
\newcommand{\vect}[1]{\boldsymbol{#1}}
\newcommand{\conj}[1]{\overline{#1}}
\newcommand{\mc}[1]{\mathcal{#1}}

\newcommand{\boson}[1]{a_{\vect{#1}}}
\newcommand{\dboson}[1]{a_{\vect{#1}}^{\dagger}}
\newcommand{\ad}{a^{\dagger}}
\newcommand{\bd}{b^{\dagger}}
\newcommand{\cd}{c^{\dagger}}
\newcommand{\dd}{d^{\dagger}}
\newcommand{\Bd}{B^{\dagger}}
\newcommand{\Cd}{C^{\dagger}}
\newcommand{\phid}{\phi^{\dagger}}
\newcommand{\tc}{\tilde{c}}
\newcommand{\tcd}{\tilde{c}^{\dagger}}

\newcommand{\tdd}{\tilde{d}^{\dagger}}
\newcommand{\kk}{\vect{k}_1,\vect{k}_2}
\newcommand{\kv}{\vect{k}}
\newcommand{\pv}{\vect{p}}
\newcommand{\qv}{\vect{q}}
\newcommand{\gk}{\gamma_{\kv}}
\newcommand{\ak}{\alpha,\kv}

\newcommand{\la}{\langle}
\newcommand{\ra}{\rangle}
\newcommand{\be}{\begin{equation}}
\newcommand{\ee}{\end{equation}}
\newcommand{\ba}{\begin{align}}
\newcommand{\ea}{\end{align}}
\newcommand{\non}{\nonumber}

\newcommand{\hsat}{$h_{\text{sat}}$}
\newcommand{\hsa}{h_{\text{sat}}}
\newcommand{\loghsat}{|\log{(\hsa -h)}|}
\newcommand{\iu}{{i\mkern1mu}}

\newcommand{\opemu}[1]{{#1}_{\mu,\kv}}
\newcommand{\openu}[1]{{#1}_{\nu,\kv}}
\newcommand{\barj}{\bar{\jmath}}
\newcommand{\Uone}{\text{U}(1)}

\newcommand{\InvLog}{\frac{1}{|\log\mu|}}
\newcommand{\Jcri}{J_{2\text{c}}}

\allowdisplaybreaks

\begin{document}
\title{Quantum phase transitions in Heisenberg $J_1-J_2$ triangular antiferromagnet in a magnetic field}
\author{Mengxing Ye}
\author{Andrey V. Chubukov}
\affiliation{School of Physics and Astronomy, University of Minnesota, Minneapolis, MN 55455, USA}

\date{\today}
\begin{abstract}
We present the zero temperature phase diagram of a Heisenberg antiferromagnet on a frustrated triangular lattice with  nearest neighbor ($J_1$) and next nearest neighbor ($J_2$) interactions, in a magnetic field. We show that the classical model has an accidental degeneracy for all $J_2/J_1$ and all fields, but the degeneracy is lifted by quantum fluctuations. We show that at large $S$, for $J_2/J_1 <1/8$, quantum fluctuations select the same sequence of three sublattice co-planar states in a field as for $J_2 =0$, and for $1/8<J_2/J_1 <1$ they select the canted stripe state for all non-zero fields. The transition between the two states is first order in all fields, with the hysteresis width set by quantum fluctuations. We study the model with arbitrary $S$, including $S=1/2$, near the saturation field by exploring the fact that near saturation the density of bosons is small for all $S$. We show that for $S >1$, the transition remains first order, with a finite hysteresis width, but for $S=1/2$ and, possibly, $S=1$, there appears a new intermediate phase without a quasi-classical long-range order.
\end{abstract}
\maketitle
\section{Introduction}
Frustrated magnetic systems have been extensively studied for decades~\cite{Balents2010,IntroFM,ChernyshevRMP,Starykh2015}, and the Heisenberg antiferromagnet on a triangular lattice is considered as one of the paradigmatic model. Frustration is believed to weaken the system's tendency to form conventional long range orders. Quite a few models of frustrated magnetic systems on a triangular lattice have been proposed as candidates to possess exotic quantum phases, both magnetically ordered and disordered, such as spin nematic phase~\cite{Penc2011,Shannon2013}, magnetization plateau state~\citep[and references therein]{Svistov2006,Shannon2011,Zhitomirsky2011,Chubukov2013,Balents2013,Richter2013,Mila2016}, valence bond solid phase~\cite{Kato2007}, spin density wave phase~\cite{Starykh2014,Balents2013}, and quantum spin liquid phase~\cite{Sachdev1992,Moessner2001,Wang2006}.

In this work, we study Heisenberg antiferromagnet on a triangular lattice with nearest neighbor ($J_1$) and next nearest neighbor ($J_2$) interactions in the regime $J_2<J_1$. This system is highly frustrated in two aspects. First, triangular lattice is geometrically frustrated and does not support a simple antiferromagnetic order. This generally increases the strength of quantum fluctuations. Indeed, although the ground state of the nearest neighbor Heisenberg model on a triangular lattice is magnetically ordered, the order structure ($120^{\circ}$ Neel order) is non-collinear~\cite{Huse1988,Bernu1992,Capriotti1999,White2007}, and the magnetization is substantially suppressed from its classical value due to quantum fluctuations (by about $50\%$ for $S=1/2$ ~\cite{Jolicoeur1989,Fazekas1999lecture,Capriotti1999,White2007}). Second, as the next nearest neighbor coupling $J_2$ increases to around $J_1/8$, the spin order in zero field changes from the $120^{\circ}$ Neel order to stripe order. At large spin $S$, the transition between $120^{\circ}$ state and stripe state is first order~\cite{Chubukov1992,Korshunov1993}, but for $S=1/2$ recent numerical studies~\cite{Campbell2015, Sheng2015,White2015,Imada2014,Iqbal2016} based on coupled cluster method, density matrix renormalization group (DMRG), and variational Monte Carlo, found that, at least for $S=1/2$, there exists an intermediate quantum-disordered state in between the two ordered states, though the nature of the non-magnetic phase is not yet fully determined. The width of the quantum-disordered phase was identified numerically as $0.06 \lesssim J_2/J_1\lesssim 0.17$~\cite{White2015}.

Possible spin liquids have been discovered in various materials, identified as nearly isotropic triangular antiferromagnets. Examples include inorganic magnet $\text{YbMgGaO}_4$~\cite{Mourigal2016,Shen2016}, and organic Mott insulators $\OMi$~\cite{Shimizu2003}, $\text{EtMe}_3 \text{Sb}[\text{Pd}(\text{dmit})_2 ]_2$~\cite{Kato2008,Yamashita2010}. As the nearest neighbor Heisenberg antiferromagnetic model turns out to exhibit $120^{\circ}$ Neel order~\cite{Huse1988,Bernu1992,Capriotti1999,White2007}, enhanced frustration such as further neighbor interactions and multiple spin exchange couplings have been introduced to explain the experiments~\cite{Jolicoeur1990,Chubukov1992,Korshunov1993,Manuel1999,Phillip2010,Campbell2015, Sheng2015,White2015,Imada2014,Iqbal2016,Motrunich2005,CenkeXu2013}. In particular, recent experiments on $\text{YbMgGaO}_4$ (Ref. \cite{Mourigal2016}) have been interpreted based on the idea that the second neighbor exchange $J_2$ is sizable in this system.

In this work we study $J_1-J_2$ model on a triangular lattice in an external magnetic field. The goal of these studies is three-fold. First, we want to understand what kind of spin order emerges in the large $S$ model in a finite field, and, in particular, how the stripe order, detected at $J_2 > J_1/8$ in zero field, evolves as field increases. As we will see, there is an infinite set of classically degenerate ordered states in a finite field, and the selection of the actual order is done by quantum fluctuations via order from disorder mechanism~\cite{Chubukov1991,Chubukov1992}. Second, we want to understand whether the first order transition between quasiclassically selected ordered states in a field remains a first order, like in zero field, or occurs via an intermediate mixed phase. Third, near a saturation field we take advantage of the fact that spins are almost polarized in a direction selected by the field and the density of Holstein-Primakoff bosons is small at arbitrary $S$~\cite{Batyev1984,Batyev1986}, and search for a possible spin state without a spontaneous long-range order for $S=1/2$, and possibly, larger spins.

The Hamiltonian of the triangular $J_1-J_2$ Heisenberg antiferromagnet model in a field is
\begin{align}
\mathcal{{\tilde H}}&={\tilde J}_1\sum_{\langle i,j\rangle}\vect{S}_i\cdot\vect{S}_j+{\tilde J}_2\sum_{\langle\langle i,j\rangle\rangle}\vect{S}_i\cdot\vect{S}_j-\vect{{\tilde h}}\cdot\sum_{i} \vect{S}_i
\label{eq:modelH}
\end{align}
where $\langle i , j\rangle$ and $\langle\langle i , j \rangle\rangle$ run over all the nearest and next nearest neighbor bonds. Due to global spin-rotational symmetry of the Heisenberg exchange terms, the field $\vect{{\tilde h}}$ can be taken in any direction, and we choose $\vect{{\tilde h}}={\tilde h} \, \hat{e}_z$ in the following. For simplicity, we measure the energy in the units of ${\tilde J}_1$ and re-define $ \mathcal{ H} = \mathcal{{\tilde H}}/{\tilde J}_1,~ J_2 = {\tilde J}_2/{\tilde J}_1$. We also define $h = {\tilde h}/{\tilde J}_1 S$.

In high enough field, all spins are polarized in a direction selected by the field. Below a critical field, which depends on $J_2$ (see Eq.~\ref{ch_1} below), an order different from the ferromagnetic one starts to develop. The identification of the orders and the phase transition between different ordered states is the subject of this paper.
\subsection{The summary of our results}
\begin{figure}[tbp]
\centering
\subfigure[]{\includegraphics[height=1.5in]{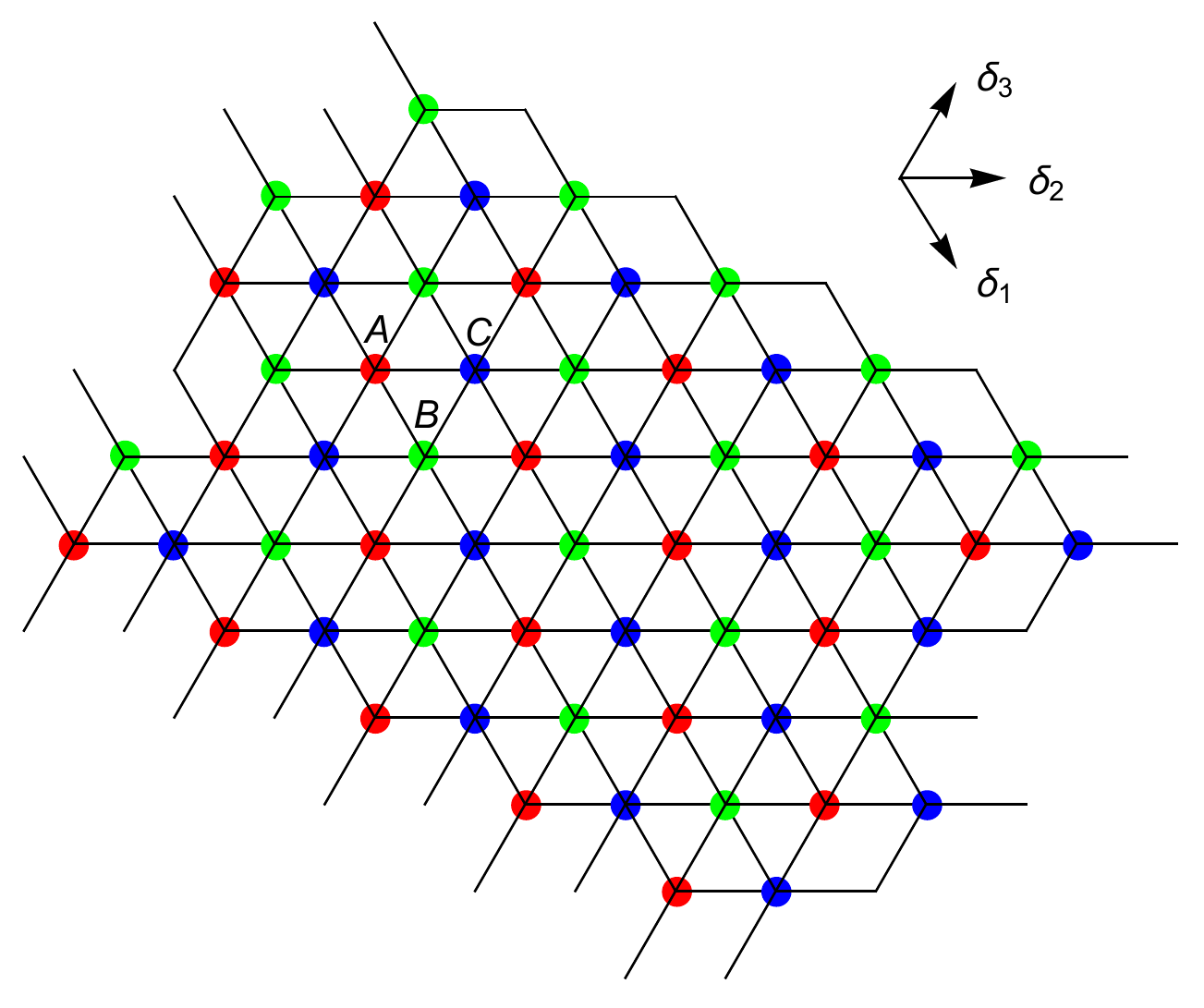}\label{fig:threesublattice}}\quad
\subfigure[]{\includegraphics[height=1.5in]{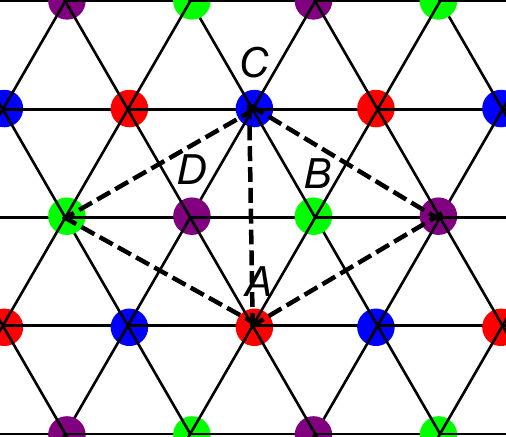}\label{fig:foursublattice}}\quad
\subfigure[]{\includegraphics[height=1.3in]{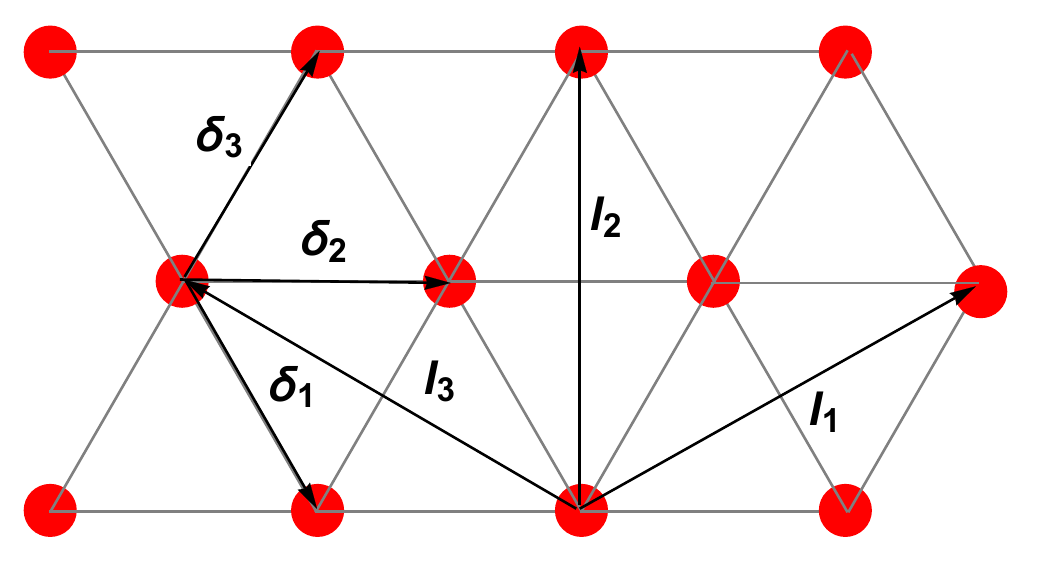}\label{fig:singlelattice}}
\caption{(a) and (b): Three-sublattice (four-sublattice) structure describes the classical ground state manifold when $J_2<1/8$ ($J_2>1/8$). The spins on the sites of the same color are within the same sublattice. (c) The nearest and the next nearest neighbor bonds, $\vect{\delta}_i$ label the nearest neighbor bonds, and $\vect{l}_i$ label the next nearest neighbor bonds.
\label{fig:sublattice}}
\end{figure}

\begin{figure}[tbp]
\centering
\subfigure[~Classical phase diagram, arbitrary field]{\includegraphics[scale=0.5]{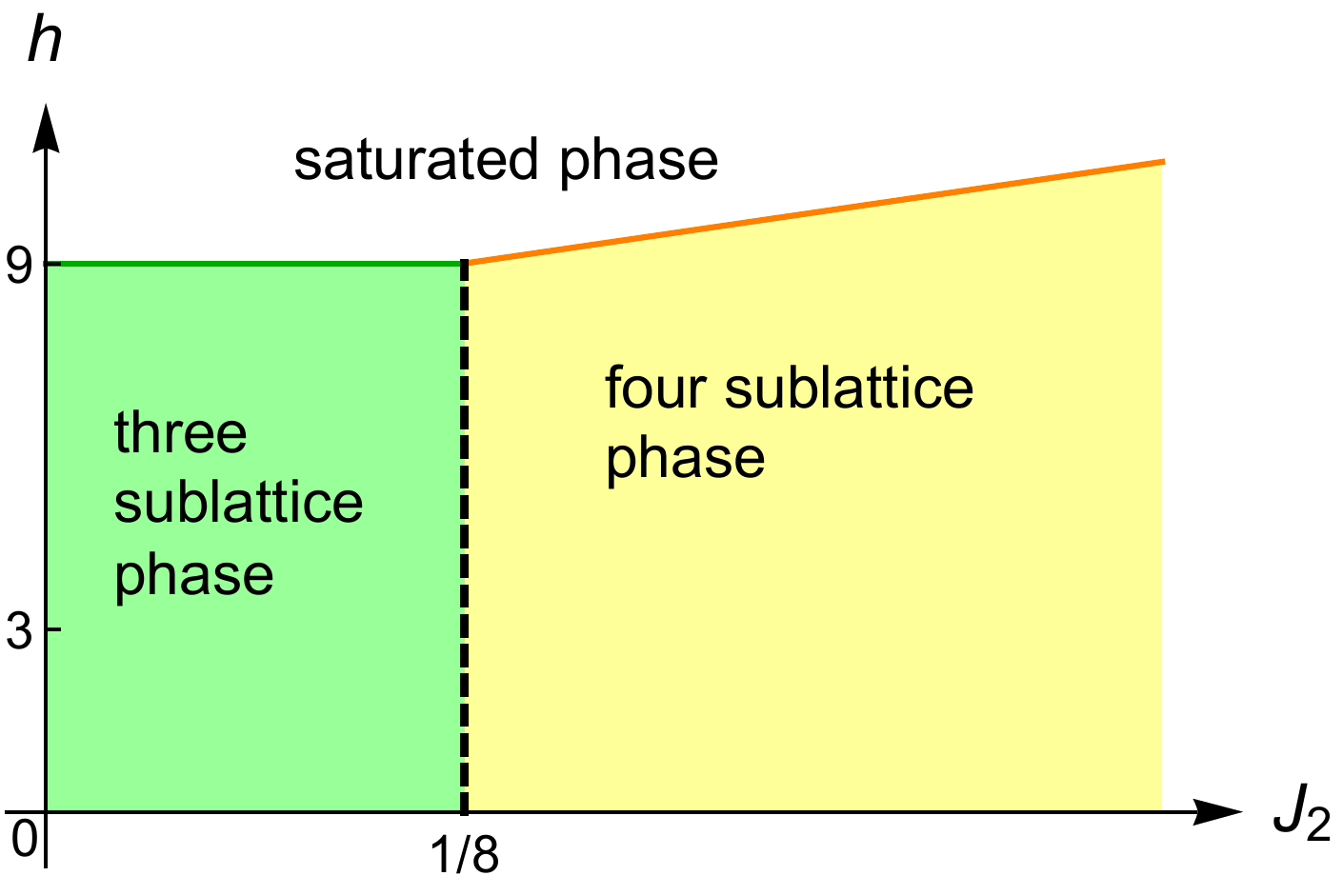}\label{fig:summaryClassical}}
\subfigure[Phase diagram near $\hsa$ at arbitrary $S$]{\includegraphics[scale=0.5]{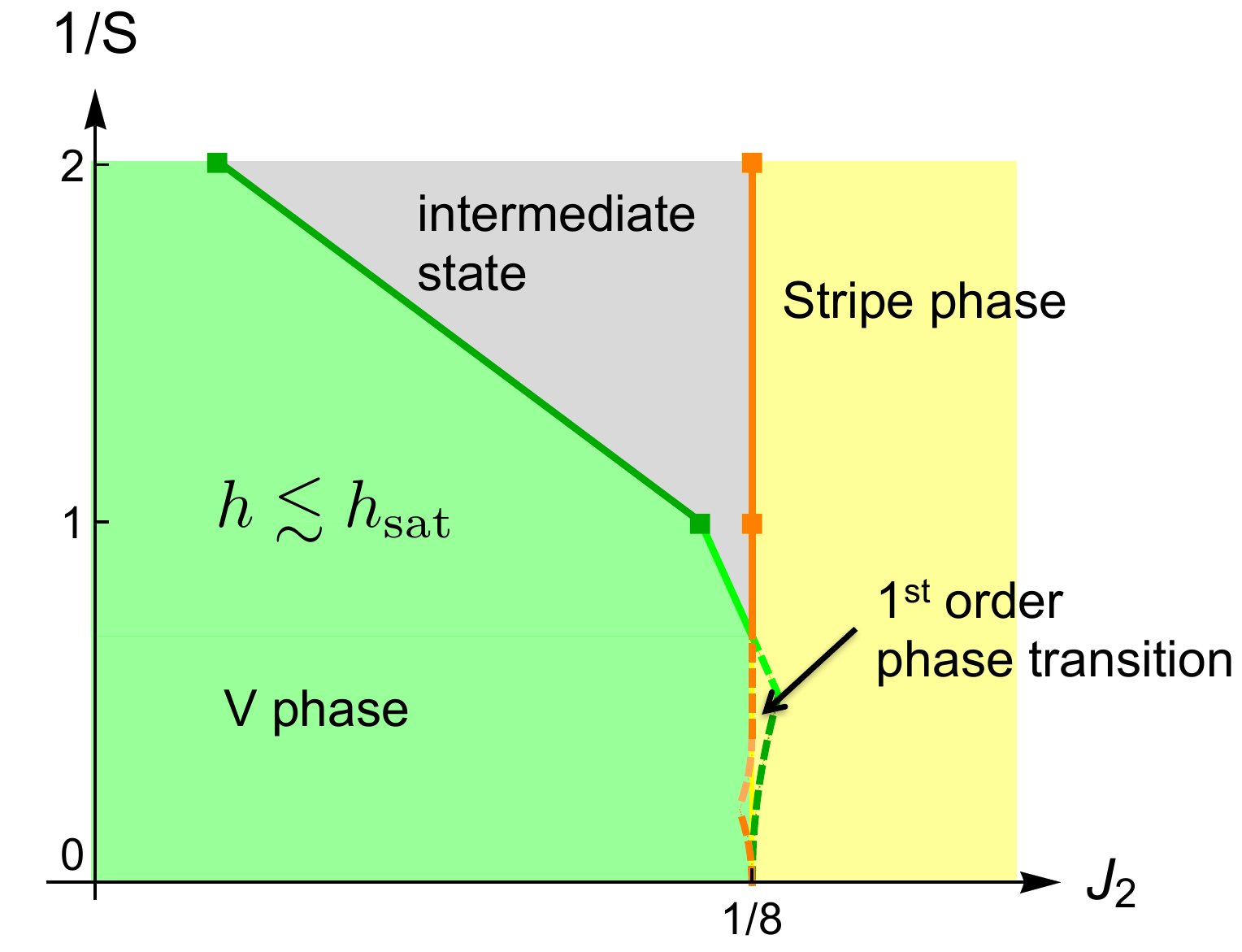}\label{fig:summaryArbitraryS}}\\
\subfigure[~Semiclassical phase diagram, arbitrary field]{\includegraphics[scale=0.7]{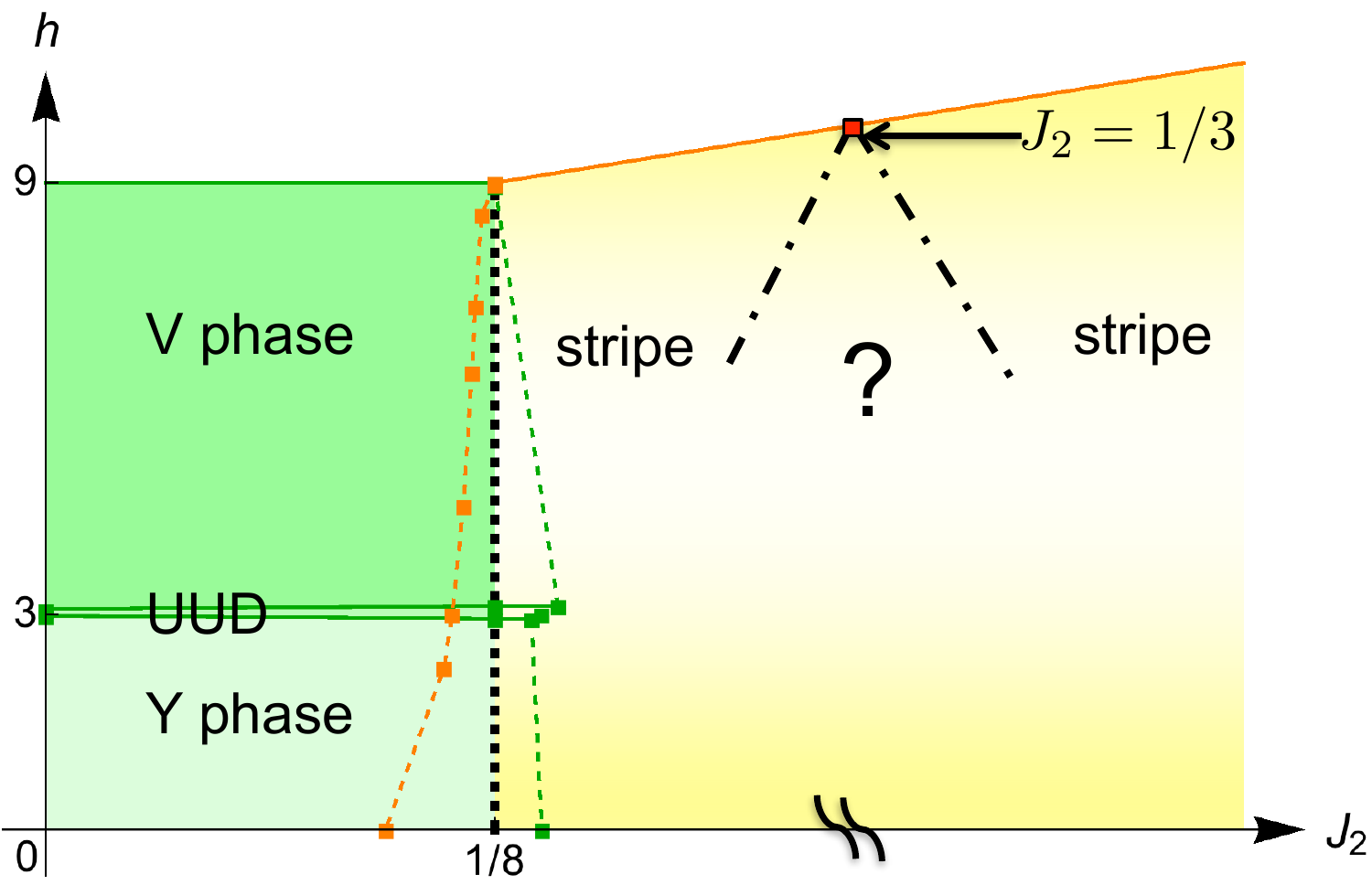}\label{fig:summarySemiClassical}}\\
\caption{(Color online) Phase diagrams of the triangular $J_1-J_2$ model in a field. Green and orange dashed lines indicate the boundaries of the stability regions of the Y(UUD)V and the stripe states (Sec.~\ref{sec:LargeSSW}). Solid lines indicate second order phase transitions. (a) Classical phase diagram for all fields (Sec.~\ref{sec:ClassicalPD}). A first order phase transition (black dashed line) seperates three and four sublattice states defined in Fig.~\ref{fig:sublattice}. There is no hysteresis. Both three sublattice and four sublattice states are infinitely degenerate. The degeneracy is accidental and is lifted by quantum fluctuations. (b) Phase diagram near $\hsa$ for arbitrary spin $S$ (Sec.~\ref{sec:HighField}) Phase boundaries of the $V$ phase and the stripe phase right below $h_{\text{sat}}$ are obtained without a simplifying assumption that $S$ is large. Dashed lines in light color (light green and light orange) interpolate between finite S and large S data. At $S=1/2$, the spin wave stability regions of the $V$ and the stripe phase don't overlap, indicating an intermediate state (Gray) in between.  The intermediate state has non-quasi-classical long-range magnetic order. (c) Semiclassical phase diagram for all fields (Sec.~\ref{sec:LargeS}). When $J_2>1/8$, stripe phase is stable in most of the phase diagram, except for the range near $J_2 =1/3$.  The instability of stripe phase near $\hsa$ around $J_2=1/3$ is discussed in Sec.~\ref{sec:specialcase}. The phase transition between the Y(UUD)V state at smaller $J_2$ and the stripe state at larger $J_2$ is first order (black dashed line) with finite hysteresis width set by the quantum fluctuations (the region between the green and orange dashed lines around $J_2=1/8$). Quantum corrections are of order $1/S$. The data points are for $S=10$.
}
\end{figure}

At small $J_2$, when the nearest neighbor exchange is dominant, the classical ground state configurations are determined by analyzing the nearest neighbor spins on sites $A,\,B,\,C$ in Fig.~\ref{fig:threesublattice}. We call it a three sublattice structure. The constraint, which minimizes the classical energy, is
\begin{align}
 \vec{S}_A+\vec{S}_B+\vec{S}_C=\frac{\vect{h}S}{3}
 \label{ch_2}
 \end{align}
At large $J_2$, the classical ground state configuration is determined by analyzing four spins on sites $A,\,B,\,C,\,D$ in Fig.~\ref{fig:foursublattice}. We call it a four sublattice structure. The constraint, that minimizes the classical energy, is~\cite{Jolicoeur1990}
\begin{align}
\vec{S}_A+\vec{S}_B+\vec{S}_C+\vec{S}_D=\frac{\vect{h}S}{2(1+J_2)}
\label{ch_3}
\end{align}
The value of $J_2$, at which the classical ground states of the three-sublattice and the four-sublattice structures are degenerate in energy, is $\Jcri=1/8$, independent of the field. The critical field, up to which the transverse order exists, is
\begin{equation} \label{ch_1}
 h_{\text{sat}} =
 \begin{cases}
9 & \text{at $J_2 < 1/8$}  \\
8\left(1+J_2\right)  &\text{at $J_2 > 1/8$}
 \end{cases}
 \end{equation}
We show the classical phase diagram of the model in Fig.~\ref{fig:summaryClassical}.

\begin{figure}[htb!]
\centering
\begin{minipage}[b]{.44\textwidth}
\subfigure[\qquad\qquad\qquad]{\includegraphics[scale=0.9]{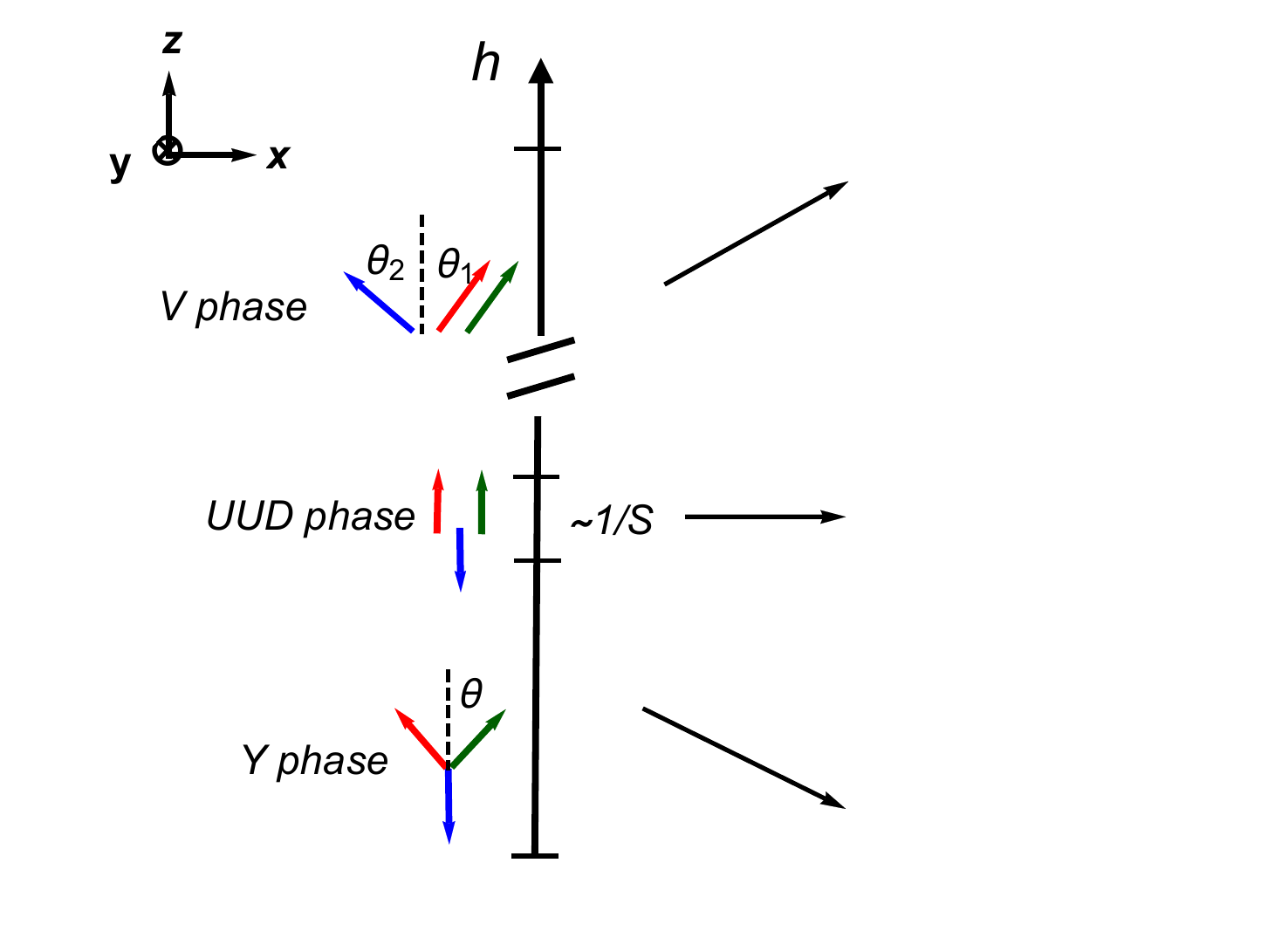}\label{fig:J20PD}}
\hspace{3in}
\end{minipage}
\qquad\quad
\begin{minipage}[b]{.47\textwidth}
\centering
\subfigure[]{\includegraphics[scale=0.6]{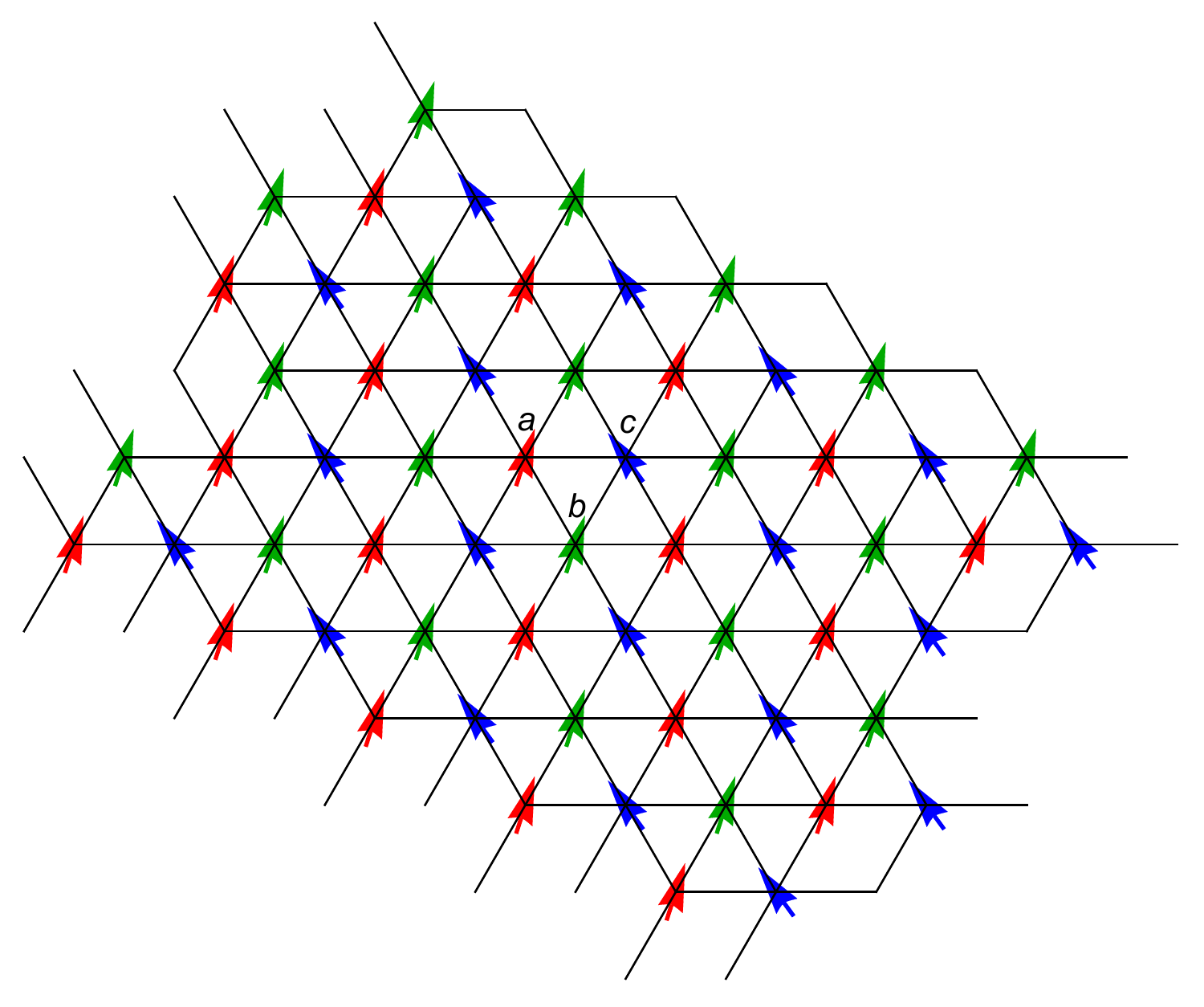}\label{fig:latticeV}}\newline
\subfigure[]{\includegraphics[scale=0.6]{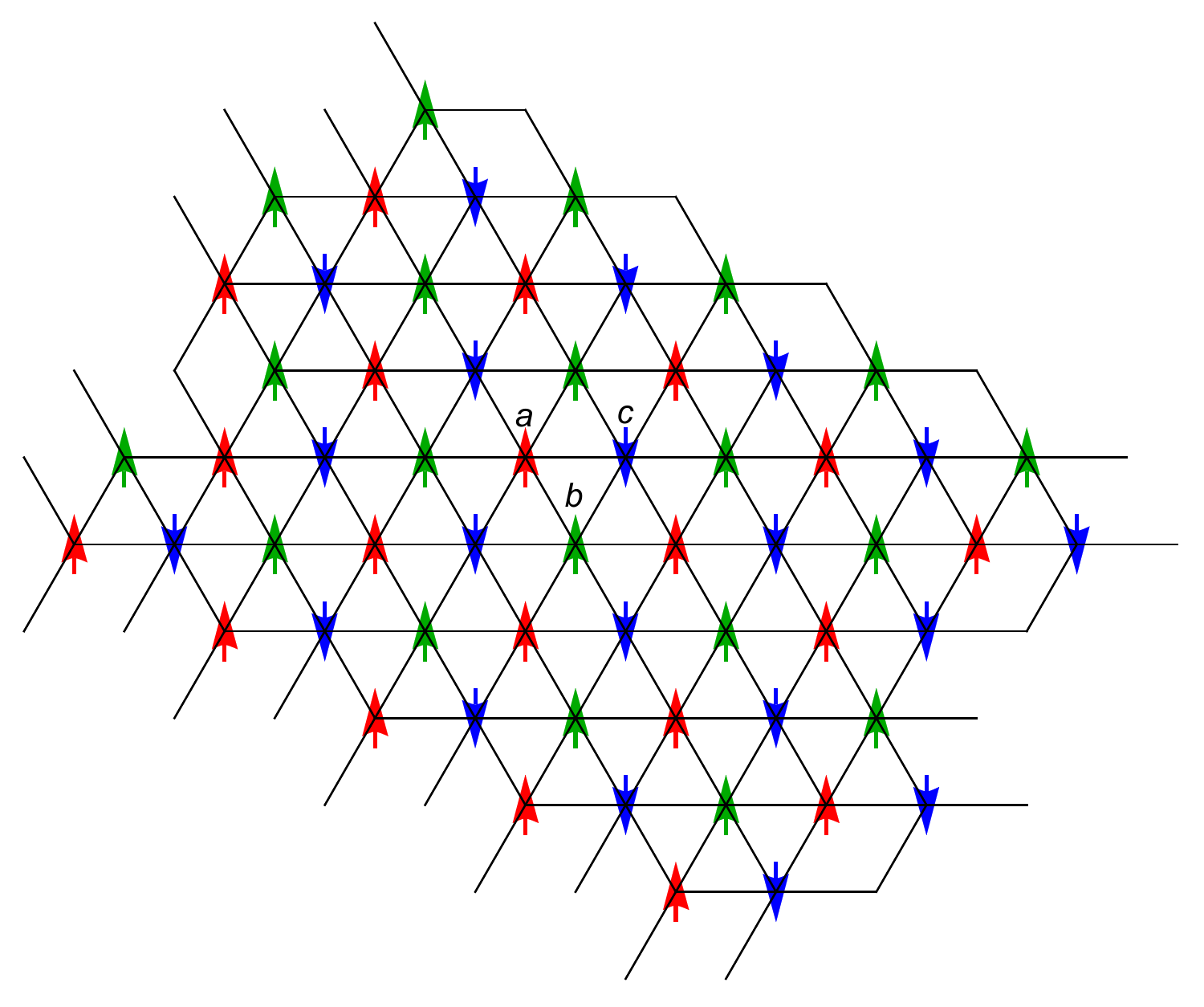}\label{fig:latticeUUD}}\newline
\subfigure[]{\includegraphics[scale=0.6]{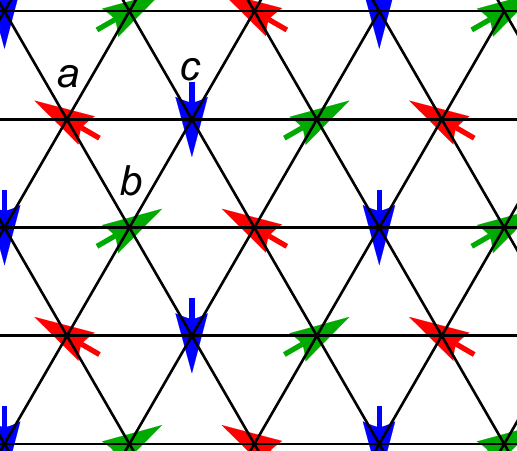}\label{fig:latticeY}}\newline
\end{minipage}
\caption{(a) Quantum phase diagram of the nearest neighbor model in a field. The magnetic field is along the $\vect{z}$ direction in our convention. (b)-(d): Spin order in the real space. (b) coplanar structure with V-type order; (c) collinear structure with up-up-down (UUD) order; (d) coplanar structure with Y-type order. The letters $a,\,b,\,c$ label the sublattices. \label{fig:orderstructure}}
\end{figure}

\begin{figure}[htb!]
\centering
\includegraphics[height=1.4in]{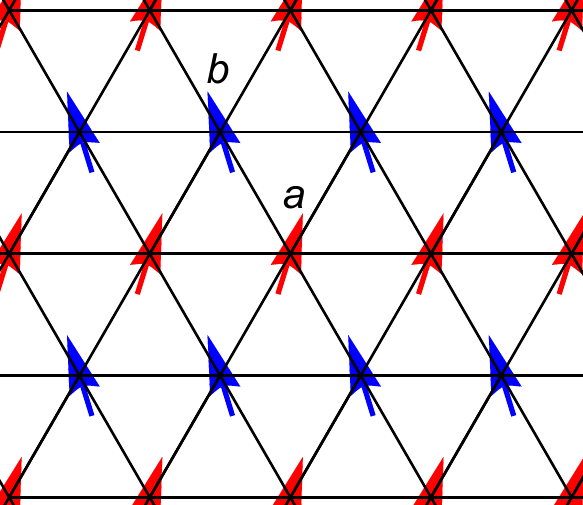}
\caption{Spin order in the canted stripe phase.
\label{fig:latticeStripe}}
\end{figure}

The classical ground state manifolds at both small and large $J_2$ are highly degenerate for any non-zero $h$ (and also at $h=0$ when $J_2>1/8$). We found that quantum fluctuations lift these degeneracies and select the ordered states via the order from disorder mechanism. We analyzed this mechanism at $S\gg 1$ in all fields from $h=0^+$ up to the saturation value. At $J_2 < 1/8$, we found the same sequence of coplanar ordered states as when $J_2=0$ (Fig.~\ref{fig:J20PD}). Namely, at $h=0$ the quantum ground state is the $120^{\circ}$ Neel state, at $0<h< h_{\text{sat}}/3$ the ordered phase has a $Y$ shape (Fig.~\ref{fig:latticeY}), at $h\approx h_{\text{sat}}/3$ it is the collinear UUD phase (Fig.~\ref{fig:latticeUUD}), and at $h_{\text{sat}}/3 < h< h_{\text{sat}}$ the ordered state has a $V$ shape (Fig.~\ref{fig:latticeV}). Below we label the set of $Y$, UUD, and $V$ states as Y(UUD)V state. In the four sublattice regime, quantum fluctuations stabilize the phase, in which the spins remain stripe ordered along a spontaneously chosen direction transverse to the field and gradually develop longitudinal ferromagnetic component along the field as  $h$ increases.  Below we will be calling  this phase a canted stripe phase, or, simply, a stripe phase. The canted stripe phase is stable in most of the phase diagram at $J_2 > 1/8$, except the region around $J_2 =1/3$ at large fields (see Fig.~\ref{fig:latticeStripe}), which we discuss in Sec.~\ref{sec:specialcase}. To be precise,  the word "stable" in our spin wave analysis implies that the canted stripe phase is a local minimum. Whether it is a global minimum or there exist other states that have lower energy, is to be investigated. We do not explore here a possibility that the canted phase, when stable, may be only a local minimum. Around the critical coupling $J_2=1/8$, we identified the phase boundary between the Y(UUD)V and the stripe phases in a generic field. We found that at $S \gg 1$ the stability regions of the two phases overlap at all fields, i.e., the phase transition between the Y(UUD)V and the stripe phases is \textit{first order} with a finite hysteresis width. Fig.~\ref{fig:summarySemiClassical} shows the semiclassical phase diagram of the model.

Another regime, where the controlled perturbative calculation can be achieved, is right below the critical field \hsat. The phase transition from ferromagnet to a nearly ferromagnetic spin ordered phase can be described by magnon condensation at a certain momenta, and the order structure depends on the structure of the magnon condensates. We found that for all $S$, the ferromagnet becomes unstable towards the $V$ phase when $J_2<1/8$, and towards stripe phase at $J_2>1/8$. To determine the nature of the phase transition between the $V$ and the stripe phases, we studied the spin wave spectrum of the two states and obtained the phase boundaries at arbitrary $S$. We found that the phase transition is \textit{first order} when $S\gg 1$, but for $S=1/2$ and, possibly, $S=1$, the spin-wave stability regions of the $V$ phase and the stripe phase do not overlap. In this situation, right below $h_{\text{sat}}$ there exists a state in which a spontaneous long range magnetic order likely does not develop. We note that there is apparently no such state near a saturation field in $J_1-J_2$ model on a square lattice~\cite{Mila2013,Morita2016}.

The structure of the paper is as follows: In Sec.~\ref{sec:ClassicalPD}, we briefly discuss the classical phase diagram and spin-wave spectrum without $1/S$ corrections. They can be obtained by minimizing the ground state energy and performing linear spin wave expansion. In Sec.~\ref{sec:LargeS} we obtain the semiclassical phase diagram at large but finite $S$. We first discuss, in Sec.~\ref{sec:LargeSaction}, quantum selection of the ordered phases right below \hsat, then obtain the spin-wave spectrum with $1/S$ corrections and show that the transition between the $V$ phase and the stripe phase is first-order with a finite hysteresis width. We then extend the calculations to smaller $h$ and obtain the semiclassical phase diagram in all fields. In Sec.~\ref{sec:HighField} we use dilute Bose gas approximation near \hsat$\,$to obtain the ground state configurations at small and large $J_2$, and the transition between them, at arbitrary $S$, including $S=1/2$. We present our conclusions in Sec.~\ref{sec:conclusion}.
\section{Classical phase diagram}\label{sec:ClassicalPD}
The classical analysis at $h=0$ has been presented in Ref~\cite{Jolicoeur1990}.
The key result of the classical analysis at a non-zero field is that the phase transition between the three sublattice phase and four sublattice phase remains first-order in all fields up to \hsat. Moreover, the critical coupling $\Jcri=1/8$ is independent of the field.

This last result follows from the analysis of classical ground state energies in a field. We have
 \begin{equation}
 E_{\text{cl}}=
 \begin{cases}
N_{\text{tot}}S^2\big(-\frac{3}{2}+3J_2-\frac{h^2}{18}\big) &\text{for the three-sublattice phase}\\
N_{\text{tot}}S^2\big(-1-J_2-\frac{h^2}{16(1+J_2)}\big) &\text{for the four-sublattice phase}
\end{cases}
\end{equation}
 The two energies obviously become equal at $J_2 =1/8$.

The spin-wave spectrum to order $S$ is obtained by performing a linear spin wave analysis around a spin order that satisfies the classical constraints, Eqs.~\ref{ch_2},~\ref{ch_3}). The Y(UUD)V order at small $J_2$ and the stripe order at large $J_2$ do satisfy the constraints, along with many other states. We assume and then verify that these states will be selected by quantum fluctuations and expand around them. To do this, we express spin operators in terms of the Holstein-Primakoff (HP) bosons as $S_{\vect{r}}^z =S-a_{\vect{r}}^{\dagger}a_{\vect{r}},\,S_{\vect{r}}^+ \approx\sqrt{2S}a_{\vect{r}},\,S_{\vect{r}}^- \approx\sqrt{2S}a_{\vect{r}}^{\dagger}$. The Hamiltonian in terms of bosons is ${\mathcal H} = E_{\text{cl}} + {\mathcal H}_1 + {\mathcal H}_2$, where ${\mathcal H}_1$ and ${\mathcal H}_2$ are terms linear and quadratic $a$ and $a^\dag$. The condition $\mathcal{H}^{(1)}=0$ sets the $z-$ component of the magnetization (the same result is obtained by minimizing $E_{\text{cl}}$). The quadratic part yields, after diagonalization, the spin-wave spectrum. From this spectrum one can find the boundaries of the stability of the classical phases and the classical ground state degeneracy.

From general symmetry analysis, the order parameter manifold in the Y(UUD)V and the stripe phases has $\text{U}(1)$ freedom to rotate the coplanar spin order around the field direction.  The only exception is the UUD state at $h=h_{\text{sat}}/3$. This state is collinear, with spins along $h$, and has no transverse magnetization. For all other states, $U(1)$ symmetry is spontaneously broken and there must be one Goldstone zero mode.  The fact that the constraints Eq.~\ref{ch_2} and Eq.~\ref{ch_3} are satisfied by a continuum of states implies that, at a classical level, there should be zero energy cost to transform from either the Y(UUD)V state or the stripe state into a neighboring state from the continuum. Accordingly, at the classical level, the excitation spectrum should contain additional, ``accidental'' zero modes.
\begin{figure}[tbp]
\centering
\subfigure[\qquad\quad]{\includegraphics[width=2.5in]{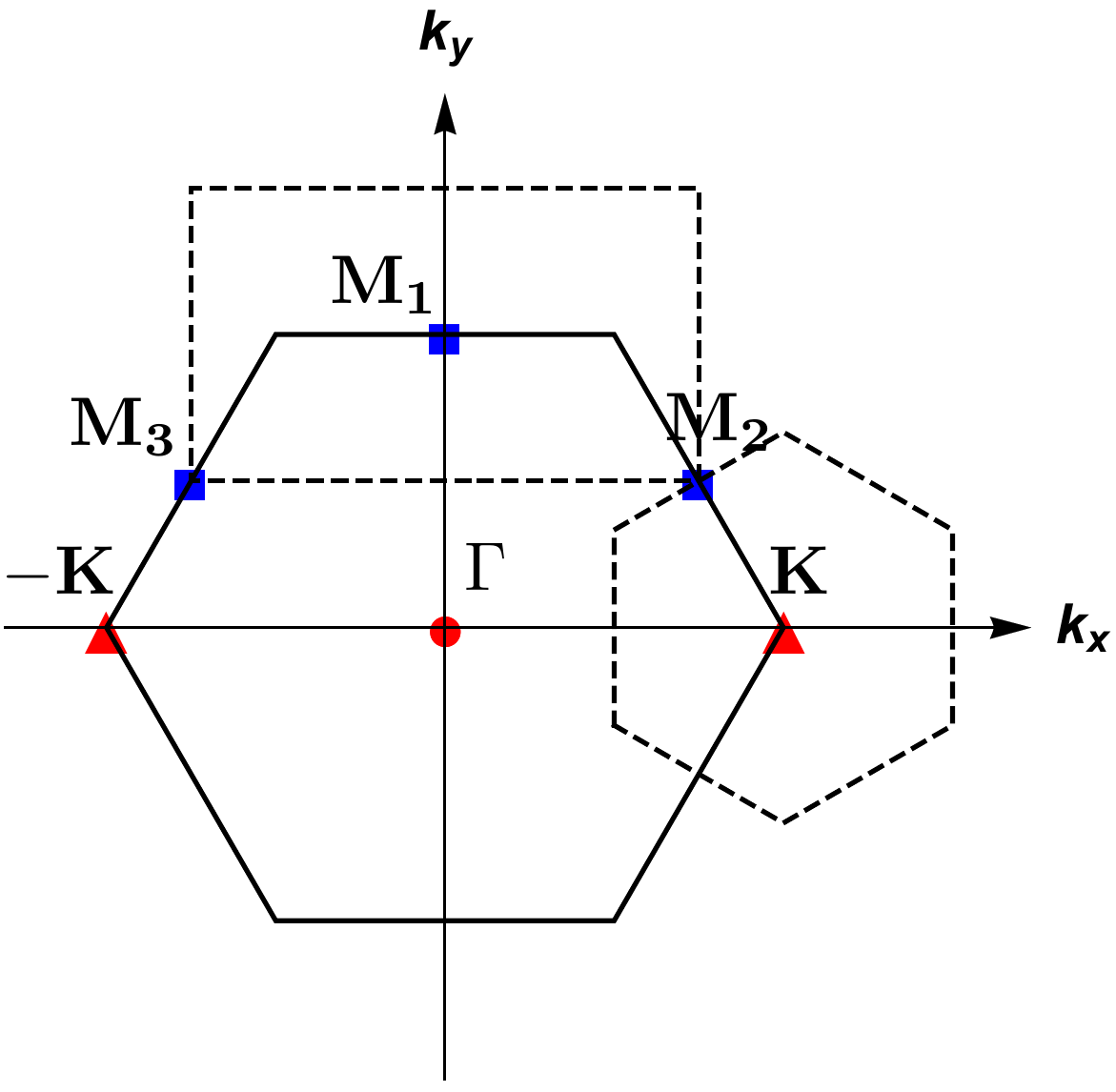}\label{fig:singleSLBZ}}
\subfigure[]{\includegraphics[width=2in]{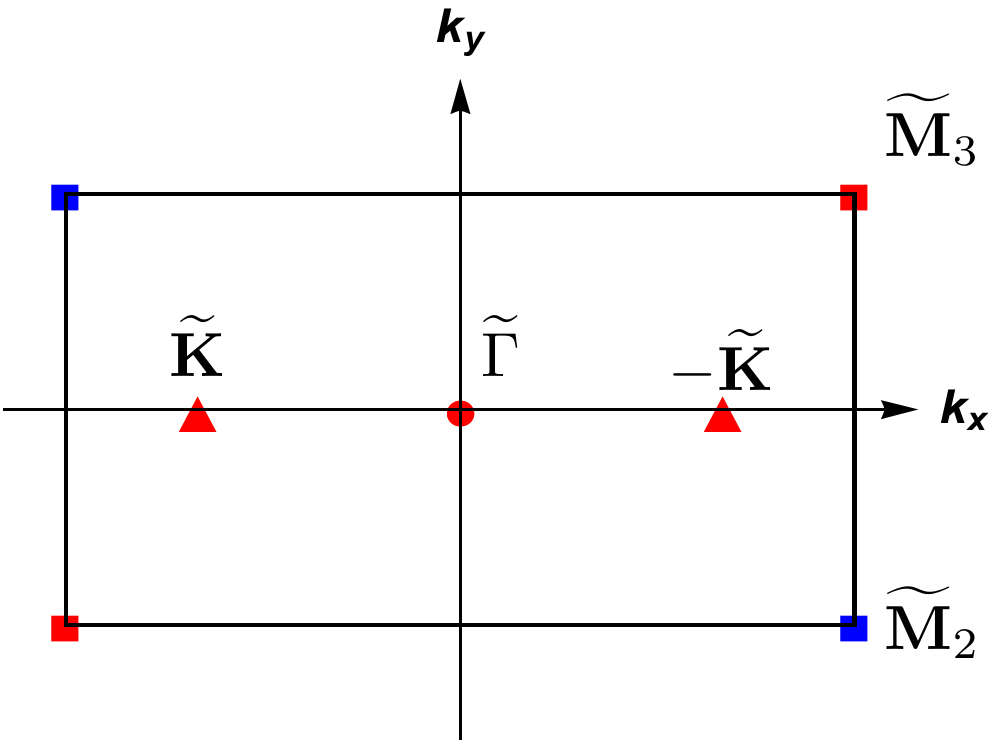}\label{fig:fourSLBZ}}~
\subfigure[]{\includegraphics[width=1.8in]{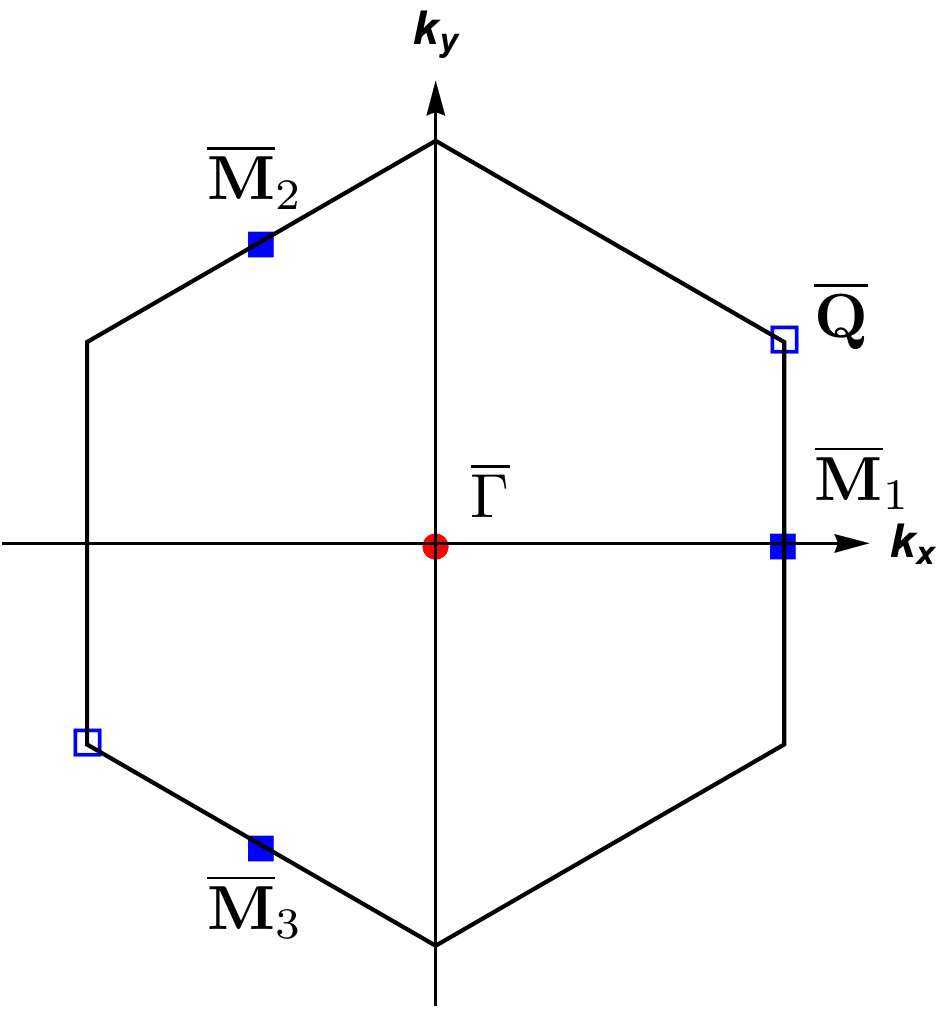}\label{fig:threeSLBZ}}
\caption{ (a) The full Brillouin zone. (b) Reduced Brillouin zone for the two-sublattice stripe phase. (c) Reduced Brillouin zone for three-sublattice states. The high symmetry points are labeled such that $\widetilde{\Gamma}$ in panel (b) coincides with $\vect{M}_1$ in panel (a), and $\overline{\Gamma}$ in panel (c) coincides with $\vect{K}$ in panel (a).}
\end{figure}
\subsection{Three sublattice phase}
To obtain the linear spin wave spectrum in the three-sublattice state, we expand the spin Hamiltonian in terms of HP bosons around $Y$, UUD, and $V$ states, shown in Fig.~\ref{fig:orderstructure}. At the classical level UUD state occurs at just one value of the field $h = h_{\text{sat}}/3$, and the classical excitation spectrum of the UUD phase is the same as those of $Y$ and $V$ phases at $h = h_{\text{sat}}/3$.

We define HP bosons on sublattices A, B, C as $a,\,b,\,c\,$. The linear term $\mc{H}^{(1)}$ in the $Y$ and $V$ phases are:
\begin{align}
\mathcal{H}_{\text{Y}}^{(1)} & =\frac{\iu \sin{\theta} S\sqrt{S}}{\sqrt{2}} \sqrt{N}\, (a_{\vect{0}}-b_{\vect{0}})(h+3-6\cos{\theta})+h.c.\\
\mathcal{H}_{\text{V}}^{(1)} & =\frac{\iu S\sqrt{S}}{\sqrt{2}} \sqrt{N}\, \big((a_{\vect{0}}+b_{\vect{0}})(\sin{\theta_1}h-3\sin{\theta_{1-2}})+c_{\vect{0}}(\sin{\theta_2}h+6\sin{\theta_{1-2}})\big)+h.c.
\end{align}
where the angles $\theta$ and $\theta_{1,2}$ are shown in Fig.~\ref{fig:J20PD}, and $\theta_{1-2}=\theta_1-\theta_2$. $a_{\vect{0}},\,b_{\vect{0}},\,c_{\vect{0}}$ are H-P bosons at $\kv=0$.

Setting $\mathcal{H}_{\text{Y}}^{(1)} = \mathcal{H}_{\text{V}}^{(1)} =0$ we obtain in the $Y$ phase
\be
\cos{\theta} = \frac{h+3}{6}
\ee
and in the $V$ phase
\be
\cos{\theta}_1 = \frac{h^2+27}{12h},~~\cos{\theta}_2 = \frac{h^2-27}{6h}
\label{eq:coplanarH}
\ee
The transverse order in the $V$ phase vanishes at $h=9$, which is the saturation field $h_{\text{sat}}$, and at $h=3$, when $Y$ phase transforms into UUD phase.
The quadratic part $\mathcal{H}^{(2)}$ for $Y$ and $V$ phases can be written in a matrix form as:
\be
\mathcal{H}^{(2)}=\frac{S}{2}\sum_{\vect{k}}\Psi_{\vect{k}}^{\dagger}H_{\vect{k}}\Psi_{\vect{k}},
\ee
where
 \be\label{eq:coplanarConvention}
 \Psi_{\vect{k}}  =(a_{\vect{k}},b_{\vect{k}},c_{\vect{k}},a^{\dagger}_{-\vect{k}},b^{\dagger}_{-\vect{k}},c^{\dagger}_{-\vect{k}})^T.
  \ee
and
\begin{equation}\label{eq:Yphase}
H_{\vect{k}}=
    \begin{pmatrix}
A_{\vect{k}} & B_{\vect{k}}\\
B_{\vect{k}} & A_{\vect{k}}
    \end{pmatrix}
  \end{equation}
$A_{\kv},\,B_{\kv}$ for the Y phase is:
\begin{equation}\label{eq:Yphase}
A_{\text{Y},\vect{k}}=
    \begin{pmatrix}
      \epsilon _{a\vect{k}} & \alpha_+ \gamma_{\vect{k}} &\beta_- \conj{\gamma}_{\vect{k}} \\
      \alpha_+ \conj{\gamma}_{\vect{k}} & \epsilon _{a\vect{k}} &\beta_-\gamma_{\vect{k}} \\
      \beta_- \gamma_{\vect{k}} & \beta_- \conj{\gamma}_{\vect{k}} & \epsilon _{b\vect{k}}
    \end{pmatrix}\qquad
    B_{\text{Y},\vect{k}}=
    \begin{pmatrix}
      0 & \alpha_- \gamma_{\vect{k}} & \beta_+ \conj{\gamma}_{\vect{k}} \\
       \alpha_- \conj{\gamma}_{\vect{k}} & 0 & \beta_+ \gamma_{\vect{k}} \\
      \beta_+ \gamma_{\vect{k}} & \beta_+ \conj{\gamma}_{\vect{k}} & 0
    \end{pmatrix}
  \end{equation}
$A_{\kv},\,B_{\kv}$ for the V phase is:
\begin{equation}\label{eq:Yphase}
A_{\text{V},\vect{k}}=
    \begin{pmatrix}
       \epsilon _{a\vect{k}} & \conj{\gamma}_{\vect{k}} & \tilde{\alpha}_+\gamma_{\vect{k}} \\
     \gamma_{\vect{k}} &  \epsilon _{a\vect{k}} &\tilde{\alpha}_+\conj{\gamma}_{\vect{k}} \\
       \tilde{\alpha}_+ \conj{\gamma}_{\vect{k}} & \tilde{\alpha}_+ \gamma_{\vect{k}} &  \epsilon _{b\vect{k}}
    \end{pmatrix}\qquad
    B_{\text{V},\vect{k}}=
    \begin{pmatrix}
       0 & 0 & \tilde{\alpha}_- \gamma_{\vect{k}} \\
      0 & 0 & \tilde{\alpha}_- \conj{\gamma}_{\vect{k}} \\
     \tilde{\alpha}_- \conj{\gamma}_{\vect{k}} & \tilde{\alpha}_- \gamma_{\vect{k}} & 0
    \end{pmatrix}
  \end{equation}
where $\conj{\gamma}_{\vect{k}}\equiv \gamma^*_{\vect{k}}$. For the $Y$ phase, we defined $\alpha_{\pm}=\frac{1\pm\cos {2 \theta}}{2}$, $\beta_{\pm}=\frac{1\pm\cos {\theta}}{2}$ and
\be
   \epsilon_{a\vect{k}} =3\,(\cos\theta-\cos 2\theta)+\cos\theta\,h+2J_2(\mu_{\kv}-3), ~~\epsilon_{b\vect{k}} =6\cos\theta-h+2J_2(\mu_{\kv}-3)
   \ee
For the $V$ phase, we defined $\tilde{\alpha}_{\pm}=\frac{1\pm\cos{\theta_{1-2}}}{2}$ and
      \be
    \epsilon_{a\vect{k}} =-3\,(1+\cos \theta_{1-2})+\cos\theta_1\,h+2J_2(\mu_{\kv}-3),~  \epsilon_{b\vect{k}} =-6\cos\theta_{1-2}+\cos\theta_2\,h+2J_2(\mu_{\kv}-3)
          \ee
For both phases
\be\label{eq:th_ch_2}
 \gamma_{\vect{k}} =(\mathrm{e}^{i\,k_x}+2\cos\frac{\sqrt{3}}{2}k_y\,\mathrm{e}^{-i\,k_x/2}), ~~ \mu_{\kv} =\cos\sqrt{3}k_y+2\cos\frac{\sqrt{3}k_y}{2}\cos{\frac{3}{2}k_x}
 \ee
 To obtain the spin wave spectrum and the canonical eigenmodes, one can either use equation of motion for operators or solve for eigenvalues of the matrix $\tau_3H_{\vect{k}}$, where $\tau_3\equiv\sigma_3\otimes I_3$~\cite{Colpa1978}. $\sigma_3$ is the z-component of the Pauli matrix that acts on the particle-hole conjugate space, and $I_3$ is the identity matrix of size three that acts on the three sublattice space (see Appendix~\ref{app:Classical}).

\begin{figure}[tbp]
\centering
\includegraphics[width=3.2in]{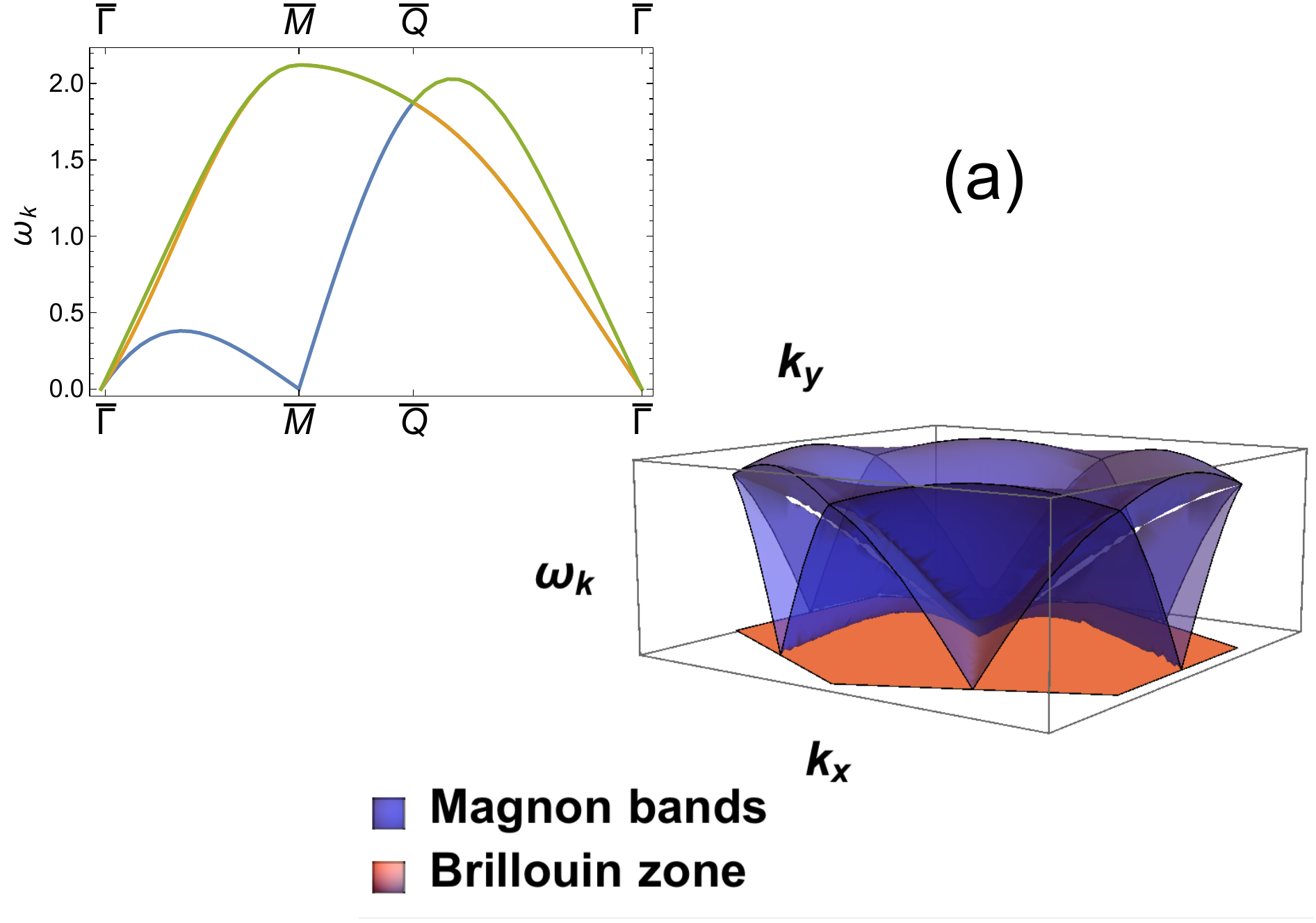}
\includegraphics[width=3.2in]{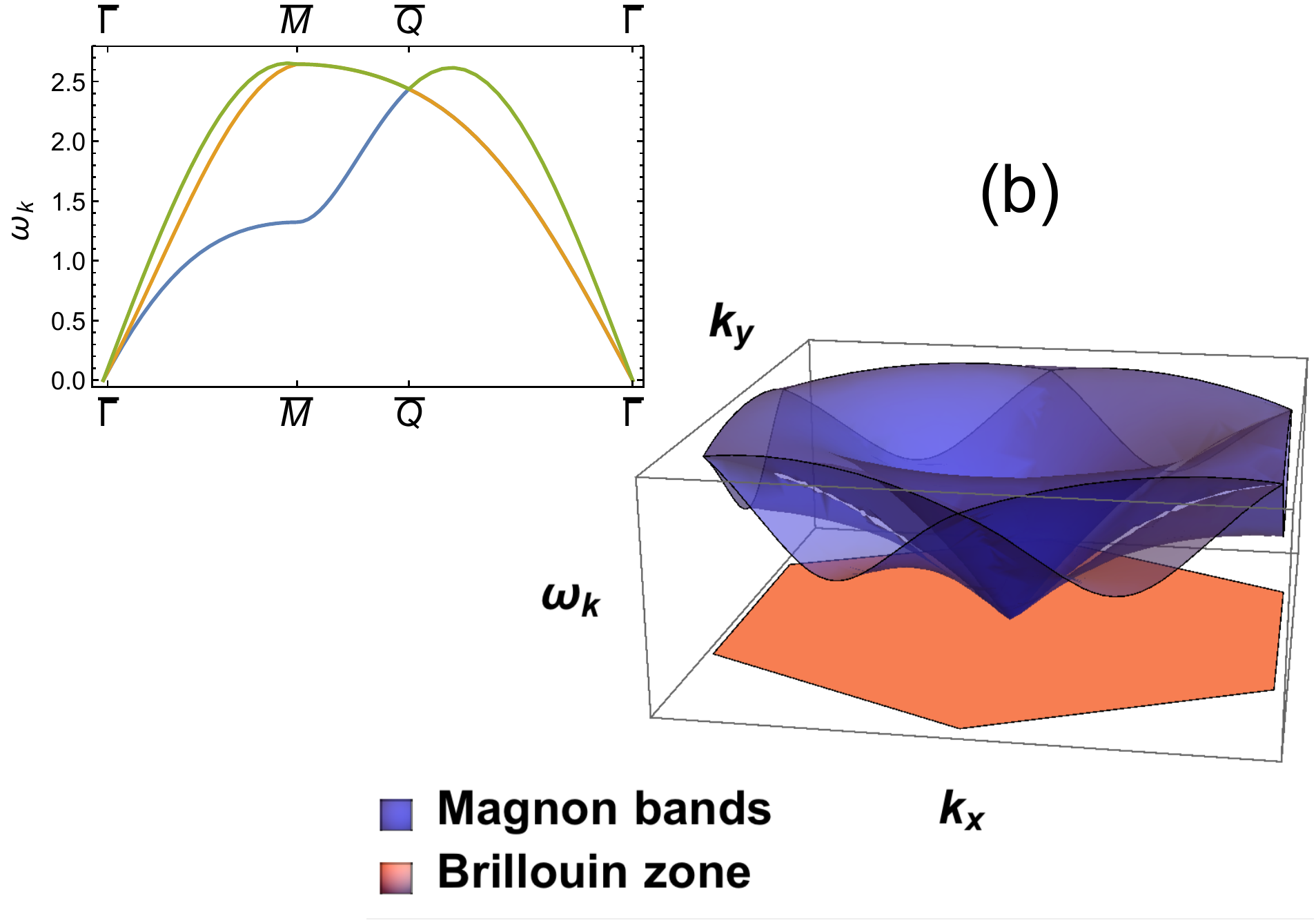}
\includegraphics[width=3.2in]{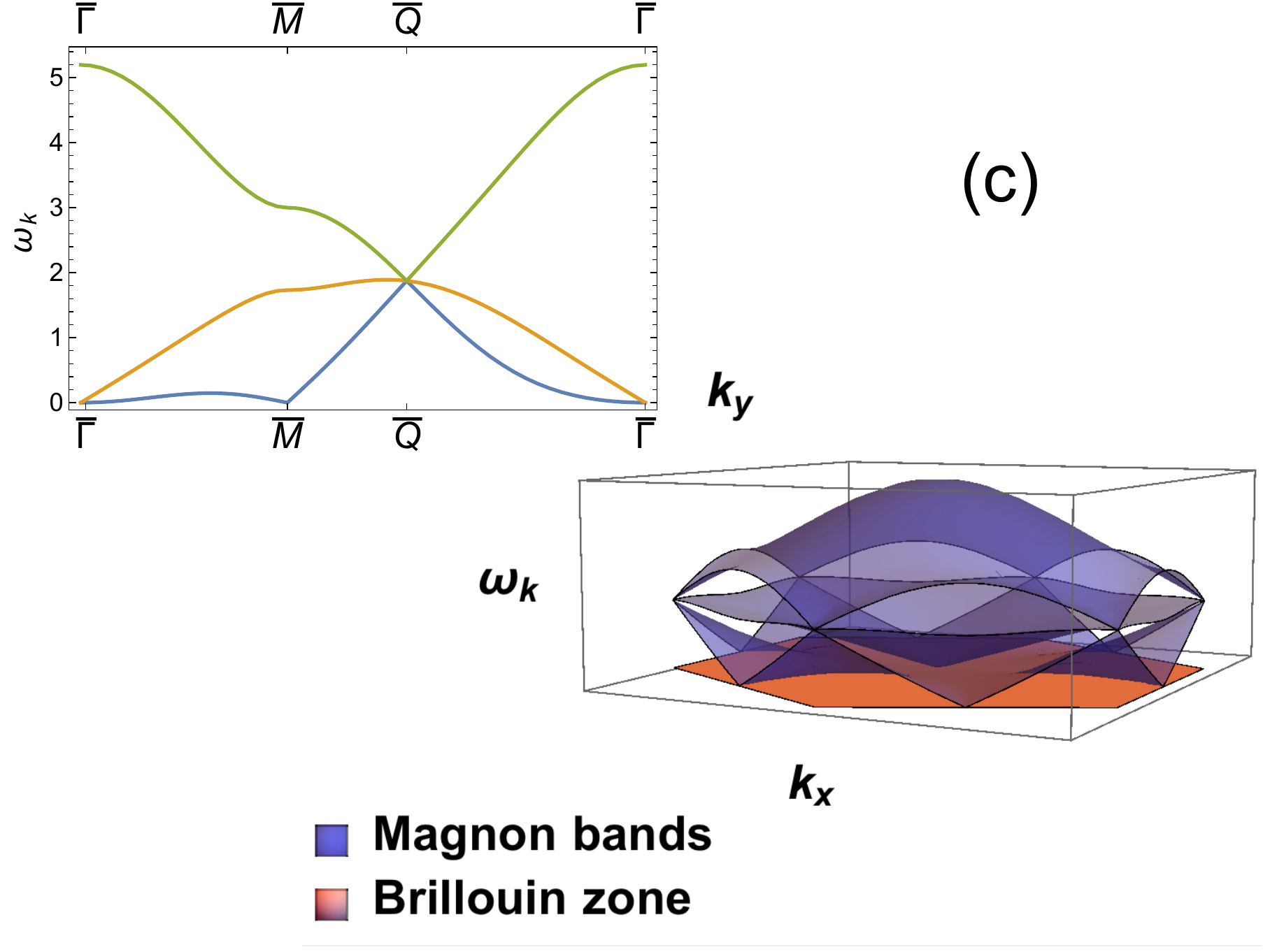}
\includegraphics[width=3.2in]{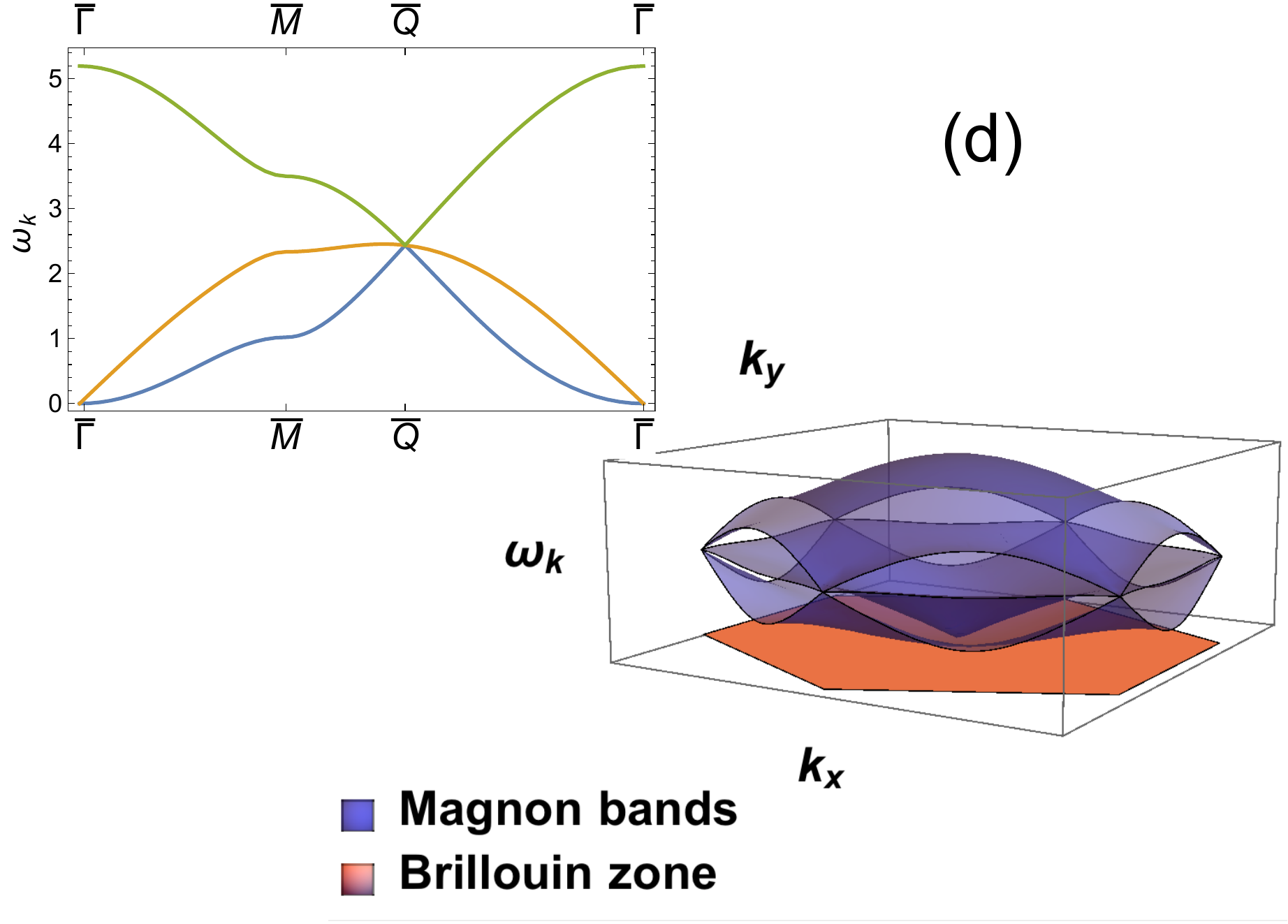}
\includegraphics[width=3.2in]{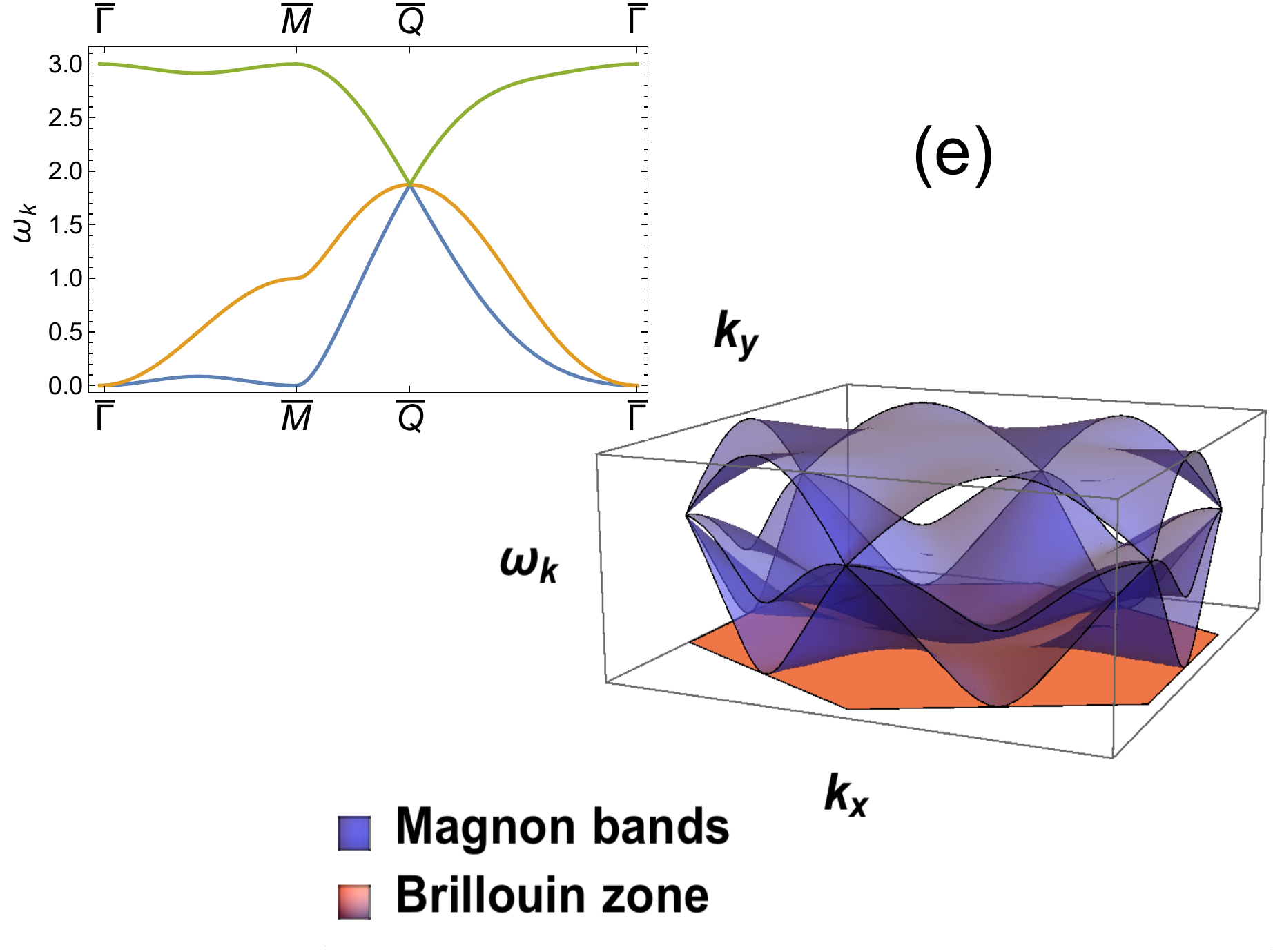}
\includegraphics[width=3.2in]{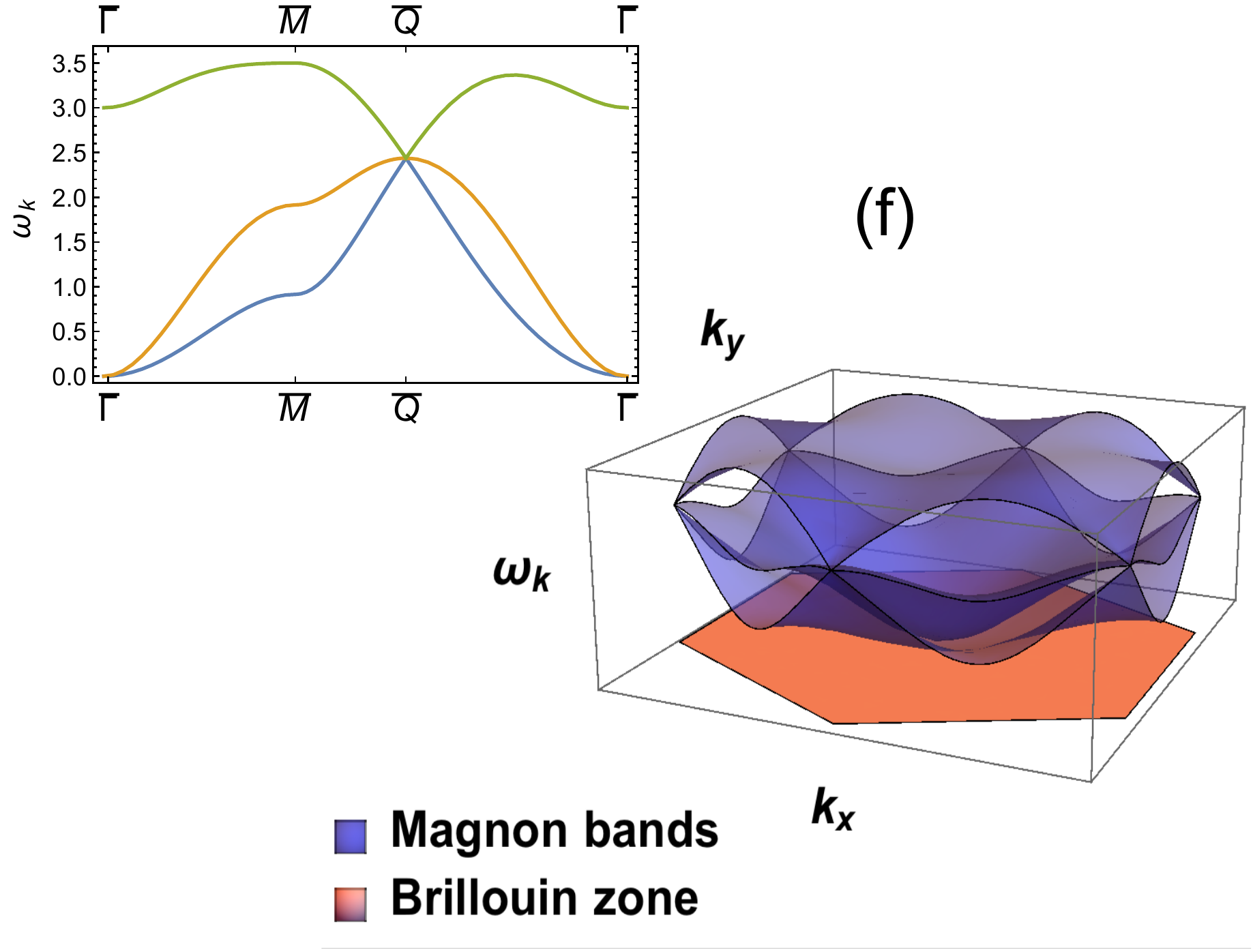}
\caption{Classical magnon spectrum of the Y(UUD)V phase at $J_2=1/8$ (left) and $J_2=1/16$ (right) in different fields. The spectrum is unit of $J_1 S$. At $J_2=1/8$, the zero modes at the $\vect{\overline M}$ points imply that this $J_2$ is the end point of stability region of the Y(UUD)V phase. (a) and (b): Spectrum in zero field. There are three linear dispersing Goldstone modes at $\overline\Gamma$; (c) and (d): Spectrum of the $V$-phase in a field $h=5.2>\hsa/3$. There is one linear dispersing Goldstone mode and one quadratic zero mode due to the accidental degeneracy; (e) and (f): Spectrum of UUD phase, there are two accidental zero modes at $\overline\Gamma$.
\label{fig:classicaldispersion2}}
\end{figure}

The magnon spectra at different fields for $J_2 =1/16$ and $J_2 =1/8$ are shown in Fig.~\ref{fig:classicaldispersion2}. The linear dispersing zero mode at the $\overline{\Gamma}$ point is a classical Goldstone mode. The zero energy modes with the quadratic dispersion are accidental zero modes associated not with $U(1)$ symmetry breaking but with the fact that an infinite set of classical grounds states satisfies the constraint Eq.~\ref{ch_2}. At $J_2 = 1/8$ additional zero modes appear at the three points along the Brillouin zone (BZ) boundary $\overline{\vect{M}}_1 = (2\pi/3,\,0),\,\overline{\vect{M}}_2 =(-\pi/3,\,\pi/\sqrt{3}),\,\overline{\vect{M}}_3 = (-\pi/3,\,-\pi/\sqrt{3})$. When $J_2>1/8$, the excitation near $\overline{\vect{M}}_1, ~\overline{\vect{M}}_2$, and $\overline{\vect{M}}_3$ becomes complex and the three-sublattice state becomes unstable.

\subsection{Four sublattice phase}
When $J_2\ge 1/8$, the configurations minimizing the classical ground state energy satisfy Eq.~\ref{ch_3}, and they are four sublattice states in general. The stripe phase is one of these classical ground state configurations. In distinction to more complex states which satisfy Eq.~\ref{ch_3}, it can be described by only two sublattices. We perform the linear spin wave analysis around the stripe phase using the same logic as in the three sublattice case.

The spin-wave operators for the HP expansion around the stripe order are defined as $a$ and $b$. The term of the Hamiltonian linear in $a$ and $b$ is
\begin{equation}
\mathcal{H}_{\text{stripe}}^{(1)} =\frac{\sin{\theta} S\sqrt{S}}{\sqrt{2}}\sqrt{N}\big((a_{\vect{0}}-b_{\vect{0}})(h-8(1+J_2)\cos{\theta})\big)+h.c.
\label{eq:stripe1}
\end{equation}
From the condition $\mathcal{H}_{\text{stripe}}^{(1)}=0$ we relate $\theta$ and $h$:
\begin{equation}
\cos \theta = \frac{h}{8(1+J_2)}
\end{equation}
At $\theta=0$, the transverse order disappears and spins are polarized along the field. The corresponding $h_{\text{sat}} = 8(1+J_2)$ is the same as in Eq.~\ref{ch_1}.

The quadratic term is
\be\label{eq:stripe2}
\mathcal{H}_{\text{stripe}}^{(2)}=\frac{S}{2}\sum_{\vect{k}}\psi_{\vect{k}}^{\dagger}H_{\text{stripe},\vect{k}}\psi_{\vect{k}}
\ee
where we introduced $\psi_{\vect{k}}=(a_{\vect{k}},b_{\vect{k}},a^{\dagger}_{-\vect{k}},b^{\dagger}_{-\vect{k}})^T$. The matrix $H_{\text{stripe},\vect{k}}$ is
\begin{equation}
H_{\text{stripe},\vect{k}}=\\
 \begin{pmatrix}
    \epsilon_{1,\vect{k}} &  \alpha \epsilon_{2,\vect{k}} & 0 & \beta \epsilon_{2,\vect{k}} \\
      \alpha \epsilon_{2,\vect{k}} & \epsilon_{1,\vect{k}} & \beta \epsilon_{2,\vect{k}} & 0 \\
      0 & \beta \epsilon_{2,\vect{k}} & \epsilon_{1,\vect{k}} & \alpha \epsilon_{2,\vect{k}} \\
      \beta \epsilon_{2,\vect{k}} & 0 & \alpha \epsilon_{2,\vect{k}} & \epsilon_{1,\vect{k}} \\
  \end{pmatrix}=
    \begin{pmatrix}
      \epsilon_{1}\mathcal{I}+\alpha \epsilon_{2}\sigma_1 & \beta \epsilon_{2}\sigma_1 \\
      \beta \epsilon_{2}\sigma_1 & \epsilon_{1}\mathcal{I}+\alpha \epsilon_{2}\sigma_1\\
     \end{pmatrix}\nonumber
\end{equation}
and
\begin{align}\label{eq:stripestrucutre}
&\alpha=(1+\cos 2\theta),\quad \beta= (-1+\cos 2 \theta) \\
& \epsilon_{1\vect{k}}=\cos \theta \,h+2\big(\zeta_{\text{a}\vect{k}}-(2\cos 2\theta+1)(1+J_2)\big),\quad
\epsilon_{2\vect{k}}=\zeta_{\text{b}\vect{k}}\non\\
&\zeta_{\text{a}\vect{k}}=\cos k_x+J_2 \cos \sqrt{3}k_y \non\\
&\zeta_{\text{b}\vect{k}}=\cos (\frac{1}{2}k_x+\frac{\sqrt{3}}{2}k_y)+\cos (\frac{1}{2}k_x-\frac{\sqrt{3}}{2}k_y)+J_2\big(\cos (\frac{3}{2}k_x+\frac{\sqrt{3}}{2}k_y)+\cos (\frac{3}{2}k_x-\frac{\sqrt{3}}{2}k_y)\big)\non
\end{align}
We solve for the eigenvalues of the matrix $\tau_3H_{\text{stripe},\vect{k}}$ to obtain the spin wave spectrum. The matrix can be diagonalized analytically. We first introduce new magnon operators ${c}_{\kv},\,{d}_{\kv}$ by
\begin{align}\label{eq:stripedecouple}
\begin{pmatrix}
{ c}_{\vect{k}}\\
{ d}_{\vect{k}}\\
\end{pmatrix}=
\begin{pmatrix}
1/\sqrt{2} & -1/\sqrt{2}\\
1/\sqrt{2} & 1/\sqrt{2}\\
\end{pmatrix}
\begin{pmatrix}
a_{\vect{k}}\\
b_{\vect{k}}\\
\end{pmatrix}
\end{align}
In terms of new
\be
  \phi_{\vect{k}}=({ c}_{\vect{k}},{ d}_{\vect{k}},{ c}^{\dagger}_{-\vect{k}},{ d}^{\dagger}_{-\vect{k}})
\ee
we have
\be
\mathcal{H}_{\text{stripe}}^{(2)}=\frac{S}{2}\sum_{\vect{k}}\phi_{\vect{k}}^{\dagger}H_{\text{stripe},\vect{k}}\phi_{\vect{k}}\non
\ee
where now
\begin{equation}
H_{\text{stripe},\vect{k}}=\\
  \begin{pmatrix}
      \epsilon_{1,\vect{k}}-\alpha \epsilon_{2,\vect{k}} &  0 & -\beta \epsilon_{2,\vect{k}} & 0 \\
      0 & \epsilon_{1,\vect{k}}+\alpha \epsilon_{2,\vect{k}} & 0 & \beta \epsilon_{2,\vect{k}} \\
      -\beta \epsilon_{2,\vect{k}} & 0 & \epsilon_{1,\vect{k}}-\alpha \epsilon_{2,\vect{k}} & 0 \\
      0 & \beta \epsilon_{2,\vect{k}} & 0 & \epsilon_{1,\vect{k}}+\alpha \epsilon_{2,\vect{k}} \\
  \end{pmatrix}
\end{equation}
 decouples between ${ c}$ and ${ d}$.

\begin{figure}[tbp]
\centering
\subfigure[]{\includegraphics[scale=0.5]{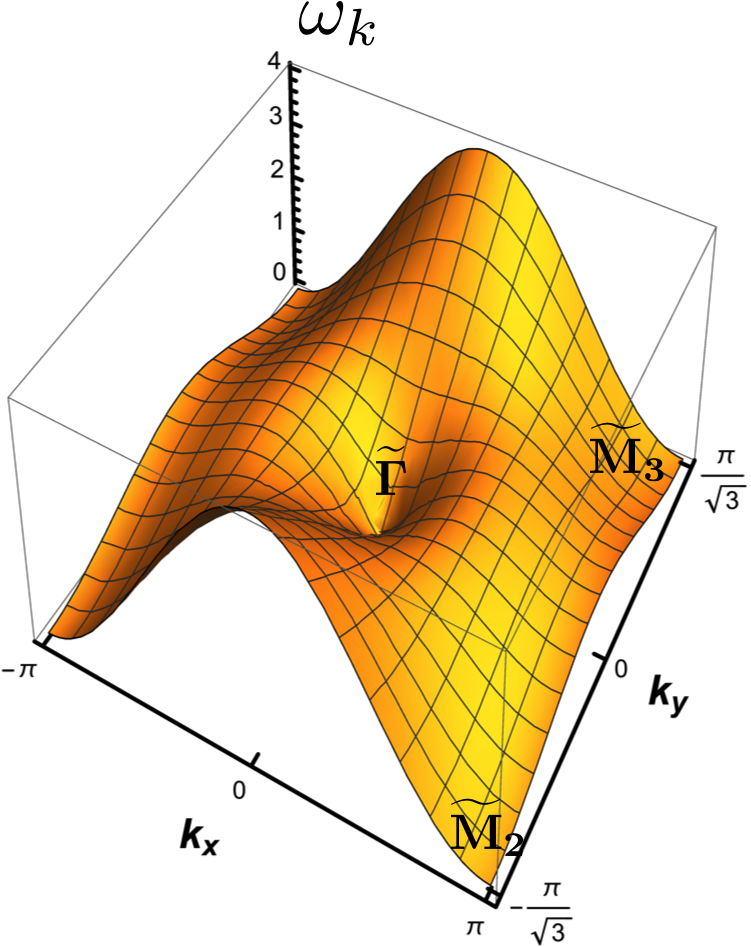}}\qquad
\subfigure[]{\includegraphics[scale=0.5]{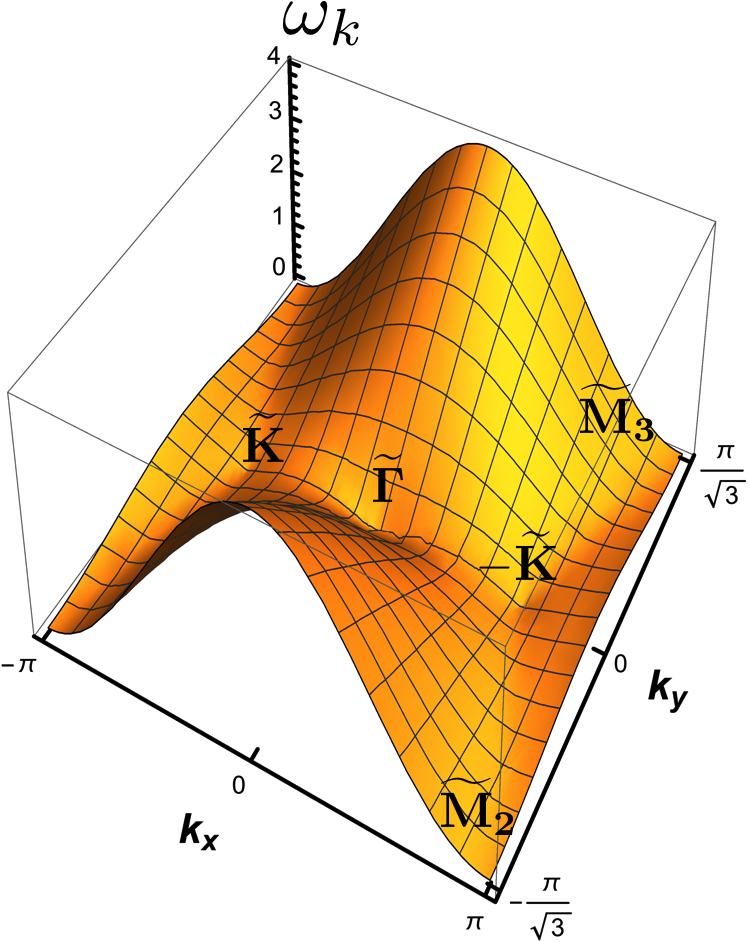}}
\caption{Classical spectrum of the stripe phase in a field $h=5.7(1+J_2)$, for different $J_2$. The spectrum is in unit of $J_1 S$. Only the low energy branch is shown. (a) $J_2=1/4$. (b) $J_2=1/8$. The Goldstone mode is at the $\widetilde\Gamma$ point, the quadratic zero modes at $\vect{\widetilde M}_2$ and $\vect{\widetilde M}_3$ are due to accidental degeneracy. At $J_2=1/8$, there are two additional zero modes at $\pm\vect{\widetilde K}=(\mp\frac{2\pi}{3},0)$ indicating that $J_2=1/8$ is the boundary of the stability region of the stripe phase.
\label{fig:stripedispersion}
}
\end{figure}

We then diagonalize separately $2 \times 2$ matrices for ${ c}_{\kv}$ and ${ d}_{\kv}$ modes using the standard Bogoliubov transformation. We obtain
 \begin{align}
\mathcal{H}_{\text{stripe}}^{(2)}=\sum_{\vect{k}} S\big(\omega_{\text{c},\kv} {\tilde c}^{\dagger}_{\kv} {\tilde c}_{\kv}+\omega_{\text{d},\kv} {\tilde d}^{\dagger}_{\kv} {\tilde d}_{\kv}\big)\\
\omega_{\text{c/d},\kv}=\sqrt{ \epsilon_{1,\vect{k}}^2\mp2\alpha \epsilon_{1,\vect{k}}\epsilon_{2,\vect{k}}+(\alpha^2-\beta^2) \epsilon_{2,\vect{k}}^2}\non
\end{align}
The spectrum of the $\tilde c$-mode at $J_2=1/4$ and $J_2=1/8$ is shown in Fig.~\ref{fig:stripedispersion}. The linear dispersing zero mode at the $\widetilde{\Gamma}$ point is the Goldstone mode associated with the U(1) symmetry breaking of the stripe order. The two quadratic zero mode at the corners of the BZ, at $\vect{\widetilde M}_2 = (\pi,\,-\pi/\sqrt{3}),$ and $\vect{\widetilde M}_3 = (\pi,\,\pi/\sqrt{3})$, are due to classical accidental degeneracy. At $J_2=1/8$, there are two additional linear dispersing zero modes at $\vect{\widetilde K}=(-2\pi/3,\,0)$ and $-\vect{\widetilde K}$. The spin-wave spectrum becomes complex at the $\pm\vect{\widetilde K}$ points when $J_2<1/8$, indicating the instability of the four sublattice phase. The $\tilde d$-mode is gapped, but it also has accidental zero modes at $\vect{\widetilde M}_2$ and $\vect{\widetilde M}_3$, where its dispersion coincides with that of the $\tilde c-$mode.

To summarize this section, the three-sublattice state is stable for $J_2 < 1/8$, the four-sublattice state is stable for $J_2 > 1/8$. At $J_2 = 1/8$ the system undergoes a first-order transition between the two phases. The critical coupling $\Jcri =1/8$ is independent on $h$. The transition has no hysteresis, i.e., each phase becomes unstable right at the transition point (see Fig.~\ref{fig:summaryClassical}). The excitation spectrum in each phase contains true Goldstone zero modes and additional zero modes associated with the fact that in each case the constraint on the ground state configuration is satisfied by infinite number of classical states. The three-sublattice Y(UUD)V state and the two-sublattice stripe state belong to these classes.
\section{Phase diagram at large spin $S$}\label{sec:LargeS}
\subsection{High field phase diagram}\label{sec:LargeSaction}
We first study the phase diagram at large S right below the saturation field \hsat. In the fully polarized state at $h > h_{\text{sat}}$ an exact elementary excitation is a gapped magnon with spin quantum number $S_z=1$. The magnon excitation gap decreases as the field reduces and vanishes at $h = h_{\text{sat}}$. A magnon condensation below $h_{\text{sat}}$ leads to transverse magnetic order, whose structure can be identified by analyzing condensate fields. Below we derive an effective Ginzburg-Landau functional to describe magnon condensates near $h_{\text{sat}}$ and show that for $J_2<1/8$ quantum fluctuations, acting at order $1/S$, select the same $V$ phase as when $J_2=0$ (Fig.~\ref{fig:latticeV}), and for $J_2>1/8$, these fluctuations select the stripe phase (Fig.~\ref{fig:latticeStripe}).

\begin{figure}[tbp]
\centering
\includegraphics[scale=0.5]{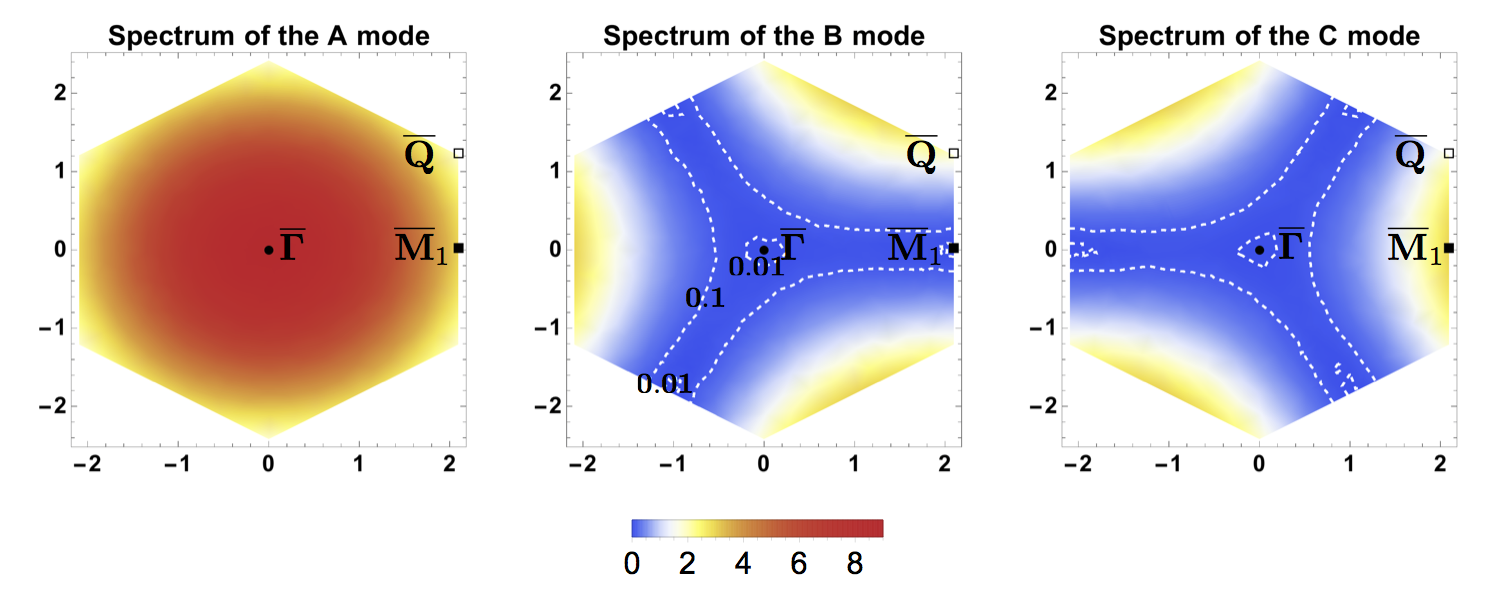}
\caption{(Color online) The magnon spectrum in three sublattice representation at $J_2=1/8$, and $h=\hsa$. In the regions between the dashed lines
 the magnon energy is small and the dispersion is almost flat.
\label{fig:ABCspectrum}}
\end{figure}

The quadratic part of the spin wave Hamiltonian at $h > h_{\text{sat}}$ is
\begin{align}
\label{eq:QuadraticHF}
 \mathcal{H}^{(2)} &= \sum_{\vect{k}\in B.Z.}\,(S\omega_{\vect{k}}-\mu)\dboson{k}\boson{k}\non\\
 \omega_{\vect{k}} &= J_{\vect{k}}-J_{\vect{Q}_{\text{min}}}\qquad \mu = S(J_0-J_{\vect{Q}_{\text{min}}})-S h=S(h_{\text{sat}}-h)\non\\
 J_{\vect{k}}& = \sum_{\pm\vect{\delta}_i}\mathrm{e}^{\pm i\vect{k}\cdot\vect{\delta}_i}+J_2\sum_{\pm\vect{l}_i}\mathrm{e}^{\pm i\vect{k}\cdot\vect{l}_i}\non\\
&= 2(\cos k_x+2\cos \frac{k_x}{2} \cos \frac{\sqrt{3}k_y}{2})+2J_2(\cos \sqrt{3}k_y+2\cos\frac{\sqrt{3}k_y}{2}\cos \frac{3k_x}{2})
\end{align}
The interaction terms (the ones we will need below) are, keeping corrections from normal ordering at order $1/S$,
\begin{align}
\mathcal{H}^{(4)}& = \frac{1}{2N}\sum_{\kk,\vect{q}\in B.Z.}V_{\vect{q}}(\vect{k_1},\vect{k_2})\dboson{k_1+q}\dboson{k_2-q}\boson{k_2}\boson{k_1}\\
\mathcal{H}^{(6)}& = \frac{1}{16SN^2}\sum_{\kk,\kv_3,\vect{q},\qv'\in B.Z.}U_{\vect{q},\qv'}(\vect{k_1},\vect{k_2},\kv_3)\dboson{k_1+q+q'}\dboson{k_2-q}\dboson{k_3-q'}\boson{k_3}\boson{k_2}\boson{k_1}
\end{align}
where
\begin{align}\label{eq:ptvertexexp}
V_{\vect{q}}(\vect{k_1},\vect{k_2}) =& \frac{1}{2}[(J_{\vect{q}}+J_{\vect{k_2}-\vect{k_1}-\vect{q}})-\frac{1}{2}(1+\frac{1}{8S})(J_{\vect{k_1}}+J_{\vect{k_1+q}}+J_{\vect{k_2}}+J_{\vect{k_2-q}})]\\
 U_{\vect{q},\qv'}(\vect{k_1},\vect{k_2},\kv_3) =& \frac{1}{9}(1+\frac{1}{4S})\big(J_{\kv_1+\qv}+J_{\kv_3+\qv}+J_{\kv_1+\kv_3-\kv_2+\qv}+J_{\kv_1+\qv'}+J_{\kv_2+\qv'}+J_{\kv_1+\kv_2-\kv_3+\qv'}\non\\
&+J_{\kv_2+\kv_3-\kv_1-\qv-\qv'}+J_{\kv_2-\qv-\qv'}+J_{\kv_3-\qv-\qv'}\big)\non\\
&-\frac{1}{6}(1+\frac{3}{4S})(J_{\vect{k_1}}+J_{\vect{k_2}}+J_{\kv_3}+J_{\kv_1+\qv+\qv'}+J_{\kv_2-\qv}+J_{\kv_3-\qv'})]
\end{align}
Lowering the magnetic field below \hsat$\,$ makes the quadratic spectrum negative in some momentum range and drives Bose-Einstein condensation of magnons at the minima of the dispersion. At $J_2<1/8$, the minima are at $\pm\vect{K}=\pm (4\pi/3,\,0)$; at $J_2>1/8$, the minima are at $\vect{M}_1=(0,2\pi/\sqrt{3})$, $\vect{M}_2=(\pi,\pi/\sqrt{3})$, $\vect{M}_3=(-\pi,\pi/\sqrt{3})$; at $J_2=1/8$, the minima are at all the five momenta $\vect{K}$, $-\vect{K}$, $\vect{M}_1$, $\vect{M}_2$ and $\vect{M}_3$ (see Fig.~\ref{fig:ABCspectrum}). At even larger $J_2>1$, which we will not discuss here, the magnon condensates are at incommensurate momenta. The magnon operator in the condensate background can be written as
\begin{equation}
\begin{cases}
\boson{k}=\sqrt{N}\Delta_1\delta_{\vect{k},\vect{K}}+\sqrt{N}\Delta_2\delta_{\vect{k},-\vect{K}}+\tilde{a}_{\vect{k}} & J_2<1/8\\
\boson{k}=\sqrt{N}\Phi_1\delta_{\vect{k},\vect{M}_1}+\sqrt{N}\Phi_2\delta_{\vect{k},\vect{M}_2}+\sqrt{N}\Phi_3\delta_{\vect{k},\vect{M}_3}+\tilde{a}_{\vect{k}} & J_2>1/8
\end{cases}
\end{equation}
 When  $J_2<1/8$, the ground state energy in terms of the uniform condensate fields $\Delta$ is:
\begin{align}\label{eq:Eden10}
E_{\Delta}/N=
-\mu(|\Delta_1|^2+|\Delta_2|^2)+\frac{1}{2}\Gamma_1(|\Delta_1|^4+|\Delta_2|^4)+\Gamma_2\,|\Delta_1|^2|\Delta_2|^2+\Gamma_\text{u}(\bar{\Delta}_1^3\Delta_2^3+h.c.)
\end{align}
When when $J_2>1/8$, the ground state energy in terms of the uniform condensate fields $\Phi$ is:
\begin{align}\label{eq:Eden11}
E_{\Phi}/N=&
-\bar{\mu}(|\Phi_1|^2+|\Phi_2|^2+|\Phi_3|^2)+\frac{1}{2}\bar{\Gamma}_1(|\Phi_1|^4+|\Phi_2|^4+|\Phi_3|^4)+
\bar{\Gamma}_2(|\Phi_1|^2|\Phi_2|^2+|\Phi_1|^2|\Phi_3|^2\non\\&+|\Phi_2|^2|\Phi_3|^2)+
\bar{\Gamma}_\text{u}(\bar{\Phi}_1^2\Phi_2^2+\bar{\Phi}_2^2\Phi_3^2+\bar{\Phi}_3^2\Phi_1^2+h.c.)
\end{align}
where $\mu,~\bar{\mu} \sim S(h_{\text{sat}}-h)$. The selection of the condensates depends on the values of the quartic coefficients $\Gamma_i$ and ${\bar \Gamma}_i$ ($i=1,2$), which are determined from the analysis of the four-point vertex function. $\Gamma_{\text{u}}$ and $\bar{\Gamma}_{\text{u}}$ are from the Umklapp process. As we will see, $\Gamma_{\text{u}}$ term determines the relative phase between $\Delta_1$ and $\Delta_2$. Similarly, $\bar{\Gamma}_{\text{u}}$ term determines relative phases between  ${\Phi}_i$.

 In the classical $S \to \infty$ limit, $\Gamma_1^{(0)}=\Gamma_2^{(0)}=9$, and $\Gamma_{\text{u}}^{(1)}=0$ when $J_2<1/8$; $\bar{\Gamma}_1^{(0)}=\bar{\Gamma}_2^{(0)}=8(1+J_2)$, and $\bar{\Gamma}_{\text{u}}^{(0)}=0$  when $J_2>1/8$. The superscript $(i)$ labels the order of perturbative expansion in power of $1/S$. The minimization of the energy then yields $|\Delta|_1^2+|\Delta|_2^2\equiv \mu/\Gamma_1$ when $J_2<1/8$, and $|\Phi|_1^2+|\Phi|_2^2+|\Phi|_3^2\equiv \mu/\bar{\Gamma}_1$ when $J_2>1/8$. Neither of these conditions specifies the ratio of $|\Delta_1|/|\Delta_2|$ or $|\Phi_2|/|\Phi_1|$ and $|\Phi_3|/|\Phi_1|$. In other words,  in the $S\to\infty$ limit, the condensed phases retain the accidental degeneracy.

When quantum fluctuations of order $1/S$ are included, $\Gamma_i$ and ${\bar \Gamma}_i$ acquire additional contributions, which are not necessarily equal for different $\Gamma_i$. We evaluated these contributions following the computation steps in Ref. ~\cite{Chubukov2014}, where a similar problem has been considered. Because our calculations parallel the ones in ~\cite{Chubukov2014}, we don't show the details of the derivations and just present the results. For $J_2<1/8$, $\Delta\Gamma=\Gamma_2-\Gamma_1=\Delta\Gamma^{(0)}+\Delta\Gamma^{(1)}=\Delta\Gamma^{(1)}$ is:
\be
\Delta\Gamma=\frac{1}{SN}\sum_{\kv} \big(\frac{V^2_{\kv}(\vect{K},\,\vect{K})}{\omega_{\vect{K}+\kv}+\omega_{\vect{K}-\kv}} -\frac{2\,V^2_{\kv}(\vect{K},\, -\vect{K})}{\omega_{\vect{K}+\kv}+\omega_{-\vect{K}-\kv}} \big)+\frac{3-2J_2}{8S}
\label{tu_ch_1}
\ee
In the thermodynamic limit, $\frac{1}{N}\sum_{\kv}\rightarrow \frac{1}{\mc{A}_{B.Z.}}\int_{\kv}$. The first term in $\Delta \Gamma$ is the second-order perturbation contribution from $\Delta^2_{1,2}\, \ad \ad+h.c.$ and $\Delta_{1}\Delta_2 \,\ad \ad+ h.c.$ terms in the Hamiltonian, the second term comes from the corrections to the quartic vertex associated with normal ordering of boson operators. Each of the two integrals in Eq.~\ref{tu_ch_1} is logarithmically divergent as the denominator in each integrand behaves as $k^2$ at small $k$. The difference between the two terms is, however, finite. We evaluated $\Delta \Gamma$ at different $J_2$ numerically and found that $\Delta\Gamma<0$ for all $J_2<1/8$, i.e., $\Gamma_1 > \Gamma_2$. An elementary analysis then shows that it is energetically favorable for the system to develop both condensates $\Delta_1$ and $\Delta_2$ with equal amplitudes $\rho=\mu/(\Gamma_1+\Gamma_2)$. To understand the structure of such an order in real space we set $\Delta_1=\sqrt{\rho}\mathrm{e}^{i\theta_1}$, $\Delta_2=\sqrt{\rho}\mathrm{e}^{i\theta_2}$, and define $\phi=(\theta_1+\theta_2)/2$ and $\psi=(\theta_1-\theta_2)/2$. The magnetic order $\la \vect{S}_{\vect{r}} \ra$ is then
\be\label{eq:CPstructure}
\la \vect{S}_{\vect{r}} \ra=(S-2\rho\cos^2{[\vect{K}\cdot\vect{r}+\psi]})\hat{z}+\sqrt{4S\rho}\cos{[\vect{K}\cdot\vect{r}+\psi]}\times(\cos{\phi}~\hat{x}+\sin{\phi}~\hat{y})
\ee
This order parameter has only two components, one along $\hat{z}$ and the other along $\cos{\phi}~\hat{x}+
\sin{\phi}~\hat{y}$ in XY plane, i.e., the order is co-planar. The ground state manifold has $\Uone\times \Uone$ symmetry. One of the $\Uone$, associated with $\phi$, is the freedom to select the direction of $\la \vect{S}_{\vect{r}} \ra$ in the XY plane, another $\Uone$, associated with $\psi$, is the freedom to select the origin of the coordinate. A choice of some $\phi$ and some $\psi$ spontaneously breaks $U(1) \times U(1)$ symmetry. Beyond the order $\Delta^4$, the $U(1)$ translational symmetry is explicitly broken if $\Gamma_{\text{u}}$ is non-zero. Within $1/S$ expansion, a non-zero $\Gamma_{\text{u}}$ emerges at order $1/S^2$. There are three contributions to $\Gamma_{\text{u}}$ at this order. One, $\Gamma_{\text{u}}^{(\text{n})}$, comes from normal ordering of the term of sixth order in bosons; another, $\Gamma_{\text{u}}^{(\text{a})}$ comes from second order perturbation in cross-products of representatives $\Delta^2_{1,2}\, \ad \ad+h.c$ and $1/S(\Delta^3_{1}\overline{\Delta}_{2} \ad \ad + h.c)$, $1/S(\Delta^3_{2}\overline{\Delta}_{1} \ad \ad + h.c)$; and third contribution, $\Gamma_{\text{u}}^{(\text{b})}$, comes from third order terms in $\Delta^2_{1,2}\, \ad \ad+h.c$ and $\Delta_{1}\overline{\Delta}_2\, \ad a+h.c$. In explicit form we have
\be
\Gamma_{\text{u}}^{(\text{n})}=\frac{9(1-2J_2)}{32S^2}
\ee
\be
\Gamma_{\text{u}}^{(a)}=-\frac{1}{2S^2}\sum_{\kv}\frac{V_{\kv}(\vect{K},\,\vect{K})\big(3 U_{\kv+2\vect{K},\,2\vect{K}}(\vect{K},\,\vect{K},\,\vect{K})/4+V_{-2\vect{K}+\kv}(0,\,-\vect{K})\big)}{\omega_{\vect{K}+\kv}+\omega_{\vect{K}-\kv}}
\ee
\be
\Gamma_{\text{u}}^{(b)}=-\frac{2}{S^2}\sum_{\kv}\frac{V_{\kv}(-\vect{K},\,-\vect{K})V_{\kv+\vect{Q}}(\vect{K},\,\vect{K})V_{-\vect{K}}(-\kv,\,\vect{K})}{(\omega_{-\vect{K}+\kv}+\omega_{-\vect{K}-\kv})(\omega_{-\vect{K}+\kv}+\omega_{-\kv})}
\ee
where $ V_{\vect{q}}(\vect{k_1},\vect{k_2})$ and $U_{\vect{q}}(\vect{k_1},\vect{k_2},{\vect k}_3)$ are defined in Eqs.~\ref{eq:ptvertexexp}, and $J_{\qv}$ is defined in Eq.~\ref{eq:QuadraticHF}. We verified that the total $\Gamma_{\text{u}}=\Gamma_{\text{u}}^{(\text{n})}+\Gamma_{\text{u}}^{(a)}+\Gamma_{\text{u}}^{(b)}+\mc{O}(\frac{1}{S^3})$ is non-singular (potential logarithmical terms cancel out), and computed $\Gamma_{\text{u}}$ numerically for several $J_2<1/8$ and found that it is non-zero and negative (see Table~\ref{tab:HFVphase}). A negative $\Gamma_{\text{u}}$ breaks the $\Uone$ translational symmetry down to $\mathbb{Z}_3$ and reduces the continuum set of $\psi$ to the discrete subset $\psi=\frac{l\pi}{3},\,l=0,\,1,\,2$. The order parameter in each of three possible spin states has a V-type shape with two spins in each triad pointing in one direction and the remaining spin in the other direction~\cite{Balents2013,Chubukov2014}.
\begin{table}
\begin{center}
  \begin{tabular}{|c|c|c|c|}
    \hline
    $J_2$ & \quad\quad\quad0\quad\quad\quad & \quad\quad\quad$0.1$\quad\quad\quad & \quad\quad\quad$1/8$\quad\quad\quad \\
    \hline
 $\Delta \Gamma~(1/S)\quad$   &   -1.6           &    -6.9       &  -247.7 \\\hline
 $\Gamma_{\text{u}}~(1/S^2)$ & $-0.68$ &   -0.81       & -0.85\\
  \hline
  \end{tabular}
\end{center}
\caption{The parameters of the Ginzburg-Landau functional of the $V$ phase at different $J_2$, to leading order in $1/S$.
$\Delta \Gamma=\Gamma_2-\Gamma_1$ is the difference between the prefactors of the two quartic terms, and $\Gamma_{\text{u}}$ is the prefactor for the sixth order term.
\label{tab:HFVphase}}
\end{table}

We did similar analysis for $J_2>1/8$ and found that logarithmical singularities from individual contributions to $\Delta {\bar \Gamma}=\bar{\Gamma}_2-\bar{\Gamma}_1$ do not cancel.  To logarithmic accuracy,
\begin{equation}
\Delta {\bar \Gamma} = \frac{8(1+J_2)^2}{\pi} \left[\frac{1}{\sqrt{4J_2 - (1-3J_2)^2}} - \frac{1}{\sqrt{4J_2}} \right]  \frac{|\log{\bar \mu}|}{S}
\label{th_ch_1}
\end{equation}
This formula is valid up to $J_2 =1$, which, as we said, is the upper boundary (in $J_2$) of the stripe phase. We see that $\Delta\bar{\Gamma}=\bar{\Gamma}_2-\bar{\Gamma}_1>0$ everywhere, except for $J_2 =1/3$. A positive $\bar{\Gamma}_2-\bar{\Gamma}_1$ implies that it is energetically favorable for the system to develop just one condensate, either $\Phi_1$, or $\Phi_2$, or $\Phi_3$ (i.e., to develop order parameter with one out of three possible momenta ${\bf M}_i$, ($i =1-3$)). Setting $\Phi_1=\sqrt{\rho}\mathrm{e}^{i\phi},\Phi_2=\Phi_3=0$, we obtain spin configuration in real space
\be
\la \vect{S}_{\vect{r}} \ra=(S-\rho)\hat{z}+\sqrt{2S\rho}(\cos{[\vect{M}_1\cdot\vect{r}+\phi]}\hat{x}+\sin{[\vect{M}_1\cdot\vect{r}+\phi]}\hat{y})
\ee
For a generic ${\bf M}$, such an order would be a non-coplanar cone phase. In our case, however, $\vect{M}_i$ are special points for which $\vect{M}\cdot\vect{\delta}_{\alpha}=0 ~\text{or}~ \pi$. One can easily verify that in this situation the spins order in a stripe manner in XY plane -- parallel in one direction and anti-parallel in the other. Such an order is co-planar and is termed as canted stripe. We show such spin ordering in real space in Fig.~\ref{fig:latticeStripe}.

\subsubsection{Special case around $J_2=1/3$}\label{sec:specialcase}
For $J_2 =1/3$ the prefactor for $|\log{\bar \mu}|$ cancels out. We computed the remaining regular piece in $\Delta\bar{\Gamma}$ numerically and found that it is actually negative. This implies that at $J_2=1/3$ quantum fluctuations select a different phase, in which all three condensates $\Phi_1$, $\Phi_2$ and $\Phi_3$ are non-zero and have equal amplitudes. To determine the relative phases between the condensates, we need to compute  $\bar{\Gamma}_{\text{u}}$ at  $J_2 =1/3$.   This term is given by
 \be
\bar{\Gamma}_{\text{u}}=-\frac{1}{SN}\sum_{\kv} \big(\frac{V_{\kv}(\vect{M}_1,\,\vect{M}_1)V_{\vect{M}_1+\kv-\vect{M}_2}(\vect{M}_2,\,\vect{M}_2)}{\omega_{\vect{M}_1+\kv}+\omega_{\vect{M}_1-\kv}} \big)+\frac{1+J_2}{4S}
\label{eq:Umkapp2}
\ee
Evauating the lattice integrals, we found
$\bar{\Gamma}_{\text{u}}=-0.69/S <0$.  To minimize the last term in Eq.~\ref{eq:Eden11}, the relative phase between $\Phi_{i}$ and $\Phi_{j}$ ($i,\,j=1,\,2,\,3$) then should be 0 or $\pi$. An elementary analysis shows that the corresponding spin structure is similar to the V phase found near $\hsa$ when $J_2<1/8$ -- it has 
  three out of four spins pointing in the same direction, and the fourth spin pointing in the opposite direction in the XY plane.
We didn't study in this paper the phase transition between this V-type phase and the stripe phase at high field and the evolution of the  V-type phase as the field decreases.
\subsection{Phase transition near \hsat}\label{sec:HFLargeSSW}
To analyze the nature of the phase transition between the $V$ and the stripe phase near \hsat~we obtain the stability boundaries of the two phases by analyzing the spin wave spectrum. Near \hsat~there are two small parameters -- $1/S$ and the magnitude of a magnon condensate $\rho$ in each of the two phases. In this section, we study the limit when $1/S$ is small enough such that $|\log{\rho}|/S\ll 1$. In the next section we explore another limit when $S = \mc{O}(1)$ and $|\log{\rho}|/S \gg 1$.

We first calculate the spin wave spectrum in the $V$ phase near \hsat. The structure of the $V$ phase is shown in Fig.~\ref{fig:J20PD}. Near the saturation field the angles between sublattice magnetizations and the direction of the magnetic field (the $z$ axis) are small. In the classical limit, we obtain from Eq.~\ref{eq:coplanarH}: $\theta_1 = (h_{\text{sat}} -h)^{1/2}/3$, $\theta_2 = -2 (h_{\text{sat}} -h)^{1/2}/3$. The leading order quantum corrections to the tilt angles and to magnon self-energy are of order $(h_{\text{sat}}-h) |\log(h_{\text{sat}} -h)|/S$.

We expand the Hamiltonian up to the quartic order in terms of the magnons $a,\,b,\,c$ defined in the local coordinates of $\vect{S}_a,\,\vect{S}_b,\,\vect{S}_c$ ( see Appendix~\ref{app:HighFieldSW} for details) and keep terms of order $h_{\text{sat}}-h$ (modulo logarithms). The quadratic term is
\begin{equation}
\mathcal{H}^{(2)}=\mathcal{H}_{2,0}+ \delta \mathcal{H}_2
\end{equation}
where $\mathcal{H}_{2,0}$ is the same as in fully polarized state at $h = h_{\text{sat}}$ and
 $\delta \mathcal{H}_{2} \sim (h_{\text{sat}}-h)$ is the perturbation to $\mathcal{H}_{2,0}$ due to the transverse magnetic order. We diagonalize $\mathcal{H}^{(2)}$ in two steps. First, we diagonalize $\mathcal{H}_{2,0}$ and find the eigenmodes $\phi_{\mu,\kv}=\{A_{\kv},\,B_{\kv},\,C_{\kv}\}$. Then we express the whole $\mathcal{H}^{(2)}$ with quantum corrections in the new basis $\Phi_{\mu,\kv}=\{\phi_{\mu,\kv},\,\phid_{\mu,-\kv}\}$ and diagonalize $\mathcal{H}^{(2)}$ in this basis. The diagonalization of $\mathcal{H}_{2,0}$ is elementary and is achieved by simply rotating the original basis $(a_{\kv},b_{\kv},c_{\kv})$ to $(A_{\kv}, B_{\kv}, C_{\kv})$ as
\begin{align}\label{eq:FerroBasis}
\begin{pmatrix}
a_{\vect{k}}\\
b_{\vect{k}}\\
c_{\kv}\\
\end{pmatrix}=
\begin{pmatrix}
1 & \jmath & \bar{\jmath}\\
1 & \bar{\jmath} & \jmath\\
1 & 1 & 1\\
\end{pmatrix}
\begin{pmatrix}
A_{\vect{k}}\\
B_{\vect{k}}\\
C_{\kv}
\end{pmatrix}
\end{align}
where $\jmath=e^{\iu 2\pi/3}, \bar{\jmath}=e^{-\iu 2\pi/3}$. $A,~ B$ and $C$ bands are
\begin{align}\label{eq:FMspectrum}
\omega^{(0)}_{A}(\kv)&=3+2J_2\,\mu_{\kv}+2\Re[\gk]\non\\
\omega^{(0)}_{B}(\kv)&=3+2J_2\,\mu_{\kv}-\Re[\gk]+\sqrt{3}\Im[\gk]\non\\
\omega^{(0)}_{C}(\kv)&=3+2J_2\,\mu_{\kv}-\Re[\gk]-\sqrt{3}\Im[\gk]
\end{align}
where $\gamma_{\kv},\,\mu_{\kv}$ are defined in Eq.~\ref{eq:th_ch_2}. The Brillouin zone for three sublattice description is shown in Fig.~\ref{fig:singleSLBZ}.
At $J_2<1/8$, when we expect the $V$ phase to be stable right below $h_{\text{sat}}$, the dispersions of $B$ and $C$ modes have zeros at the $\overline{\Gamma}$ point. At $J_2=1/8$, $\omega^{(0)}_B$ has additional zero modes at the $\vect{\overline{M}}_1=(2\pi/3,0),\,\vect{\overline{M}}_2=(-\pi/3,\sqrt{3}\pi/3),\,\vect{\overline{M}}_3=(-\pi/3,-\sqrt{3}\pi/3)$ and $\omega^{(0)}_C$ has zero modes at $-\vect{\overline{M}}_1,\,-\vect{\overline{M}}_2,\,-\vect{\overline{M}}_3$. And when $J_2>1/8$, the spectrum near $\vect{\overline{M}}$ becomes complex and the $V$ phase is unstable. We are interested in how the modes near the $\vect{\overline{M}}$ points become unstable right below \hsat. Accordingly, we set $J_2$ to be near $\Jcri$, expand in momentum near, say ${\bf \overline{M}}_1$ as $\vect{k}=\vect{\overline{M}}_1+\vect{q},\,|\qv|<<1$, and keep only the soft B and C modes (It has been checked explicitly that the inclusion of the gapped A mode does not change the conclusions below). With this, we computed the $1/S$ corrections to the relation between $\theta_1$ and $(h_{\text{sat}} -h)^{1/2}$ from cubic terms in the Hamiltonian, expressed in $A$, $B$ and $C$ bosons, and corrections to the classical dispersion from three-boson and four-boson terms. Collecting all $1/S$ contributions and combining them with classical result for $\mathcal{H}^{(2)}$ to order $(h_{\text{sat}} -h)$ we obtain
\begin{align}
\label{eq:coplanarLElargeS}
&\mathcal{H}^{(2)}=\frac{S}{2}\sum_{\qv}\non\\
&\begin{pmatrix}
\Bd_{\vect{M}_1+\qv}~
C_{-\vect{M}_1-\qv}
\end{pmatrix}
\begin{pmatrix}
\omega_{\qv}+(\frac{1}{3}+\delta_1) (h_{\text{sat}} -h) & (-\frac{1}{3}+\delta_2) (h_{\text{sat}} -h)\\
(-\frac{1}{3}+\delta_2)(h_{\text{sat}} -h) & \omega_{-\qv}+(\frac{1}{3}+\delta_1) (h_{\text{sat}} -h)\\
\end{pmatrix}
\begin{pmatrix}
B_{\vect{M}_1+\qv}\\
\Cd_{-\vect{M}_1-\qv}\\
\end{pmatrix}
\end{align}
where $\omega_{\qv}=1-8 J_2+\frac{1}{16}(q_x^2+21q_y^2)$, and $\delta_1$ and $\delta_2$ are $1/S$ quantum corrections to the normal and anomalous self-energy at $\qv=0,\,\omega=0,\,J_2=1/8$. Note that other terms like $\Bd_{\vect{M}_1+\qv}\Bd_{-\vect{M}_1-\qv}$, $\Bd_{\vect{M}_1+\qv}C_{\vect{M}_1+\qv}$ do not contribute to the spectrum near $\vect{M}_1$ to first order in $1/S$. A simple algebra shows that the critical coupling of $J_2$, at which spin-wave excitations in the $V$ phase becomes complex (and, as the consequence, the phase becomes unstable) is $J_{2\text{V}}=1/8+\frac{(1/3+\delta_1)-|1/3-\delta_2|}{8} (h_{\text{sat}} -h)$.  In the limit $\frac{\loghsat}{S}\ll 1$, we found that to logarithmic accuracy,
 $\delta_1=\frac{0.22}{S} |\log {(h_{\text{sat}}-h)}|$ and $\delta_2= \frac{1.58}{S} |\log{(h_{\text{sat}}-h)}|$, thus $J_{2V}=1/8+ \frac{0.22}{S} (h_{\text{sat}}-h) |\log{(h_{\text{sat}}-h)}|$.

Using a similar analysis for the stripe phase, we found $J_{2\text{stripe}}=1/8-\frac{0.07}{S} (h_{\text{sat}}-h)|\log{(h_{\text{sat}} -h)}|$. We show more details of calculations
 in Appendix~\ref{app:LargeS}.

By looking at the sign of the corrections to the critical $J_2$, we see that the phase boundary of the $V$ phase shifts to the right of $1/8$ by $\mc{O}(1/S)$, while that of the stripe phase shift to the left of $1/8$, thus the stability regions of the two phases overlap near $J_2=1/8$. This implies that the transition between the $V$ and stripe phase is \textit{first order} with finite hysteresis in the large S limit near \hsat.
\subsection{Phase diagram in a generic field}\label{sec:LargeSSW}
In Sec.~\ref {sec:LargeSaction} we found that in fields near \hsat~quantum fluctuations select the $V$ phase and the canted stripe phase at small and large $J_2$, respectively. As the $V$ phase is the same as the one found at $J_2=0$, and the canted stripe phase can be regarded as the stripe phase found in zero field with additional longitudinal ferromagnetic component along the field, it is tempting to assume that the Y(UUD)V phase and the canted stripe phase are stable at a generic field. To test this, in this section we study the spin wave spectrum of the Y(UUD)V state with $1/S$ quantum corrections as a function of $J_2$ at various fields between $h=0$ and $h = h_{\text{sat}}$ and obtain the stability region of the Y(UUD)V state at a generic field. At larger $J_2 $, we evaluate the spin wave spectrum of the canted stripe phase in a generic field and find the stability region of this phase. We show that the regions, where the Y(UUD)V phase and the canted stripe phase are both stable, overlap near $J_2=1/8$ in all fields (Fig.~\ref{fig:summarySemiClassical}). As a result, the phase transition between the two states remains \textit{first order} with a finite, $\mc{O}(1/S)$ hysteresis width in all fields.

In the spin wave framework, the state selected by quantum fluctuations via order from disorder mechanism has symmetry related zero modes (at $\overline{\Gamma}$ and $\widetilde{\Gamma}$ in the reduced BZ), but the accidental zero modes, present at the classical level, are all lifted by the quantum fluctuations. We verify this for both phases by calculating quantum corrections to the spin wave spectrum to the leading order in $1/S$ and at the momenta where there are additional zero modes at present in the classical analysis. We then use the same strategy as near the saturation field and analyze at what $J_2 \approx 1/8$ the spin-wave spectrum softens at some other finite momenta, and a given state becomes unstable. The details of the calculations are presented in Appendix~\ref{app:LargeS}, below we present the summary of the results.
\subsubsection{Stripe phase}\label{sec:Hstripe}
We computed quantum corrections to the spin wave spectrum at three momenta: $\widetilde{\Gamma}=(0,\,0)$, where the true Goldstone mode is located, and at $\vect{\widetilde{M}}_{2,3}=(\pm \pi,\,-\pi/\sqrt{3})$, where spin-wave spectrum without $1/S$ corrections has zeros (see Fig.~\ref{fig:stripedispersion}). We found that the Goldstone mode and the linear spectrum around it survive, as they should. However at $\vect{M}_{2,3}$ we found a positive gap $\delta m = \mc{O}(1/S)$. We list the values of $\delta m_{\vect{\widetilde{M}}_{2,3}}$ for various fields in Table II. Near $J_2=1/8$, we calculated quantum corrections to the spectrum at momenta $\pm\vect{\widetilde{K}}=(\mp 2\pi/3,\,0)$, at which the classical spin-wave spectrum becomes unstable at $J_2=1/8$. The spectrum with $1/S$ correction can be expressed as $\omega^{(1)}_{\kv}\simeq \sqrt{\omega_{\kv}^2+\delta m}$. As $\omega_{\tilde{\vect{K}}}^2\sim (J_2-1/8)$ and $\delta m=2(A\delta A-B\delta B)\big|_{\tilde{\vect{K}}}$, $\omega^{(1)}_{\kv}$ vanishes at $J_{2\text{stripe}}\sim 1/8-\delta m<1/8$, at which spin-wave spectrum of the canted stripe phase becomes unstable. We present the results in the last column in Table~\ref{tab:stripe}. We see that for all fields $J_{2\text{stripe}} <1/8$, i.e., the stability region of the stripe phase extends to the left of the classical transition line $\Jcri=1/8$.
\begin{table}
\begin{center}
 \begin{tabular}{|c|c|c|}
    \hline
    $h$ & \quad$\delta m |_{\kv=\vect{\widetilde{M}}_{2,3}}$ in unit $1/S$\quad\quad & \quad$J_{2\text{stripe}}-1/8$ in unit $1/S$\quad\quad \\
    \hline
 9   &       0       &  0 \\\hline
 8.56 &    0.00       & -0.055 \\\hline
 7.28 &    0.04     & -0.082 \\\hline
 6.36 &  0.14 & -0.096\\\hline
 4.5 &  0.40 & -0.13\\\hline
 3.3 &   0.78 & -0.19\\\hline
 0  & 1.99 & -0.46\\
  \hline
  \end{tabular}
\end{center}
\caption{Properties of the stripe phase at $J_2 \approx 1/8$ in a magnetic field. $\delta m$ and $J_{2\text{stripe}}$ are expressed in units $1/S$. The field value in first column is between the saturation field $h_{\text{sat}}=9$ and $0^+$. Second column -- the mass of would be accidental zero mode at $\vect{\widetilde{M}}_{2,3}$ at order $\mc{O}(1/S)$. A positive $\delta m |_{\kv=\vect{\widetilde{M}}_{2,3}}\,(1/S)$ implies the stripe phase is selected among the states from a generate classical ground state manifold. Third column -- the critical $J_{2\text{stripe}}$. A negative sign implies that the stripe phase remains stable up to $J_2 < 1/8$.
\label{tab:stripe}}
\end{table}
\subsubsection{Y(UUD)V phase}
The semiclassical spin wave analysis of the Y(UUD)V phase is more involved as the three bose fields $a,\,b,\,c$, defined on sublattices $A,\,B,\,C$, do not decouple in a generic field. One needs to diagonalize the $6\times6$ matrix, express the canonical eigenmodes as linear combinations of $a,\,b,$ and $c$, and then follow the same procedure as in the stripe phase to calculate quantum corrections.

We performed spin wave analysis in the Y(UUD)V phase at two fields: $h=0$, when the order is the $120^{\circ}$ Neel phase, and at $h=h_{\text{sat}}/3$, when system is in UUD phase. Compared to a generic field, these two configurations are relatively easy to handle because $120^{\circ}$ phase has the $\mathbb{Z}_3$ symmetry, and magnon branches get decoupled after a global rotation of the basis, and UUD phase is collinear, and therefore there are no cubic terms in the Hamiltonian. We found that in both cases quantum fluctuations introduce a positive mass of order $1/S$ at momenta where the classical spectrum has accidental zeros, and shift the phase boundary of either $120^{\circ}$ phase or UUD phase to the right of $J_2= 1/8$. We show the results of $1/S$ calculation in Table~\ref{tab:Y(UUD)Vphase}. The sign of the shift is the same as of $J_{2\text{V}}$ at fields near $h_{\text{sat}}$, which shows that, most likely, the stability region of the Y(UUD)V phase shifts to the right of $J_1=1/8$ for all fields. As the consequence, at large $S$, the transition between the Y(UUD)V phase and the canted stripe phase remains first order for all fields, with the finite hysteresis width of order $1/S$.

For completeness, we also calculated critical $J_2$ of the UUD phase at the upper and the lower critical fields $h_{u} = 3 + 1.12/S$ and $h_l = 3-0.66/S$, when this phase become unstable either towards the $V$ phase or the $Y$ phase. We again found that the critical $J_2$ shift to larger values than $1/8$. We show the results in Table~\ref{tab:UUDphase}.
\begin{table}[tbp]
\begin{center}
 \begin{tabular}{|c|c|}
    \hline
    $h~$ & \quad\quad$J_{2\text{Y(UUD)V}}-1/8$ in unit $1/S$\quad\quad  \\
    \hline
 $h=0$             &    0.13     \\\hline
 $h=h_{\text{sat}}/3$ &    0.13\\\hline
 $h=h_{\text{sat}}-0^+$ &   $0.22(h_{\text{sat}}-h)|\log (h_{\text{sat}}-h)|$ \\
  \hline
  \end{tabular}
\end{center}
\caption{Critical $J_2$ at which the Y(UUD)V state becomes unstable. $J_{2\text{Y(UUD)V}}-1/8$ is expressed in unit $1/S$. The values of $J_{2\text{Y(UUD)V}}$ have been obtained by analyzing the spin-wave excitations at $\vect{\overline M}=(0,\,2\pi/\sqrt{3})$ to order $1/S$.
\label{tab:Y(UUD)Vphase}}
\end{table}
\begin{table}[tbp]
\begin{center}
  \begin{tabular}{|c|c|c|}
    \hline
    ~ & \quad\quad $h-h_{\text{sat}}/3$ in unit $1/S$\quad\quad &  \quad\quad$J_{2\text{UUD}}-1/8$ in unit $1/S$\quad\quad  \\
    \hline
 upper             &    + 1.12 & (3.09+1.12)/24=0.18     \\\hline
 lower  &   -0.66 & (3.09-0.66)/24=0.10\\\hline
  \end{tabular}
\end{center}
\caption{Phase boundary of the UUD phase near $J_2=1/8$. The numbers are expressed in units $1/S$. The phase boundary has been obtained by analyzing the spin wave spectrum at $\overline\Gamma$ and $\vect{\overline M}$ points to order $1/S$.\label{tab:UUDphase}}
\end{table}

To summarize this section, the semiclassical spin wave analysis at large but finite $S$ shows that in all fields the Y(UUD)V state is stable as $J_2$ increases from zero to $1/8+\mc{O}(1/S)$, and the stripe phase at larger $J_2$ is stable down to $1/8-\mc{O}(1/S)$. The stability regions of the two ordered phases overlap around $J_2=1/8$, and the phase transition in all fields is \textit{first order} with a finite hysteresis width of order $1/S$.
\section{High field region for a model with a generic spin}\label{sec:HighField}
We now discuss the phase diagram of the model with an arbitrary spin $S= \mc{O}(1)$, with particular interest to $S=1/2$. In a generic field, there is no small parameter to justify perturbative calculations for $S=\mc{ O}(1)$. However, right below \hsat, the density of magnon condensates is small, as we pointed out in Sec.~\ref{sec:LargeS}. In this situation, one can perturbatively expand in powers of magnon condensates (or, equivalently, in terms of the tilt angle between a sublattice magnetization and the $z$ axis). The coefficients of this expansion can be obtained at arbitrary $S$, and this gives us an opportunity to study the transition between $V$ and stripe phases outside of semiclassical limit.

Below we first identify the orders at small and large $J_2$ near \hsat, and find that the same $V$ and stripe phases are selected for an arbitrary spin, as in the large S limit. Then we analyze the nature of the phase transition between the $V$ phase and the stripe phase for a generic S.

\begin{figure}
\centering
\subfigure[]{\includegraphics[scale=0.5]{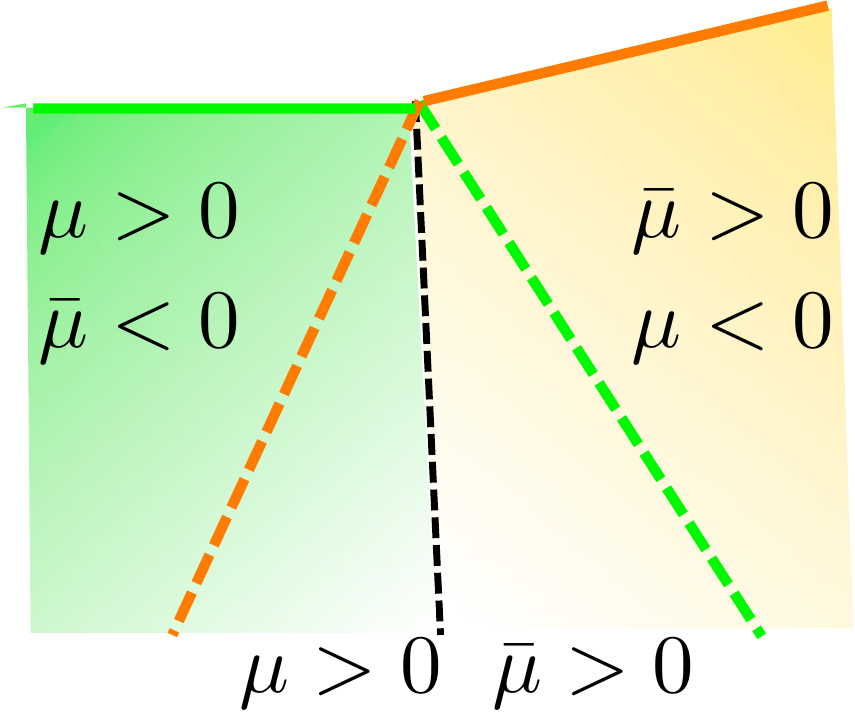}\label{fig:HFposs1}}\qquad\qquad
\subfigure[]{\includegraphics[scale=0.5]{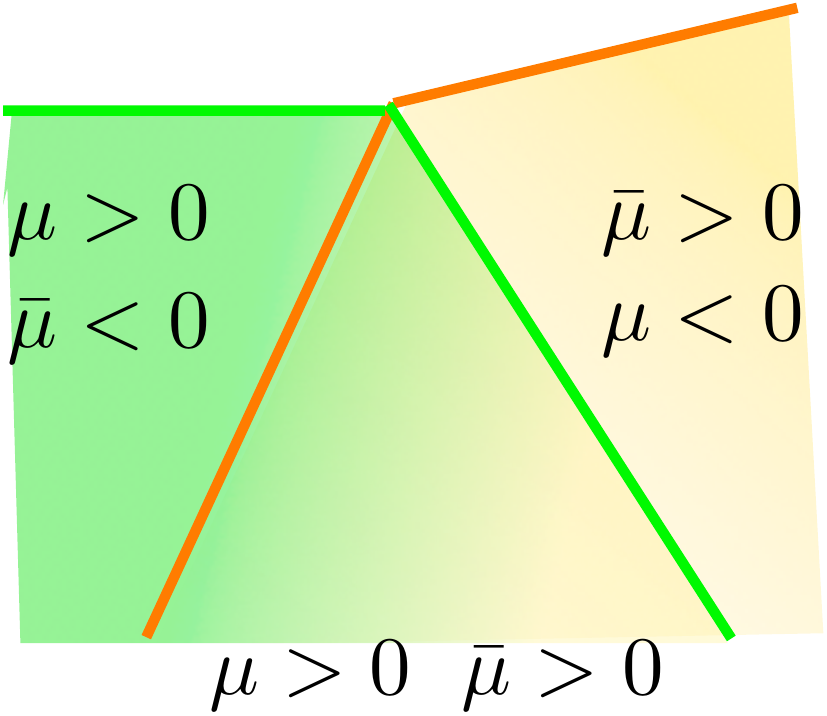}\label{fig:HFposs2}}\qquad\qquad
\subfigure[]{\includegraphics[scale=0.5]{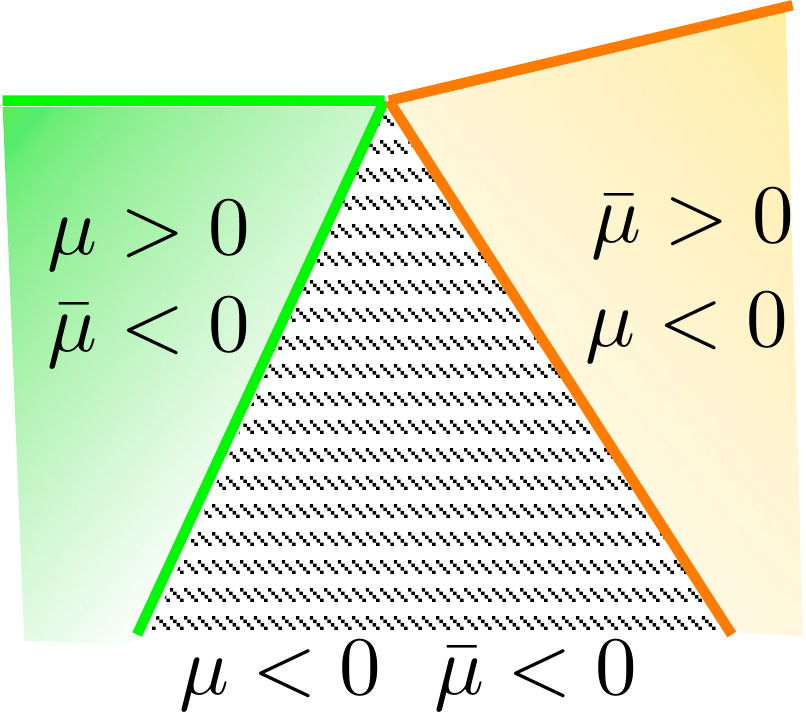}\label{fig:HFposs3}}
\caption{Three possibilities of the phase transition between the $V$ phase and the canted stripe phase near \hsat. (a) and (b): Condensates associated with both the $V$ and the stripe phase are stable over a range around $J_2=1/8$ (the region between the green and orange dashed lines). The phase transition can be either (a) first order or (b) involve an intermediate co-existence phase, depending on the interplay between quartic couplings $\Gamma_i$. (c) \textit{Neither} of the two condensates are stable over a finite range around $J_2=1/8$ (shaded region). The transition between the $V$ and the stripe phase then necessarily occurs via an intermediate state, which either has some non-quasi-classical long-range order with or without a continuous symmetry breaking, or has no spontaneous order.
\label{fig:HFposs}}
\end{figure}

In general, there are three options for the phase transition. It can be a first order transition with or without hysteresis, as in the classical and the large S cases. Or there can be an intermediate co-existence phase, in which both orders are present simultaneously. Or, one order looses it stability before the other becomes stable. In the latter case there is a intermediate region in which neither the $V$ phase nor the canted phase are stable. This intermediate state may have some non-quasi-classical long-range order with or without a continuous symmetry breaking, or may have no spontaneous order. We illustrate these possibilities in Fig.~\ref{fig:HFposs}. For the first two possibilities the prefactors $\mu$ and $\bar{\mu}$ for the quadratic terms in $\Delta$ and $\Phi$ in the Free energy of the $V$ and the stripe phases respectively (see Eq.~\ref{eq:Eden2}) are both positive over some range of $J_2$ at a given $h \lesssim \hsa$. Whether the phase transition is first order or occurs via a co-existence phase is determined by the interplay between the prefactors of the fourth-order terms in the Ginzburg-Landau model, which includes both fields~\cite{Schmalian2012} (Fig.~\ref{fig:HFposs1} and Fig.~\ref{fig:HFposs2}). The third scenario occurs when both $\mu$ and $\bar{\mu}$ are negative in a finite range near $J_2=1/8$, i.e., neither of the two orders develop (Fig.~\ref{fig:HFposs3}).

We present Ginzburg-Landau analysis in~Sec. \ref{sec:HighFieldGL} below and present the analysis of spin-wave dispersion with quantum corrections in Sec. \ref{sec:HighFieldSW}. We show that the fields $\Delta$ and $\Phi$ don't coexist for arbitrary $S$. For $S >1$, the regions where $\mu >0$ and ${\bar \mu} >0$ overlap. Then the system remains ordered at all $J_2$, and the transition between the $V$ and the stripe phases is first order. However, when $S=1/2$ and, most likely, also $S=1$, the two phases don't overlap near $J_2=1/8$. In this case, there exists an intermediate phase without a quasi-classical long-range magnetic order. We emphasize that {\it this happens near $h =\hsa$, where the density of magnons is small}. To identify the nature of this intermediate state one needs to go beyond the spin wave framework, and we leave this for future studies.
\subsection{Ginzburg-Landau formalism}\label{sec:HighFieldGL}
Like we discussed in Sec.~\ref{sec:LargeS} the transition at $h=\hsa$ can be described as magnon condensation, and the condensation energy at $T=0$ can be expandeded in powers of the condensate fields. For arbitrary $S$, the condensation energy in the $V$ phase and in the stripe phase has the same form as in Eq.~\ref{eq:Eden10} and Eq.~\ref{eq:Eden11}, but the quartic couplings $\Gamma_{1,2}$ and ${\overline \Gamma}_{1,2}$ are proportional to the fully renormalized four-point vertex function $\Gamma_{\vect{q}}(\vect{k_1},\vect{k_2})$, taken at certain momenta. In our case
\begin{align}\label{th_ch_3}
\Gamma_1&=\Gamma_{\vect{q}=0}(\vect{K},\vect{K})\non\\
\Gamma_2&=\Gamma_{\vect{q}=0}(\vect{K},-\vect{K})+\Gamma_{-2\vect{K}}(\vect{K},-\vect{K})\non\\
\bar{\Gamma}_1&=\Gamma_{\vect{q} =0}(\vect{M_1},\vect{M_1})\non\\
\bar{\Gamma}_2&=\Gamma_{\vect{q} =0}(\vect{M_1},\vect{M_2})+\Gamma_{\vect{M_2-M_1}}(\vect{M_1},\vect{M_2})
\end{align}

\begin{figure}[tbp]
\centering
\includegraphics[scale=0.8]{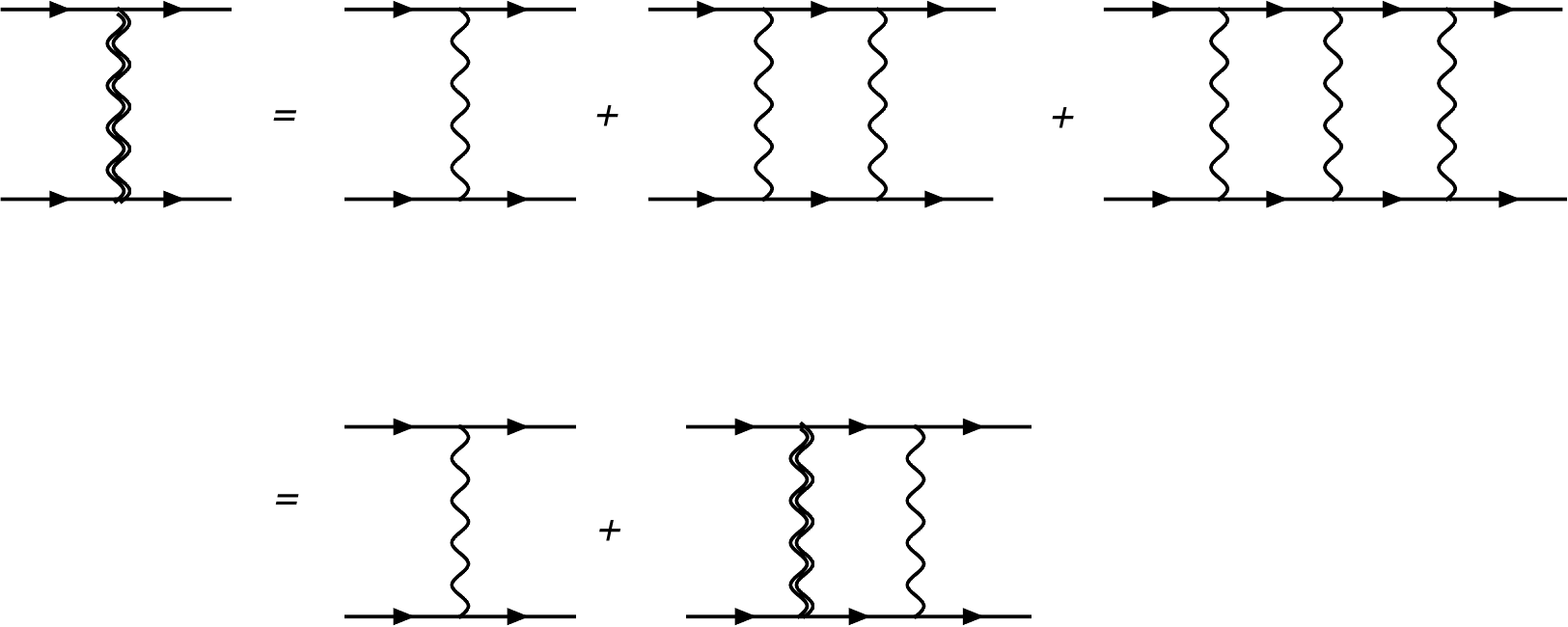}
\caption{Self-consistency equation for the fully renormalized four-point vertex function $\Gamma_{\vect{q}}(\vect{k_1},\vect{k_2})$, (double wavy line). A single wavy line is the four-boson interaction potential $V_{\vect{q}}(\vect{k_1},\vect{k_2})$.
\label{fig:BSconsistency}}
\end{figure}

To find $\Gamma_{\vect{q}}(\vect{k_1},\vect{k_2})$, all orders of scattering of two excited magnons should be counted. We show this in the diagrammatic formalism in Fig.~\ref{fig:BSconsistency}. The ladder series of diagrams is equivalent to the integral Bethe-Salpeter (BS) equation:
\begin{equation}
\Gamma_{\vect{q}}(\vect{k_1},\vect{k_2})=V_{\vect{q}}(\vect{k_1},\vect{k_2})-\frac{1}{N}\sum_{\vect{q'}}\frac{\Gamma_{\vect{q'}}(\vect{k_1},\vect{k_2})V_{\vect{q-q'}}(\vect{k_1+q'},\vect{k_2-q'})}{S(\omega_{\vect{k_1+q'}}+\omega_{\vect{k_2-q'}})}
\label{eq:BSmain}
\end{equation}
where 
\begin{align}\label{eq:fourptvertexexp}
V_{\vect{q}}(\vect{k_1},\vect{k_2}) = \frac{1}{2}[(J_{\vect{q}}+J_{\vect{k_2}-\vect{k_1}-\vect{q}})+2S(K_s-1)(J_{\vect{k_1}}+J_{\vect{k_1+q}}+J_{\vect{k_2}}+J_{\vect{k_2-q}})]
\end{align}
where $K_s=\sqrt{1-1/2S}$. $K_s$ is obtained by re-expressing the H-P expansion of $\hat{S}^+/\hat{S}^-$, in terms of normally ordered bosons.  This factor can can also be obtained by matching the matrix element of spin operators $\hat{S}^+/\hat{S}^-$ and their Bose representations~\cite{Batyev1986}. We explicitly verified that for $S=1/2$ this procedure yields the same result as the one in which spin operators are mapped to hard core bosons. 

One can easily make sure that for $q$, $\vect{k_1}$ and $\vect{k_2}$, which we need in Eq.~\ref{th_ch_3}, the integrand scales as $1/(q')^2$ at small $q'$, if we evaluate it right at $h= \hsa$. The 2D integral over $q'$ then diverges logarithmically. The log-divergence is cut at $h < \hsa$ by $\hsa-h$, which then appears under the logarithm. We already used this in the calculations at large $S$. In the latter case, we used the fact that, as a function of $S$, $V = \mc{O}(1)$ and $\epsilon_{\kv}=S\omega_{\kv} = \mc{O}(S)$, and $\Gamma_{\vect{q'}}(\vect{k_1},\vect{k_2})$ is restricted with only one scattering process, i.e., replace $\Gamma_{\vect{q'}}(\vect{k_1},\vect{k_2})$ in the r.h.s. of Eq.~\ref{eq:BSmain} by $V_{\vect{q}}(\vect{k_1},\vect{k_2})$. This is how we obtained terms $(1/S) |\log{(\hsa -h)}|$. Now $S =\mc{ O} (1)$, but $\loghsat$ is still large, and all terms in the ladder series matter.

\begin{figure}[tbp]
\centering
\includegraphics[scale=0.7]{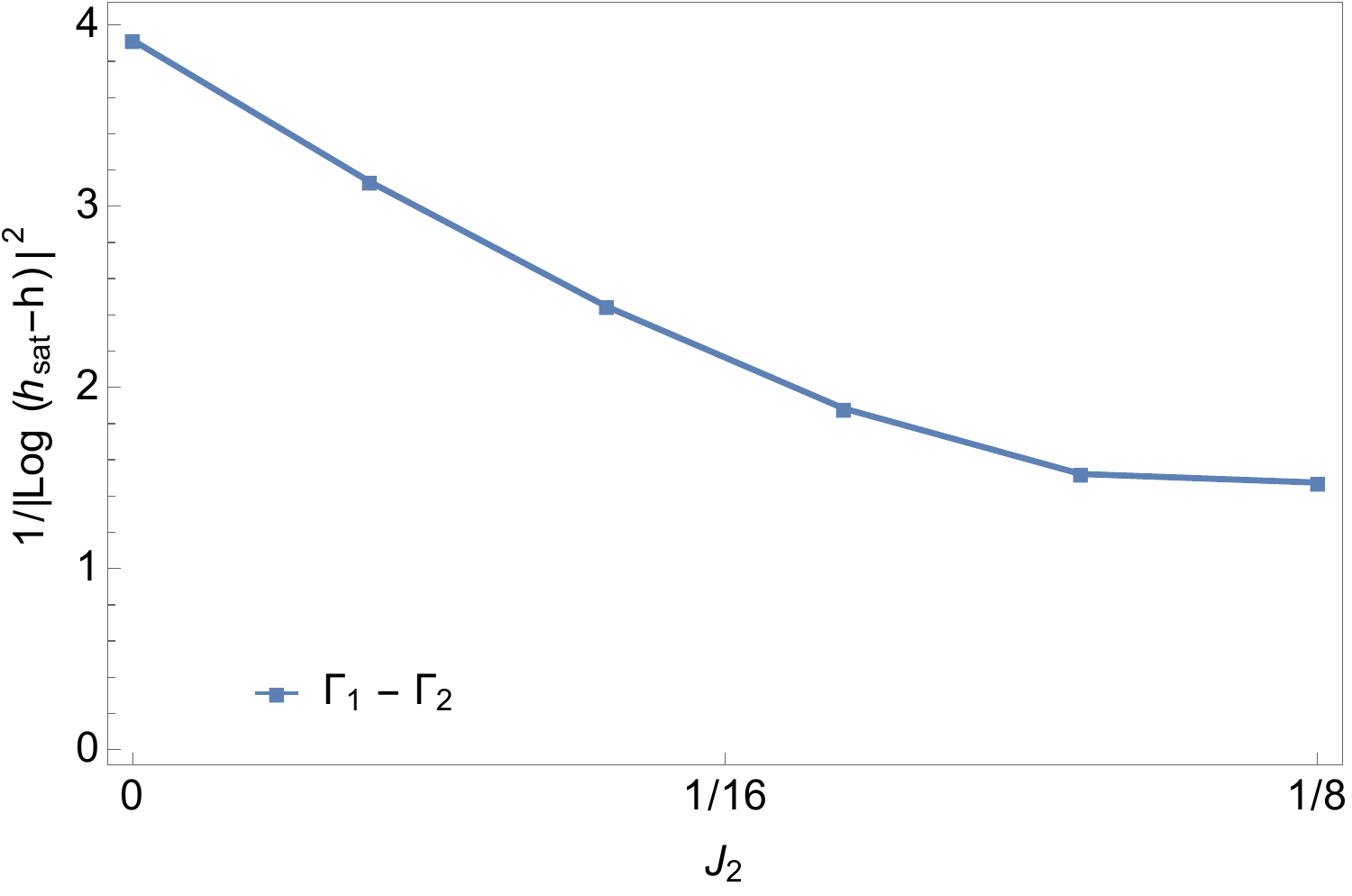}
\caption{The difference between the two quartic coefficients, $\Gamma_1-\Gamma_2$, in the Ginzburg-Landau expansion for the $V$ phase, Eq.~\ref{eq:Eden10}, for $S=1/2$. The difference scales as $\frac{1}{\loghsat^2}$ with a $J_2$ dependent prefactor.
\label{fig:Gamma2}}
\end{figure}

The ladder series for $\Gamma$ contain higher powers of $(1/S) \loghsat$. We found that the series are geometrical, to logarithmic accuracy. Because $(1/S) \loghsat \gg 1$ for $S=\mc{O} (1)$ and $h \lesssim \hsa$, the resulting $\Gamma$ and ${\bar \Gamma}$ are actually small in $1/\loghsat$. For the $V$ phase we found
\be
\Gamma_1=\Gamma_2=(1-6J_2)4\pi\sqrt{3}\frac{S}{\loghsat}
\ee
We see that, to this accuracy, $\Gamma_1=\Gamma_2$, like in the classical limit. However, the equivalence between $\Gamma_1$ and $\Gamma_2$ gets broken once we go beyond the leading term and compute contributions of order $1/\loghsat^2$. We did this numerically for $S=1/2$ and show the results in Fig.~\ref{fig:Gamma2}. We see that $\Delta \Gamma = \Gamma_1-\Gamma_2$ is positive, like in the quasiclassical limit. A positive $\Delta \Gamma$ implies that it is energetically favorable for a system to develop both condensates, $\Delta_1$ and $\Delta_2$, with equal amplitude $\rho=\mu/(\Gamma_1+\Gamma_2)$ ($\mu \propto (\hsa -h)$). This implies that the ordered state at small $J_2$ is coplanar. To determine the specific type of a coplanar order, i.e. to specify the angle $\Psi$ in Eq.~\ref{eq:CPstructure}, one would, in principle, need to obtain $\Gamma_{text{u}}$ -- the prefactor for $(\Delta_1^3\overline{\Delta}_2^3+h.c.)$ term in the condensation energy. This term selected the $V$ phase in the quasiclassical limit. The calculation of $\Gamma_{text{u}}$ at arbitrary $S$ is rather involved and we didn't do it. Rather, we use the fact that the $V$ state has been identified for $S=1/2$ in the numerical analysis at $J_2=0$~\cite{Balents2013}, and assume that the same holds for finite $J_2$, i.e., that $\Gamma_{text{u}}$ is negative at arbitrary $S$, as it is at $S \gg 1$.

As larger $J_2$, the condensation energy is expressed in terms of three $\Phi$ fields (see the second equation in Eq.~\ref{eq:Eden11}).
To leading order in $1/\loghsat$ we obtained
\begin{align}
\overline{\Gamma}_1&=8\pi\sqrt{4J_2 - (1-3J_2)^2}\frac{S}{\loghsat}\non\\
\overline{\Gamma}_2&=8\pi\sqrt{4J_2}\frac{S}{\loghsat}
\end{align}
We see that $\bar{\Gamma}_1\le\bar{\Gamma}_2$, the equality holds only when $J_2=1/3$. This matches the result that we obtained in the large $S$ limit. For $\bar{\Gamma}_2 > \bar{\Gamma}_1$, only one out of three order parameters $\Phi_i$ develops a non-zero value, and the resulting state is the canted stripe phase, same as at large $S$. Like we already said, the case $J_2= 1/3$ requires a separate analysis.

We now use Ginzburg-Landau expansion to analyze the phase transition between the $V$ and the stripe phases at arbitrary $S$. We introduce $\Delta$ field for the order parameter in the $V$ phase $(\Delta_1 = \Delta_2 = \Delta/\sqrt{2})$ and $\Phi$ field for the order parameter in the stripe phase and derive the form of the condensation energy $E_{\text{cri}}$ up to fourth order in the coupled $\Delta$ and $\Phi$ fields.
 The most generic form of $E_{\text{cri}}$ is
\begin{equation}\label{eq:Eden2}
E_{\text{cri}}/N=-\mu|\Delta|^2-\bar{\mu}|\Phi|^2+\frac{1}{4}(\Gamma_1+\Gamma_2)|\Delta|^4+\frac{1}{2} \bar{\Gamma}_1|\Phi|^4+\Gamma_{\Delta,\Phi}|\Delta|^2|\Phi|^2+\bar{\Gamma}_{\Delta,\Phi}|\Delta|^2|\Phi|^2\cos{2\phi}
\end{equation}
where $\phi$ is chosen such that $E_{\text{cri}}$ is minimized. In the classical limit $\mu=\bar{\mu}=S(h_{\text{sat}}-h)$. Quantum fluctuations renormalize the slope of $(h_{\text{sat}}-h)$ dependence differently for $\mu$ and ${\bar \mu}$, and the two are generally different.

In the next Section we use spin-wave formalism to find out how $\mu$ and ${\bar \mu}$ behave near $J_2 =1/8$. Here we analyze the prefactors of the quartic terms. The calculations are similar to the ones for the $V$ and the stripe phases, and we just present the results. To leading order in $1/|\log{(h_{\text{sat}} -h)}|$ we obtained
\begin{align}
\Gamma_1&=\sqrt{3}\pi\frac{S}{\loghsat},
&\Gamma_2=\sqrt{3}\pi\frac{S}{\loghsat},\quad\quad
&\bar{\Gamma}_1=\sqrt{7}\pi\frac{S}{\loghsat}\nonumber\\
\bar{\Gamma}_2&=4\sqrt{2}\pi\frac{S}{\loghsat},
&\Gamma_{\Delta,\phi} = 2\sqrt{2}\pi\frac{S}{\loghsat},\quad\quad
&\bar{\Gamma}_{\Delta,\phi} = 0
\end{align}
We see that $\frac{1}{2}(\Gamma_1+\Gamma_2)\bar{\Gamma}_1<\Gamma^2_{\Delta,\phi}$. An elementary analysis of Eq.~\ref{eq:Eden2} shows that in this situation the $V$ and the stripe orders repel each other and repulsion is strong enough so that mutual co-existence is excluded. This leaves two possibilities: if the regions around $J_2=1/8$ where $\mu>0$ and $\bar{\mu}>0$ overlap, the phase transition between the two phases is first order, like at large $S$ (Fig.~\ref{fig:HFposs1}). If upon increasing of $J_2$, $\mu$ changes sign from positive to negative before ${\bar \mu}$ changes from negative to positive, then there is a region near $J_2 =1/8$ where neither $V$ nor stripe order develops (Fig.~\ref{fig:HFposs3}). In this situation, the transition between the $V$ and the stripe phase occurs via an intermediate phase, which is either disordered or has some non-quasi-classical long-range order, different from both the $V$ and the stripe orders. This last option is not realized at large $S$, but may develop at $S = \mc{O}(1)$. To check this we now analyze spin-wave excitations at an arbitrary $S$ and find the stability regions of the two phases.
\subsection{Spin wave calculations}
\label{sec:HighFieldSW}
We show the calculations for the $V$ phase. The analysis of the stripe phase is performed in the same way.
To study the instability of the $V$ phase as $J_2$ increases from $0$ to $1/8$, we expand the Hamiltonian in powers of the Holstein-Primakoff bosons. For an arbitrary spin, the prefactors in the expansion of $S^{+-}$ operators in powers of the density of Holstein-Primakoff bosons contain complex dependence of $S$ due to the fact that one should perform normal ordering of the bosons after expanding $\sqrt{1- a^\dag a/(2S)}$. The result of normal ordering is
\begin{align}\label{eq:HPgenericS}
S_{\vect{r}}^+ & =\sqrt{2S}(1-\frac{1}{4S}(1+\frac{1}{8S}+\frac{1}{32S^2}+...)a_{\vect{r}}^{\dagger}a_{\vect{r}})a_{\vect{r}}+\mathcal{O}(a^5)\non\\
&=\sqrt{2S}(1+(\sqrt{1-\frac{1}{2S}}-1)a_{\vect{r}}^{\dagger}a_{\vect{r}})a_{\vect{r}}+\mathcal{O}(a^5)
\end{align}

The computations of the spin-wave dispersions follows the same steps as for large $S$, but now we have to keep the explicit dependence on $S$ in the prefactors of all terms. Like at large $S$, we analyze the dispersion around, say, ${\bf \overline{M}}_1$ point in the three-sublattice Brillouin zone, where the instability develops in the large $S$ analysis. The low-energy Hamiltonian expressed in terms of soft $B$ and $C$ bosons has form similar to Eq.~\ref{eq:coplanarLE}:
\begin{align}
\label{eq:coplanarLE_1}
&\mathcal{H}^{(2)}=\frac{S}{2}\sum_{\qv}\non\\
&\begin{pmatrix}
\Bd_{\vect{M}_1+\qv}~
C_{-\vect{M}_1-\qv}
\end{pmatrix}
\begin{pmatrix}
\omega_{\qv}+(3+9\delta_1)\theta^2 &(-3+9\delta_1)\theta^2\\
(-3+9\delta_1)\theta^2 & \omega_{-\qv}+(3+9\delta_1)\theta^2\\
\end{pmatrix}
\begin{pmatrix}
B_{\vect{M}_1+\qv}\\
\Cd_{-\vect{M}_1-\qv}\\
\end{pmatrix}
\end{align}
where $\theta$ is the angle between the spin order on A sublattice and the field, $\omega_{\qv}=1-8 J_2+\frac{1}{16}(q_x^2+21q_y^2)$ is the spin wave dispersion at $h=\hsa$, and $\delta_1$ and $\delta_2$ originate from magnon-magnon interactions. In distinction to large $S$, these two parameters are no longer simply $\mc{O}(1/S)$, but have complex dependence on $S$. The relation between $\theta^2$ and $\hsa-h$ is also affected by magnon-magnon interaction.

The computation of $\delta_1$, $\delta_2$ and $\theta$ at arbitrary $S$ is somewhat involved. We show the computational steps in Appendix~\ref{app:HighFieldSW} and here present the results. With logarithmic accuracy, we found $\theta^2=\alpha_1(\hsa-h)\loghsat$, $\alpha_1>0$ (see Eq.~\ref{eq:thetah}), $\delta_1 = -1/3 + \mc{O}(1)$ and $\delta_2= 1/3 +\mc{O}(\frac{1}{\loghsat})$. Substituting these $\delta_{1,2}$ and $\theta^2$ into Eq.~\ref{eq:coplanarLE_1}, we found, to logarithmic accuracy, the dispersion near ${\bf M}_1$ in the form
\begin{equation}
\label{eq:coplanarZero}
\omega_{{\bf M}_1 + {\bf q}} = \big(1-8 J_2+ (3+9\delta_1)\alpha_1 (\hsa -h)\loghsat\big) + \frac{1}{16}(q_x^2+21q_y^2)
\end{equation}
In the large-S limit we had $(1/3+\delta_1) >0$ in Sec.~\ref{sec:HFLargeSSW}. In this situation the instability develops at $\qv=0$, i.e., at ${\bf k} = {\bf M}_1 +{\bf q} \equiv {\bf M}_1$, and the critical $J_2 = J_{2\text{V}}>1/8$, i.e., the stability region of the $V$ phase extends to the right of $J_2 =1/8$. For arbitrary $S$ we found that the sign of $(1/3+\delta_1)$ depends on $S$. For $S >1$, it is positive, like at large $S$. For $S=1/2$, however, we found that $1/3+\delta_1=-0.1<0$. As the consequence, the $V$ phase becomes unstable before $J_2$ reaches $J_2=1/8$. For $S=1$, our numerical calculation yields a slightly negative $1/3+\delta_1$. We summarize our numerical results for $1/3 + \delta_1$ in Table~\ref{tab:HFgenericS}.

\begin{table}[tbp]
\begin{center}
\begin{tabular}{|c|c|c|c|}
    \hline
    ~ & \quad$S=1/2$\quad  & \quad $S=1$\quad &  \quad$1\ll S\ll \loghsat$\quad  \\
    \hline
 $1/3+\delta_1$   &   $-0.1 $          &    $-0.02$ & $+0.03/S$    \\\hline
 $J_{2V}-J_{2\text{stripe}}\,((\hsa-h)\loghsat)$ & $-$ &   $-$  &$+$\\\hline
  \end{tabular}
\end{center}
\caption{Quantum corrections to the mass of $V$ phase spectrum at momentum $\vect{M}$ (first row), from which the width of overlap between the $V$ phase and the stripe phase can be obtained. A negative width (sign) indicates that the two states don't overlap near $J_2=1/8$.
\label{tab:HFgenericS}}
\end{table}

We analyzed the spin wave spectrum in the stripe phase, near momentum $\pm\vect{K}$. The low energy part of the quadratic Hamiltonian near $\pm\vect{K}$ can be expressed as:
\begin{align}
\label{eq:coplanarLE}
\mathcal{H}^{(2)}=&\frac{S}{2}\sum_{\qv}\\
&\begin{pmatrix}
\cd_{\vect{K}+\qv}\,
c_{-\vect{K}-\qv}
\end{pmatrix}
\begin{pmatrix}
\tilde{\omega}_{\qv}+(3/2+9/2\,\tilde{\delta}_1)\tilde{\theta}^2 & (3/2+9/2\,\tilde{\delta}_2)\tilde{\theta}^2\\
(3/2+9/2\,\tilde{\delta}_2)\tilde{\theta}^2 & \tilde{\omega}_{-\qv}+(3/2+9/2\,\tilde{\delta}_1)\tilde{\theta}^2\\
\end{pmatrix}
\begin{pmatrix}
c_{\vect{K}+\qv}\\
\cd_{-\vect{K}-\qv}\\
\end{pmatrix}\non
\end{align}
where $\tilde \theta$ is the angle between the canted stripe order and the field, $\tilde{\omega}_{\qv}=8J_2-1+\frac{3}{16}(q_x^2+q_y^2)$. Similar to the $V$ phase case, $\tilde{\theta}^2=\alpha_2(\hsa-h)\loghsat$, $\alpha_2>0$. In the large S limit in Sec.~\ref{sec:HFLargeSSW}, $(1/3+\tilde \delta_1)>0$. In this situation, the instability develops at $\pm\vect{K}$, and at $J_{2\text{stripe}}<1/8$. At arbitrary S and $h\lesssim \hsa$, we found that $3/2+9/2\,\tilde{\delta}_1$ and $3/2+9/2\,\tilde{\delta}_2$ both scale as $1/\loghsat$, and it is true as long as $S \ll \loghsat$. Hence, to logarithmic accuracy, the stripe phase becomes unstable right at $J_2=J_{2{\text{stripe}}} =1/8$.

Comparing $J_{2V}$ and $J_{2{\text{stripe}}}$, we see that for $S>1$, the stability regions of the two phases overlap. The Ginzburg-Landau analysis from the previous sub-section
 shows that the transition between the two stable phases is first order. For $S=1/2$ and, possibly, $S=1$, the situation is different because the $V$ phase becomes unstable prior to the $J_2$ at which the stripe phase becomes stable. In this situation, there exists an intermediate phase at which neither the $V$ phase nor the stripe phase is stable. We illustrate this in the inset of Fig.~\ref{fig:summarySemiClassical}.

Whether the intermediate phase at high-field is disordered or has some non-quasi-classical long-range order is not clear at the moment. If we use Eq.~\ref{eq:coplanarZero} for the dispersion, we find that at $J_2= J_{2V}$ the dispersion is quadratic at small $q$. For such dispersion, quantum corrections to sublattice magnetization logarithmically diverge in 2D and eliminate long-range order. We caution, however, that this spectrum was obtained to leading order in $1/\loghsat$. Subleading terms can potentially halt the divergence of the corrections to sublattice magnetization. Still, at $S=1/2$, subleading terms are small near $h = \hsa$, i.e., quantum corrections to sublattice magnetization are large and likely restore $U(1)$ symmetry, at least near $J_2 = J_{2V}$. A phase with a discrete, dimer-like order is another possibility. We verified that a columnar dimer phase is not an option, but this does not exclude some other dimer-like state. And yet another possibility is a disordered, spin-liquid type state, possibly the same as has been detected in numerical studies of zero-field phase diagram around $J_2 =1/8$~\cite{White2015,Sheng2015,Iqbal2016}.
\section{Conclusion}
\label{sec:conclusion}
In this paper, we studied the zero temperature phase diagram of a Heisenberg antiferromagnet on a frustrated triangular lattice with nearest neighbor ($J_1$) and next nearest neighbor ($J_2$) interactions, in a magnetic field. We analyzed the stabilization of the ordered phases at smaller and larger $J_2/J_1$ via order from disorder phenomenon and the phase transition between the ordered states at smaller and larger $J_2/J_1$.
 We first considered the limit of large but finite $S$ and obtained the semiclassical phase diagram in all fields. We found that at $J_2/J_1 < 1/8+ \mc{O}(1/S)$, quantum fluctuations select the same set of co-planar states as at $J_2=0$: the $Y$ state at fields $h < \hsa/3$, the $V$ phase at $h >\hsa/3$, and the UUD phase at $h \approx \hsa/3$. At $J_2 > 1/8 - \mc{O}(1/S)$, quantum fluctuations select the canted stripe phase. The stability regions of the two phases overlap around $J_2/J_1 = 1/8$, and semiclassical spin wave analysis shows that the transition between the two phases is first order, with a finite hysteresis width, of order $1/S$. We next analyzed the phase diagram near the saturation field at arbitrary $S$, by mapping the spin model to a dilute boson gas. We found the same $V$ and stripe phase at smaller and larger $J_2/J_1$. For $S >1$ we also found that the stability regions of the two states overlap, and the transition between them remains first order, like at large $S$. However, for $S=1/2$ and, possibly, $S=1$, we found that there exists an intermediate range near $J_2/J_1 =1/8$, where neither of the two states is stable. We emphasize that this happens already arbitrary close to the saturation field, when the density of bosons is small. In the intermediate region the system either develops a non-quasi-classical long-range order (e.g., becomes dimerized), or remains quantum disordered. We note that the intermediate phase develops for the same $J_2/J_1 \approx 1/8$ where at $h=0$ numerical calculations found evidence for a disordered, possibly spin-liquid state for $S=1/2$~\cite{White2015,Sheng2015,Iqbal2016} (but, apparently, not $S=1$~\cite{Wang2006}). Whether the state we found at $h \approx \hsa$ is the same one as found at $h=0$ remains to be seen. We call for more numerical studies of $J_1-J_2$ model in a finite field.
\section{Acknowledgement}
We acknowledge useful conversations with C. Batista, A. Chernyshev, D. Maslov, N. Perkins, and O. Starykh. The work was supported by the NSF DMR-1523036.
\newpage
\appendix

\section{Holstein-Primakoff transformation}
\label{app:Classical}
In this Appendix we review the  basics of Holstein-Primakoff transformation and spin wave formalism.
 We follow the convention presented below Eq.~\ref{eq:modelH} in the main text. In the formulas below, $N$ is defined as the number of sites in one sublattice, i.e. $N=\frac{N_{\text{tot}}}{n_\text{bands}}$. For example, for the three sublattice states, $n_\text{bands}=3$, and $N=\frac{1}{3}N_{\text{tot}}$.

Spins polarized in the positive z direction are expressed in terms of Holstein-Primakoff (H-P) bosons as:
\begin{align}
S_{\vect{r}}^z(\vect{z}) & =S-a_{\vect{r}}^{\dagger}a_{\vect{r}}\nonumber\\
S_{\vect{r}}^+(\vect{z}) & =\sqrt{2S}\sqrt{1-\frac{a_{\vect{r}}^{\dagger}a_{\vect{r}}}{2S}}a_{\vect{r}}\non\\
S_{\vect{r}}^-(\vect{z}) & =\sqrt{2S}a_{\vect{r}}^{\dagger}\sqrt{1-\frac{a_{\vect{r}}^{\dagger}a_{\vect{r}}}{2S}}
\label{eq:HP}
\end{align}
 The spin operators $S^\alpha(\vect{l})$ in a local coordinate with the local $z$-axis along a vector $\vect{l}=\vect{z}\cos{\theta}-\vect{x}\sin{\theta}$ are related with $S^\alpha(\vect{z})$ defined in the global coordinate (Fig.~\ref{fig:J20PD}) as~\cite{Jolicoeur1989}:
\begin{align}
S^x(\vect{z}) & =\cos{\theta} S^x(\vect{l})-\sin{\theta}S^z(\vect{l})\nonumber\\
S^y(\vect{z}) & =S^y(\vect{l})\non\\
S^z(\vect{z}) & =\sin{\theta} S^x(\vect{l})+\cos{\theta}S^z(\vect{l})
\label{eq:Rot}
\end{align}

To express the Hamiltonian in terms of the H-P bosons, we expand $\sqrt{1-a_{\vect{r}}^{\dagger}a_{\vect{r}}/2S}$ in powers of the bosons. For generic spin, due to the normal ordering of the bosons in the expansion, e.g. $(\ad_ra_r)^2=\ad_r\ad_r a_r a_r+\ad_r a_r$, $S_{\vect{r}}^+$ can be written as:
\begin{align}
S_{\vect{r}}^+ & =\sqrt{2S}(1-\frac{1}{4S}(1+\frac{1}{8S}+\frac{1}{32S^2}+...)a_{\vect{r}}^{\dagger}a_{\vect{r}})a_{\vect{r}}+\mathcal{O}(a^5)\non\\
&=\sqrt{2S}(1+(\sqrt{1-\frac{1}{2S}}-1)a_{\vect{r}}^{\dagger}a_{\vect{r}})a_{\vect{r}}+\mathcal{O}(a^5)
\end{align}
In the limit $S\gg 1$, keeping the leading order in $1/S$:
\begin{equation}
S_{\vect{r}}^+ \approx\sqrt{2S}(1-\frac{1}{4S}a_{\vect{r}}^{\dagger}a_{\vect{r}})a_{\vect{r}}+\mathcal{O}(a^5)
\end{equation}
The Hamiltonian in powers of the H-P bosons can be expanded as:
\be
\mathcal{H}=\mathcal{H}^{(0)}+\mathcal{H}^{(1)}+\mathcal{H}^{(2)}+...+\mc{H}^{(n)}+...
\ee
$\mathcal{H}^{(0)}$ is the ground state energy. $\mathcal{H}^{(n)}$ is the normal ordered $n$-bosons term.

The quadratic term $\mathcal{H}^{(2)}$ can be written in the matrix form as:
\be
\mathcal{H}^{(2)}=\frac{S}{2}\sum_{\vect{k}}\Psi_{\vect{k}}^{\dagger}H_{\vect{k}}\Psi_{\vect{k}}
\ee
where $\Psi_{\vect{k}}=\big(a_{\alpha,\,\kv},\,\ad_{\alpha,\,-\kv}\big)^T$. To obtain the spin wave spectrum and the canonical eigenmodes, one can solve the eigenvalue problem of a matrix defined as $\mc{M}_{\vect{k}}=\tau_3H_{\vect{k}}$. $\tau_3\equiv\sigma_3\otimes I_n$, $\sigma_3$ is the z-component of Pauli matrix that acts on the particle-hole conjugate space and $I_n$ is the identity matrix of size n that acts on the n-sublattice space. To prove, define the $\Psi'_{\vect{k}}$ as the vector formed by eigenmodes. There must exist a matrix $T$ such that $\Psi_{\vect{k}}=T\Psi'_{\vect{k}}$, the quadratic term in the Hamiltonian:
\be\label{eq:classical2}
\mathcal{H}^{(2)}=\frac{1}{2}\sum_{\vect{k}}\Psi_{\vect{k}}^{\dagger}H_{\vect{k}}\Psi_{\vect{k}}=\frac{1}{2}\sum_{\vect{k}}\Psi_{\vect{k}}^{'\dagger}T^{\dagger} H_{\vect{k}} T\Psi'_{\vect{k}}
\ee
$K_{\vect{k}}=T^{\dagger} H_{\vect{k}} T$ is diagonal matrix. On the other hand, from the commutation relation of boson operator which reads as $[a^{\dagger}_i,a_j]=\delta_{i,j}$, we have $T\tau_3 T^{\dagger}=\tau_3$. Combing the two equations, we have:
\begin{align}\label{eq:classicaldiag}
\tau_3 K_{\vect{k}}=\tau_3 T^{\dagger}H_{\vect{k}} T &=\tau_3(\tau_3 T^{-1} \tau_3)H_{\vect{k}} T\non\\
&= T^{-1} \tau_3 H_{\vect{k}} T
\end{align}
Thus solving for $T$ and the eigenenergy of $H_{\vect{k}}$ is equivalent to the eigenvalue problem of matrix $\mathcal{M}_{\kv}=\tau_3 H_{\kv}$. \textit{Q.E.D.}

Different branches of the magnon modes can decouple for certain types of ordered states, such as the stripe phase and $120^{\circ}$ Neel phase discussed in the text. $\mathcal{H}^{(2)}$ can be written as:
\begin{align}
\mathcal{H}^{(2)}&=\frac{S}{2}\sum_{\alpha}\phi^{\dagger}_{\alpha,\kv}H_{\alpha,\kv}\phi_{\alpha,\kv}\non\\
H_{\alpha,\kv}&=
\begin{pmatrix}
A_{\alpha,\kv} & B_{\alpha,\kv}\\
B_{\alpha,\kv} & A_{\alpha,-\kv}
\end{pmatrix}\non\\
\phi_{\alpha,\kv}&=(a_{\alpha,\kv},\ad_{\alpha,-\kv})^T
\end{align}
The eigenmodes of the quadratic Hamiltonian $\mc{H}$ are:
\begin{align}
\mathcal{H}^{(2)}&=\frac{S}{2}\sum_{\alpha}\omega_{\alpha,\kv}\eta^{\dagger}_{\alpha,\kv}\eta_{\alpha,\kv}\non\\
\omega_{\alpha,\kv}&=\sqrt{A_{\alpha,\kv}^2-B_{\alpha,\kv}^2}\non\\
a_{\alpha,\kv}&=u_{\alpha,\kv}\tilde{a}_{\alpha,\kv}+v_{\alpha,\kv}\tilde{a}^{\dagger}_{\alpha,-\kv}\non\\
\eta_{\alpha,\kv}&=(\tilde{a}_{\alpha,\kv},\tilde{\ad}_{\alpha,-\kv})^T
\end{align}
$u_{\alpha,\kv}$ and $v_{\alpha,\kv}$ are defined as:
\begin{align}
u_{\alpha,\kv}=\sqrt{\frac{A_{\alpha,\kv}+\omega_{\alpha,\kv}}{2\omega_{\alpha,\kv}}} \quad\quad
v_{\alpha,\kv}=-\text{sign}(B_{\alpha,\kv})\sqrt{\frac{A_{\alpha,\kv}-\omega_{\alpha,\kv}}{2\omega_{\alpha,\kv}}}
\end{align}
As $\phi_{\alpha,\kv}$ is a linear combination of creation and annihilation canonical modes, the vacuum expectation value of $\phi^{\dagger}_{\kv}\phi_{\kv}$ is non-zero:
\begin{align}\label{eq:paircondensate}
\la \ad_{\ak} a_{\ak}\ra = \big\la\frac{A_{\ak}-\omega_{\ak}}{2\omega_{\ak}} \big\ra\quad\quad
\la \ad_{\ak} \ad_{\alpha,-\kv} \ra =-\big\la\frac{B_{\ak}}{2\omega_{\ak}}\big\ra
\end{align}

$\la ... \ra$ is defined as the average over the Brillouin zone, $\la ... \ra=\frac{1}{N}\sum_{\kv}...$. To obtain the quantum corrections to the spectrum at the leading order in $1/S$, one can work in the basis of $\phi_{\kv}$ and calculate $\delta A_{\kv}$ and $\delta B_{\kv}$ at the order of $1/S$. We replace $A_{\kv}\rightarrow A_{\kv}+\delta A_{\kv}$ and $B_{\kv}\rightarrow B_{\kv}+\delta B_{\kv}$. The normal and anomalous self-energy of the canonical modes are:
\begin{align}
\delta \omega_{\kv}&=\delta A_{\kv}(u_{\kv}^2+v_{\kv}^2)+2\delta B_{\kv}u_{\kv}v_{\kv}\non\\
&=\frac{A_{\kv}\delta A_{\kv}-B_{\kv}\delta B_{\kv}}{\omega_{\kv}}\non\\
\delta \omega^{\text{off}}_{\kv}&=2\delta A_{\kv}u_{\kv}v_{\kv}+\delta B_{\kv}(u_{\kv}^2+v_{\kv}^2)\non\\
&=\frac{A_{\kv}\delta B_{\kv}-B_{\kv}\delta A_{\kv}}{\omega_{\kv}}
\end{align}
Thus the spectrum with quantum corrections is:
\begin{equation}\label{eq:semispectrum}
\omega^{(1)}_{\kv}=\sqrt{(\omega_{\kv}+\delta\omega_{\kv})^2-(\delta\omega^{\text{off}}_{\kv})^2}
=\sqrt{(A^2-B^2)+2(A\delta A-B\delta B)+(\delta A^2-\delta B^2)}\big|_{\kv}
\end{equation}

When $\omega_{\kv}\sim \mc{O}(1)$, the quantum corrections to $\omega^{(1)}_{\ak}$ is at the order of $1/S$ and they won't change the spectrum qualitatively as long as there is no singularity in $\delta \omega_{\kv}$. We have:
\begin{equation}
\omega^{(1)}_{\kv}\simeq\omega_{\kv}+\delta\omega_{\kv}
\end{equation}

When $\omega_{\kv}\sim 0$, one needs to distinguish between two situations. Suppose it is a classical zero mode at $\kv=0$. It can be $A_{\kv=0}=B_{\kv=0}=0$, which is generally the case of accidental degeneracy. Thus the dispersion around $\kv=0$ can be written as $\omega_{\kv}\sim \kv^2$. It can also be $|A_{\kv=0}|=|B_{\kv=0}|\neq0$, which is the case of linearly dispersing zero mode, such as the Goldstone mode. Thus the dispersion around $\kv=0$ can be written as $\omega_{\kv} \sim v\kv$.
We define the quantum corrections to the spectrum as $\delta m$ such that $\omega^{(1)}_{\kv}\simeq \sqrt{\omega_{\kv}^2+\delta m}$. $\delta m$ to the leading order in $1/S$ is expressed in the two cases as:
\begin{align}
\label{eq:largeSmass}
\delta m=
\begin{cases}
(\delta A^2-\delta B^2)\big|_{\kv=0}=\delta \omega^2-(\delta\omega^{\text{off}})^2\big|_{\kv=0} & \text{if } A,\, B|_{\kv=0}=0\\
2(A\delta A-B\delta B)\big|_{\kv=0}=2\omega\delta\omega\big|_{\kv=0} & \text{if }A,\,B|_{\kv=0}\sim\mc{O}(1)
\end{cases}
\end{align}
We calculate $\delta m$ at certain momentum following Eq.~\ref{eq:largeSmass}. One can calculate either $\delta A,\,\delta B$ or $\delta \omega,\,\delta \omega^{\text{off}}$, whichever way is the easiest. For the second case, as $\delta m$ is linear in $\delta A,\, \delta B$, it is most straight forward to obtain the quantum corrections from the cubic terms by calculating $\delta \omega,\,\delta \omega^{\text{off}}$, and corrections from the quartic terms by $\delta A,\,\delta B$, and sum the two contributions.
\section{Semiclassical calculation}
\label{app:LargeS}
To obtain the quantum corrections to the spin wave spectrum near $\hsa$, i.e. calculating $\delta_1,\,\delta_2$ in Eq.~\ref{eq:coplanarLElargeS}, follows the same procedure as the calculation for a generic spin presented in Appendix~\ref{app:HighFieldSW}. In the limit $\loghsat\gg 1$, $S\gg 1$, and $\loghsat/S\ll 1$, as $V_{\qv}=\mc{O}(1)$, and $S\omega_{\kv}=\mc{O}(S)$, the $1/S$ correction to the spectrum is restricted with only one scattering process, and the vertex function don't contain corrections from normal ordering. With logarithmic accuracy, only the log divergent contributions to the scattering process are included.

\begin{figure}[tbp]
\centering
\subfigure[~~$\delta\omega_{\text{c},\,\kv}$]{\includegraphics[scale=0.45]{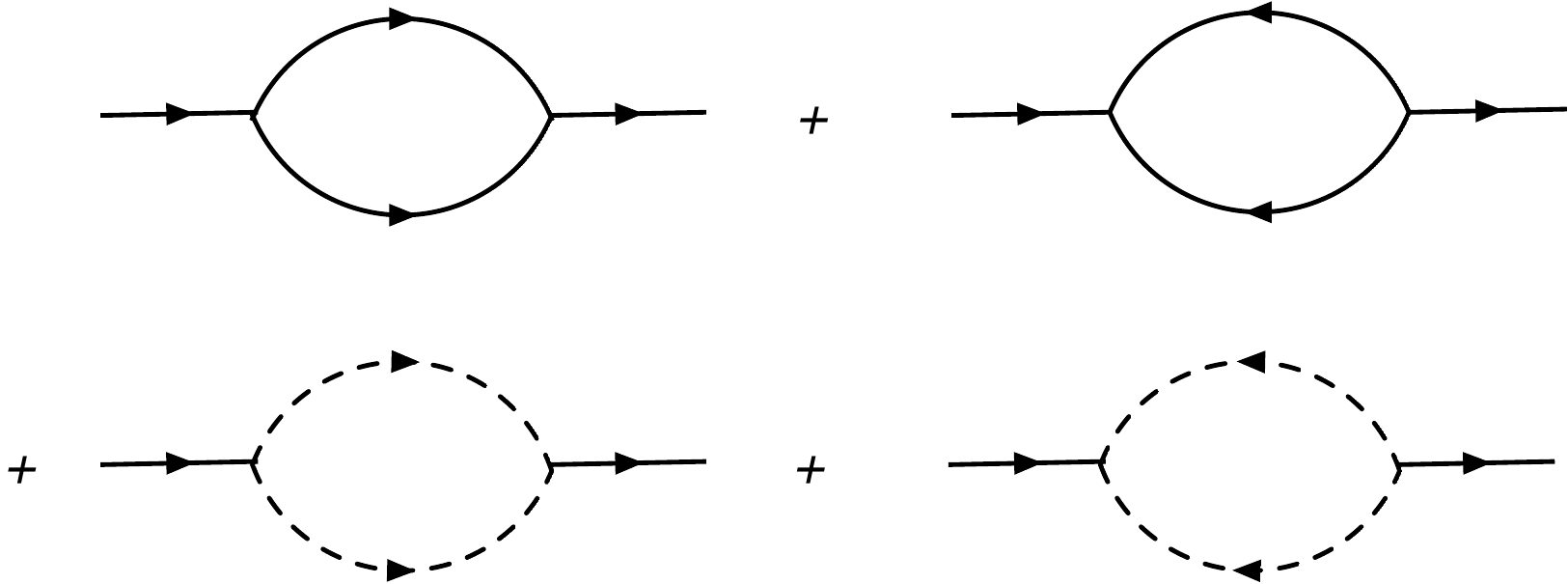}}
\subfigure[~~$\delta\omega^{\text{off}}_{\text{c},\,\kv}$]{\includegraphics[scale=0.45]{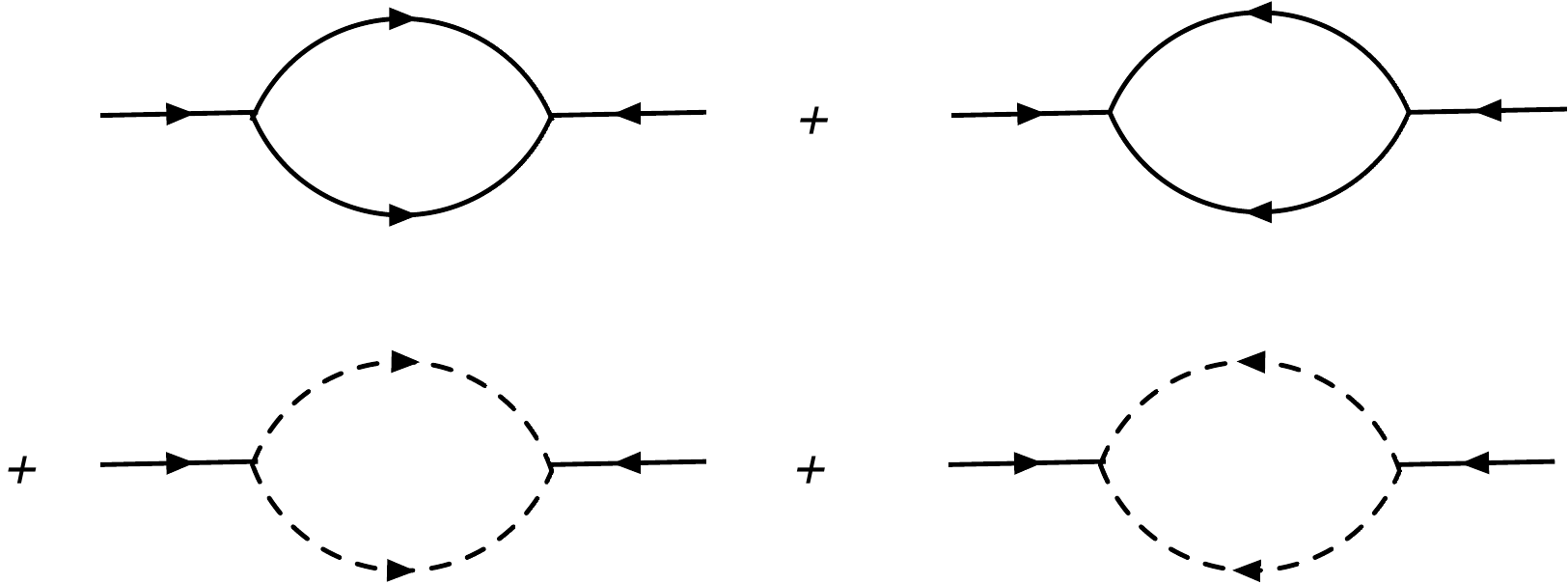}}
\caption{Solid line for the propagator of the $\tilde c$-mode; Dashed line for the propagator of the $\tilde d$-mode. (a) One loop correction to the normal Green function of the $\tilde c$-mode from the cubic vertex. (b) One loop correction to the anomalous Green function of the $\tilde c$-mode from the cubic vertex.
\label{fig:StripeCubic}}
\end{figure}

Details of obtaining the spectrum of the stripe phase with $1/S$ quantum corrections are presented here. The calculations of the coplanar phase follow the same idea and are not presented here. The number as subscript labels the momentum for magnon operator $a,\,b$ and for the structure factor $\zeta$. For example, $a_1\equiv a_{\vect{k}_1}$, $\zeta_1\equiv \zeta_{\vect{k}_1}$.The linear and quadratic terms are given in Eq.~\ref{eq:stripe1},~\ref{eq:stripe2}. The cubic and quartic terms are:
\begin{align}\label{eq:stripe3}
\mathcal{H}^{(3)}_\text{a}&=\frac{\sin{(2\theta)}\sqrt{S}}{2\sqrt{2}\sqrt{N}}\sum_{1,2}[2(1+J_2)(a^{\dagger}_1\ad_2a_{1+2}-b^{\dagger}_1\bd_2b_{1+2})+4\zeta_{\text{b}1}(a^{\dagger}_1b^{\dagger}_2b_{1+2}-b^{\dagger}_1a^{\dagger}_2a_{1+2})]+h.c.\non\\
\mathcal{H}^{(3)}_\text{b}&=-h\frac{\sin{\theta}}{4\sqrt{2}\sqrt{SN}}\sum_{1,2}[(a^{\dagger}_1\ad_2a_{1+2}-b^{\dagger}_1\bd_2b_{1+2})+h.c.]
\end{align}
\begin{align}\label{eq:stripe4}
\mathcal{H}^{(4)}_\text{a}&=-\frac{1}{2N}\sum_{1-3}\{(\zeta_{\text{a}1}-\zeta_{\text{a}\,1-3})(\ad_1\ad_2a_3a_{1+2-3}+\bd_1\bd_2b_3b_{1+2-3})\}+h.c.\\
\mathcal{H}^{(4)}_\text{b}&=\frac{1}{4N}\sum_{1-3}\{\zeta_{\text{b}1}[(1-\cos 2\theta)(\bd_1\ad_2\ad_3a_{1+2+3}+\ad_1\bd_2\bd_3b_{1+2+3})\non\\&-(1+\cos 2\theta)((\bd_1\ad_2a_3a_{1+2-3}+\ad_1\bd_2b_3b_{1+2-3})]+4\zeta_{\text{b}\,1-2}\cos 2\theta\ad_1a_2\bd_3b_{1-2+3}\}+h.c.\non
\end{align}
$\zeta_{\text{a}},\,\zeta_{\text{b}}$ are defined in Eq.~\ref{eq:stripestrucutre}. $a_1\equiv a_{\vect{k}_1}$, etc. Express $\mathcal{H}^{(1)},\,\mathcal{H}^{(3)}$ in terms of the decoupled $c_{\kv},d_{\kv}$ modes defined in Eq.~\ref{eq:stripedecouple}:
\begin{align}
\mathcal{H}^{(1)}&=\sin{\theta} S\sqrt{SN} (c_{\vect{k}}+\cd_{-\vect{k}})\delta_{\vect{k},0}(h-8(1+J_2)\cos{\theta})\non\\
\mathcal{H}^{(3)}&=\frac{\sin{(2\theta)}\sqrt{S}}{\sqrt{N}}\sum_{1,2}\zeta_{\text{b}1}\big(\cd_1\cd_2c_3+\cd_1\dd_2d_3-\dd_1\cd_2d_3-\dd_1\dd_2c_3\big)\delta_{3,1+2}+h.c.
\end{align}
The leading order quantum correction to the linear term is from the cubic term $\mathcal{H}^{(3)}$. We have:
\begin{align}
\delta\mathcal{H}^{(1)}&=\sin{(2\theta)}\sqrt{SN}\,\sum_{\kv}\cd_{\kv}\big(\zeta_{\text{b}\kv}\la\cd_{\pv}c_{\pv}\ra+\la\zeta_{\text{b}\vect{p}}\cd_{\pv}c_{\pv}\ra+\zeta_{\text{b}\kv}\la\dd_{\pv}d_{\pv}\ra\non\\
&\quad-\la\zeta_{\text{b}\vect{p}}\dd_{\pv}d_{\pv}\ra+\la\zeta_{\text{b}\vect{p}}c_{-\pv}c_{\pv}\ra-\la\zeta_{\text{b}\vect{p}}d_{-\pv}d_{\pv}\ra\big)\delta_{0,\vect{k}}+h.c.\non\\
&=\sin{(2\theta)}S\sqrt{SN}\Delta_{\text{c},\kv}c_{\kv}\delta_{0,\kv}+h.c.
\end{align}
We define $\Delta_{\text{c},\kv}$ as:
\begin{equation}
\Delta_{\text{c},\kv}=\frac{1}{S}\big(\zeta_{\text{b}\kv}\la\cd_{\pv}c_{\pv}\ra+\la\zeta_{\text{b}\vect{p}}\cd_{\pv}c_{\pv}\ra+\zeta_{\text{b}\kv}\la\dd_{\pv}d_{\pv}\ra-\la\zeta_{\text{b}\vect{p}}\dd_{\pv}d_{\pv}\ra+\la\zeta_{\text{b}\vect{p}}c_{-\pv}c_{\pv}\ra-\la\zeta_{\text{b}\vect{p}}d_{-\pv}d_{\pv}\ra\big)
\end{equation}
From the condition that the prefactor of the linear term vanishes, we have:
\begin{align}
\label{eq:LargeSfield}
\mathcal{H}^{(1)}+\delta\mathcal{H}^{(1)}&=\sin{\theta} S\sqrt{SN} (c_{\vect{k}}+\cd_{-\vect{k}})\delta_{\vect{k},0}(h-8(1+J_2)\cos{\theta}+2\cos{\theta}\Delta_{\text{c},\kv})=0\non\\
h=h_c+\delta h&=\big(8(1+J_2)-2\Delta_{\text{c},\kv}\delta_{\vect{k},0}\big)\cos{\theta}
\end{align}
To find the quantum corrections to the spectrum, the renormalized relation between the magnetic field $h$ and the angle $\theta$ at order $1/S$ should be included following Eq.~\ref{eq:LargeSfield}. The normal and anomalous self-energy of the canonical $\tilde c$-mode $\delta\omega_{\text{c},\,\kv}$ and $\delta\omega^{\text{off}}_{\text{c},\,\kv}$ from the cubic terms at order $1/S$ can be expressed diagrammatically in Fig.~\ref{fig:StripeCubic}. The canonical Bogoliubov transformations from $c_{\kv},\,d_{\kv}$ to canonical eigenmodes $\tilde{c}_{\kv},\,\tilde{d}_{\kv}$ are:
\begin{align}
c_{\kv}&=l_{\kv}(\tilde{c}_{\kv}+x_{\kv}\tilde{c}^{\dagger}_{-\kv})\non\\
d_{\kv}&=p_{\kv}(\tilde{d}_{\kv}+q_{\kv}\tilde{d}^{\dagger}_{-\kv})
\end{align}
The cubic terms Eq.~\ref{eq:stripe3} expressed in terms canonical modes are:
\begin{align}
\mathcal{H}_{\text{c}}^{(3)}&=\frac{\sin{(2\theta)}\sqrt{S}}{4\sqrt{N}}\sum_{1,2}[\Phi_1(1,2,1+2)\tcd_1\tcd_2\tc_{1+2}-\frac{1}{3}\Phi_2(1,2,-1-2)\tcd_1\tcd_2\tcd_{-1-2}]+h.c.\non\\
\mathcal{H}_{\text{d}}^{(3)}&=\frac{\sin{(2\theta)}\sqrt{S}}{4\sqrt{N}}\sum_{1,2}[\tilde{\Phi}_1(1,2,1+2)\tdd_1\tdd_2\tc_{1+2}+\tilde{\Phi}_2(1,2,-1-2)\tdd_1\tdd_2\tcd_{-1-2}]+h.c.
\end{align}
\begin{align}
\Phi_1(1,2,3)&=\frac{\zeta_{\text{b}1}f_1^-(f_2^+f_3^++f_2^-f_3^-)+\zeta_{\text{b}2}f_2^-(f_1^+f_3^++f_1^-f_3^-)-\zeta_{\text{b}3}f_3^-(f_1^+f_2^+-f_1^-f_2^-)}{\sqrt{\omega_{\text{c},\kv_1}\omega_{\text{c},\kv_2}\omega_{\text{c},\kv_3}}}\nonumber\\
\Phi_2(1,2,3)&=\frac{\zeta_{\text{b}1}f_1^-(f_2^+f_3^+-f_2^-f_3^-)+\zeta_{\text{b}2}f_2^-(f_1^+f_3^+-f_1^-f_3^-)+\zeta_{\text{b}3}f_3^-(f^1_+f^2_+-f_1^-f^2_-)}{\sqrt{\omega_{\text{c},\kv_1}\omega_{\text{c},\kv_2}\omega_{\text{c},\kv_3}}}\nonumber\\
\tilde{\Phi}_1(1,2,3)&=-\frac{\zeta_{\text{b}1}\tilde{f}_1^-(\tilde{f}_2^+f_3^++\tilde{f}_2^-f_3^-)+\zeta_{\text{b}2}\tilde{f}_2^-(\tilde{f}_1^+f_3^++\tilde{f}_1^-f_3^-)+\zeta_{\text{b}3}f_3^-(\tilde{f}_1^+\tilde{f}_2^+-\tilde{f}_1^-\tilde{f}_2^-)}{\sqrt{\omega_{\text{d},\kv_1}\omega_{\text{d},\kv_2}\omega_{\text{c},\kv_3}}}\nonumber\\
\tilde{\Phi}_2(1,2,3)&=\frac{\zeta_{\text{b}1}\tilde{f}_1^-(\tilde{f}_2^+f_3^+-\tilde{f}_2^-f_3^-)+\zeta_{\text{b}2}\tilde{f}_2^-(\tilde{f}_1^+f_3^+-\tilde{f}_1^-f_3^-)-\zeta_{\text{b}3}f_3^-(\tilde{f}_1^+\tilde{f}_2^+-\tilde{f}_1^-\tilde{f}_2^-)}{\sqrt{\omega_{\text{d},\kv_1}\omega_{\text{d},\kv_2}\omega_{\text{c},\kv_3}}}
\end{align}
where\begin{align}
f^{\pm}&=\sqrt{A_{\text{c},\kv}\pm B_{\text{c},\kv}}=\sqrt{\omega_{\text{c},\kv}}l_{\kv}(1\mp x_{\kv})\non\\
\tilde{f}^{\pm}&=\sqrt{A_{\text{d},\kv}\pm B_{\text{d},\kv}}=\sqrt{\omega_{\text{d},\kv}}p_{\kv}(1\mp q_{\kv})
\end{align}
The subscripts of $f^{\pm}$ label the momentum, for example, $f^{\pm}_1\rightarrow f^{\pm}_{\kv_1}$, etc. The correction from cubic terms to the spectrum of the low energy $\tilde c$-mode is:
\begin{align}\label{eq:cubiccorr}
&\delta_{\text{cub}}\omega_{\kv}=\\
&-\frac{\sin{(2\theta)}^2}{8S}\frac{1}{N}\sum_{\vect{q}}(\frac{|\Phi_{1}(\qv,\kv-\qv,\kv)|^2}{\omega_{\text{c},\qv}+\omega_{\text{c},\kv-\qv}-\omega_{\text{c},\kv}}+\frac{|\Phi_{2}(\qv,-\kv-\qv,\kv)|^2}{\omega_{\text{c},\qv}+\omega_{\text{c},-\kv-\qv}+\omega_{\text{c},\kv}})\non\\
&-\frac{\sin{(2\theta)}^2}{8S}\frac{1}{N}\sum_{\vect{q}}(\frac{|\tilde{\Phi}_{1}(\qv,\kv-\qv,\kv)|^2}{\omega_{\text{d},\qv}+\omega_{\text{d},\kv-\qv}-\omega_{\text{c},\kv}}+\frac{|\tilde{\Phi}_{2}(\qv,-\kv-\qv,\kv)|^2}{\omega_{\text{d},\qv}+\omega_{\text{d},-\kv-\qv}+\omega_{\text{c},\kv}})\non\\
&\delta_{\text{cub}}\omega^{\text{off}}_{\kv}=\non\\
&\frac{\sin{(2\theta)}^2}{8S}\frac{1}{N}\sum_{\vect{q}}(\frac{\Phi_{1}(\qv,\kv-\qv,\kv)\Phi_{2}^*(\qv,\kv-\qv,-\kv)}{\omega_{\text{c},\qv}+\omega_{\text{c},\kv-\qv}-\omega_{\text{c},\kv}}+\frac{\Phi_{2}^*(\qv,-\kv-\qv,\kv) \Phi_{1}(\qv,-\kv-\qv,-\kv)}{\omega_{\text{c},\qv}+\omega_{\text{c},-\kv-\qv}+\omega_{\text{c},\kv}})\non\\
&-\frac{\sin{(2\theta)}^2}{8S}\frac{1}{N}\sum_{\vect{q}}(\frac{\tilde{\Phi}_{1}(\qv,\kv-\qv,\kv)\tilde{\Phi}_{2}^*(\qv,\kv-\qv,-\kv)}{\omega_{\text{d},\qv}+\omega_{\text{d},\kv-\qv}-\omega_{\text{c},\kv}}+\frac{\tilde{\Phi}_{2}^*(\qv,-\kv-\qv,\kv) \tilde{\Phi}_{1}(\qv,-\kv-\qv,-\kv)}{\omega_{\text{d},\qv}+\omega_{\text{d},-\kv-\qv}+\omega_{\text{c},\kv}})\non
\end{align}

The quantum corrections from the quartic terms can be found in a similar way or a simpler way as shown below. The quartic terms Eq.~\ref{eq:stripe4} in terms of $c_{\kv},d_{\kv}$ is:
\begin{align}
&\mathcal{H}^{(4)}=\frac{1}{N}\sum_{1,\,2,\,3,\,4}\{\non\\
&-\frac{1}{8}\big(\zeta_{\text{b}1}(1-\cos 2\theta)(\cd_1\cd_2\cd_3c_4-\dd_1\dd_2\dd_3d_4-2\dd_1\dd_2\cd_3c_4+2\cd_1\cd_2\dd_3d_4+\cd_1\dd_2\dd_3c_4\non\\&\quad-\dd_1\cd_2\cd_3d_4)\delta_{1+2+3-4}+h.c.\big)\nonumber\\
&+\frac{1}{16}(1+\cos 2\theta)\big((\zeta_{\text{b}1}+\zeta_{\text{b}2}+\zeta_{\text{b}3}+\zeta_{\text{b}4})(\cd_1\cd_2c_3c_4-\dd_1\dd_2d_3d_4)+(\zeta_{\text{b}1}+\zeta_{\text{b}2}-\zeta_{\text{b}3}-\zeta_{\text{b}4})\nonumber\\&\quad(-\dd_1\dd_2c_3c_4+\cd_1\cd_2d_3d_4)+4(-\zeta_{\text{b}1}+\zeta_{\text{b}2}-\zeta_{\text{b}3}+\zeta_{\text{b}4})\dd_1\cd_2d_3c_4\big)\delta_{1+2-3-4}\nonumber\\
&+\frac{1}{4}\cos 2\theta\big((\zeta_{\text{b}\,1-3}+\zeta_{\text{b}\,2-3})(\cd_1\cd_2c_3c_4+\dd_1\dd_2d_3d_4-\dd_1\dd_2c_3c_4-\cd_1\cd_2d_3d_4)\nonumber\\
&\quad+4(\zeta_{\text{b}\,1-3}-\zeta_{\text{b}\,2-3})\dd_1\cd_2d_3c_4\big)\delta_{1+2-3-4}\nonumber\\
&-\frac{1}{8}(\zeta_{\text{a}1}+\zeta_{\text{a}2}+\zeta_{\text{a}3}+\zeta_{\text{a}4}-2\zeta_{\text{a}\,1-3}-2\zeta_{\text{a}\,2-3})(\cd_1\cd_2c_3c_4+\dd_1\dd_2d_3d_4+\dd_1\dd_2c_3c_4\nonumber\\
&\quad+\cd_1\cd_2d_3d_4+4\dd_1\cd_2d_3c_4)\delta_{1+2-3-4}\}
\end{align}
We contract two bosons as pair condensates following Eq.~\ref{eq:paircondensate}. Define the quantum corrections to the coefficient of $\cd c$ as $\delta A_{\kv}$, to the coefficient of $\cd \cd$ as $\delta B_{\kv}$. We have:
\begin{align}
\label{eq:quarticccorr}
\delta A_{\kv}
&=-\frac{1}{8}(1-\cos 2\theta)[(2\la\zeta_\text{b}\cd\cd\ra+\zeta_{\text{b}\kv}\la\cd\cd\ra-2\la\zeta_\text{b}\dd\dd\ra+\zeta_{\text{b}\kv}\la\dd\dd\ra)+h.c.]\nonumber\\
&\quad+\frac{1}{2}(1+\cos 2\theta)(\zeta_{\text{b}\kv}\la\cd c\ra+\la\zeta_\text{b}\cd c\ra+\zeta_{\text{b}\kv}\la \dd d\ra-\la \zeta_\text{b}\dd d \ra)\nonumber\\
&\quad+\cos 2\theta(\zeta_{\text{b}\,0}\la\cd c\ra+\la\zeta_{\text{b}\,\kv-\pv}\cd c\ra+\zeta_{\text{b}\,0}\la\dd d\ra-\la\zeta_{\text{b}\,\kv-\pv}\dd d\ra)\nonumber\\
&\quad-\big((\zeta_{\text{a}\,\kv}-\zeta_{\text{a}\,0})\la\cd c\ra+\la(\zeta_{\text{a}\,\pv}-\zeta_{\text{a}\,\kv-\pv})\cd c\ra+(\zeta_{\text{a}\,\kv}-\zeta_{\text{a}\,0})\la\dd d\ra+\la(\zeta_{\text{a}\,\pv}-\zeta_{\text{a}\,\kv-\pv})\dd d\ra\big)\non\\
\frac{1}{2}\delta B_{\kv}
&=-\frac{1}{8}(1-\cos 2\theta)[(\la\zeta_\text{b}\cd c\ra+2\zeta_{\text{b}\kv}\la\cd c\ra-\la\zeta_\text{b}\dd d\ra+2\zeta_{\text{b}\kv}\la\dd d\ra)]\nonumber\\
&\quad+\frac{1}{8}(1+\cos 2\theta)(\zeta_{\text{b}\kv}\la c c\ra+\la\zeta_\text{b} c c\ra+\zeta_{\text{b}\kv}\la d d\ra-\la \zeta_\text{b} d d \ra)\nonumber\\
&\quad+\frac{1}{4}\cos 2\theta(\la(\zeta_{\text{b}\,\kv+\pv}+\zeta_{\text{b}\,\kv-\pv}) c c\ra-\la(\zeta_{\text{b}\,\kv+\pv}+\zeta_{\text{b}\,\kv-\pv}) d d\ra)\nonumber\\
&\quad+\frac{1}{4}(\la(\zeta_{\text{a}\,\kv+\pv}+\zeta_{\text{a}\,\kv-\pv}) c c\ra+\la(\zeta_{\text{a}\,\kv+\pv}+\zeta_{\text{a}\,\kv-\pv}) d d\ra)\nonumber\\
&\quad-\frac{1}{4}(\zeta_{\text{a}\kv}\la c c\ra+\la \zeta_\text{a} c c\ra+\zeta_{\text{a}\kv}\la d d\ra+\la \zeta_\text{a} d d\ra)
\end{align}
The $1/2$ in front of $\delta B_{\kv}$ comes from the fact that in the matrix representation, term like $\cd\cd$ is counted only once, while term like $\cd c$ is counted twice. The sub-index omitted inside $\langle ... \rangle$ follows the prescription: $\cd\cd\rightarrow \cd_{\pv}\cd_{-\pv}$, $\cd c\rightarrow \cd_{\pv} c_{\pv}$, $\zeta \cd\cd\rightarrow \zeta_{\pv}\cd_{\pv}\cd_{-\pv} $, etc. Combining Eq.~\ref{eq:LargeSfield}, Eq.~\ref{eq:cubiccorr}, Eq.~\ref{eq:quarticccorr}, we obtain quantum corrections to the spectrum $\delta m$ at momenta $\vect{\widetilde M}_2$, $\vect{\widetilde M}_3$ and $\pm\vect{\widetilde K}$.
\section{Dilue bose gas approximation in high field}
\label{app:HighField}
In this section, we show details of determining the quartic couplings $\Gamma$ in the expressions of the condensate energy, i.e. Eq.~\ref{eq:Eden10}, Eq.~\ref{eq:Eden11} and Eq.~\ref{eq:Eden2}. The quartic couplings $\Gamma$ are at leading order in $1/S$ in Sec.~\ref{sec:LargeSaction}, and are summed up to all orders in $1/S$ in Sec.~\ref{sec:HighFieldGL}. In the following, we show how $\Gamma$ is obtained for a generic spin. The calculation of $\Gamma$ at leading order in $1/S$ follows the same idea and is simpler, and it will not discussed further here. One can refer~\cite{Chubukov2014} for more details.

As shown in the main text, to determine the magnetic order structure, compared with calculating the exact numerical values of $\Gamma$, the sign of the differences of $\Gamma$s (e.g. $\Gamma_1$ v.s. $\Gamma_2$, $\frac{1}{2}(\Gamma_1+\Gamma_2)\bar{\Gamma}_1$ v.s. $\Gamma^2_{\Delta,\phi}$) are more relevant. As the sign of the differences of relevant $\Gamma$s shouldn't change across the 2nd order phase transition from right above $h_{\text{sat}}$ to right below \hsat, the criterion introduced above to know the order structure slightly \textit{below} $h_{\text{sat}}$ can be determined by the fully renormalized four-point vertex function slightly \textit{above} \hsat.

The fully renormalized 2n-point vertex functions of ferromagnet can be determined exactly above $h_{\text{sat}}$ for all spins, as the single-magnon excitations of the ferromagnet are exact and the quantum corrections to the 2n-point vertex only come from magnon-magnon scattering.

The four-point and six-point bare vertex functions are defined in Eq.~\ref{eq:QuadraticHF} as $ V_{\vect{q}}(\vect{k_1},\vect{k_2})$ and $ U_{\vect{q},\qv'}(\vect{k_1},\vect{k_2},\kv_3) $:
\begin{align}\label{eq:fourptvertexexp}
V_{\vect{q}}(\vect{k_1},\vect{k_2}) = \frac{1}{2}[(J_{\vect{q}}+J_{\vect{k_2}-\vect{k_1}-\vect{q}})+2S(K_s-1)(J_{\vect{k_1}}+J_{\vect{k_1+q}}+J_{\vect{k_2}}+J_{\vect{k_2-q}})]
\end{align}
$J_{\qv}$ is defined in Eq.~\ref{eq:QuadraticHF}. The expression includes normal ordering of the magnon to all orders in $1/S$ by $K_s=\sqrt{1-\frac{1}{2S}}$. To the leading order in $1/S$, $K_s=1-1/4S$.
$ U_{\vect{q},\qv'}(\vect{k_1},\vect{k_2},\kv_3) $ keeping the $1/S$ correction from the normal ordering is:
\begin{align}\label{eq:sixptvertexexp}
 U_{\vect{q},\qv'}(\vect{k_1},\vect{k_2},\kv_3) =& \frac{1}{9}(1+1/4S)\big(J_{\kv_1+\qv}+J_{\kv_3+\qv}+J_{\kv_1+\kv_3-\kv_2+\qv}+J_{\kv_1+\qv'}+J_{\kv_2+\qv'}\non\\
& +J_{\kv_1+\kv_2-\kv_3+\qv'}+J_{\kv_2+\kv_3-\kv_1-\qv-\qv'}+J_{\kv_2-\qv-\qv'}+J_{\kv_3-\qv-\qv'}\big)\non\\
&-\frac{1}{6}(1+3/4S)(J_{\vect{k_1}}+J_{\vect{k_2}}+J_{\kv_3}+J_{\kv_1+\qv+\qv'}+J_{\kv_2-\qv}+J_{\kv_3-\qv'})]
\end{align}
We define the fully renormalized four-point vertex function as $\Gamma_{\vect{q}}(\vect{k_1},\vect{k_2})$. The quartic coefficients $\Gamma$ in the action are determined by $\Gamma_{\vect{q}}(\vect{k_1},\vect{k_2})$ at particular momenta as:
\begin{align}\label{eq:vertexdef}
\Gamma_1&=\Gamma_{\vect{0}}(\vect{K},\vect{K})\non\\
\Gamma_2&=\Gamma_{\vect{0}}(\vect{K},-\vect{K})+\Gamma_{-2\vect{K}}(\vect{K},-\vect{K})\non\\
\bar{\Gamma}_1&=\Gamma_{\vect{0}}(\vect{M_1},\vect{M_1})\non\\
\bar{\Gamma}_2&=\Gamma_{\vect{0}}(\vect{M_1},\vect{M_2})+\Gamma_{\vect{M_2-M_1}}(\vect{M_1},\vect{M_2})\non\\
\Gamma_{\Delta,\phi} &= \Gamma_{\vect{0}}(\vect{M_1},\vect{K})+\Gamma_{\vect{K-M_1}}(\vect{M_1},\vect{K})\non\\
\bar{\Gamma}_{\Delta,\phi} &= \Gamma_{\vect{K-M_1}}(\vect{M_1},\vect{M_1})+\Gamma_{\vect{-K-M_1}}(\vect{M_1},\vect{M_1})
\end{align}

 To find $\Gamma_{\vect{q}}(\vect{k_1},\vect{k_2})$, all orders of magnon-magnon scattering process should be counted (see Fig.~\ref{fig:BSconsistency}), which is equivalent to solving a consistency equation, also known as Bethe-Salpeter (BS) equation:
\begin{equation}
\Gamma_{\vect{q}}(\vect{k_1},\vect{k_2})=V_{\vect{q}}(\vect{k_1},\vect{k_2})-\frac{1}{N}\sum_{\vect{q'}}\frac{\Gamma_{\vect{q'}}(\vect{k_1},\vect{k_2})V_{\vect{q-q'}}(\vect{k_1+q'},\vect{k_2-q'})}{S(\omega_{\vect{k_1+q'}}+\omega_{\vect{k_2-q'}})}
\label{eq:BS}
\end{equation}
To solve for $\Gamma_{\vect{q}}(\vect{k_1},\vect{k_2})$, we follow the method introduced in~\cite{Balents2013}, which converts the problem of solving an integral equation to that of solving a matrix equation. As $V_{\vect{q}}(\vect{k_1},\vect{k_2})$, $V_{\vect{q-q'}}(\vect{k_1+q'},\vect{k_2-q'})$ can be expanded by lattice Harmonics, $\Gamma_{\vect{q}}(\vect{k_1},\vect{k_2})$ can also be expressed by the lattice Harmonics. We write the ansatz for $\Gamma_{\vect{q}}(\vect{k_1},\vect{k_2})$ as:
\begin{align}\label{eq:vertexansatz}
\Gamma_{\vect{q}}(\vect{k_1},\vect{k_2})&=A_0+A_{\alpha}\cos q_{\alpha}+B_{\alpha} \sin q_{\alpha}+\tilde{A}_{\alpha} \cos \tilde{q}_{\alpha}+\tilde{B}_{\alpha} \sin \tilde{q}_{\alpha}\equiv\mathcal{A}^T \Gamma_{\text{b}\vect{q}}\non\\
\Gamma_{\text{b}\vect{q}}&=\{1,\cos{q_{\alpha}},\sin{q_{\alpha}},\cos{\tilde{q}_{\alpha}},\sin{\tilde{q}_{\alpha}}\}
\end{align}
$\alpha=1,2,3\,$ and $q_{\alpha}\,,\,\tilde{q}_{\alpha}$ are defined as $q_{\alpha}=\vect{q}\cdot\vect{\delta}_{\alpha},\,\tilde{q}_{\alpha}=\vect{q}\cdot\vect{l}_{\alpha}$. $\vect{\delta}_{\alpha},\,\vect{l}_{\alpha}$ are defined in Fig.~\ref{fig:singlelattice}. Express $V_{\vect{q}}(\vect{k_1},\vect{k_2})$, $V_{\vect{q-q'}}(\vect{k_1+q'},\vect{k_2-q'})$ in the basis of $\Gamma_{\text{b}\vect{q}}$ as:
\begin{align}
V_{\vect{q}}(\vect{k_1},\vect{k_2})&=\mathcal{A}_0^T \Gamma_{\text{b}\vect{q}}\non\\
V_{\vect{q-q'}}(\vect{k_1+q'},\vect{k_2-q'})&=\Gamma_{\text{b}\qv'}^T\mc{V}_0\Gamma_{\text{b}\vect{q}}
\end{align}
Eq.~\ref{eq:BS} in the matrix form is:
\begin{equation}\label{eq:matrixBS}
\mathcal{A}^T \Gamma_{\text{b}\vect{q}}=\mathcal{A}_0^T \Gamma_{\text{b}\vect{q}}-\mc{A}^T\Big(\frac{1}{N}\sum_{\qv'}\frac{ \Gamma_{\text{b}\qv'} \Gamma_{\text{b}\qv'}^T}{S(\omega_{\vect{k_1+q'}}+\omega_{\vect{k_2-q'}})}\Big)\mc{V}_0 \Gamma_{\text{b}\qv}
\end{equation}
$\Gamma_{\qv}(\kv_1,\kv_2)$ relevant to find the quartic coupling $\Gamma$ satisfies $\omega_{\kv_1}=\omega_{\kv_2}=0$. So the sum over $\qv'$, $\frac{1}{N}\sum_{\qv'}\frac{ \Gamma_{\text{b}\qv'} \Gamma_{\text{b}\qv'}^T}{\omega_{\vect{k_1+q'}}+\omega_{\vect{k_2-q'}}}$ is logarithmically divergent. The solution to Eq.~\ref{eq:matrixBS} can be expanded order by order in $\frac{1}{|\log \mu|}$, where $\mu$ is the low energy cutoff with $\mu\rightarrow 0^+$ as $h\rightarrow \hsa^+$.

Solving for $\mc{A}$ in Eq.~\ref{eq:matrixBS} is however not efficient numerically. In the following, we show the simplification of Eq.~\ref{eq:matrixBS} to Eq.~\ref{eq:matrixBS2}. The label of incoming momenta $(\vect{k_1},\vect{k_2})$ is omitted to keep the expressions compact. First take the average of Eq.~\ref{eq:BS} with respect to $\vect{q}$:
\begin{align}\label{eq:BSave}
\langle\Gamma_{\vect{q}}\rangle_{\vect{q}}&=\langle V_{\vect{q}}\rangle_{\vect{q}}-\frac{1}{N}\sum_{\vect{q'}}\frac{\Gamma_{\vect{q'}}\langle V_{\vect{q-q'}}(\vect{k_1+q'},\vect{k_2-q'})\rangle_{\vect{q}}}{S(\omega_{\vect{k_1+q'}}+\omega_{\vect{k_2-q'}})}
\end{align}
As $\langle\cos q_{\alpha}\rangle= 0,\,\langle\sin q_{\alpha}\rangle= 0, \,\langle\cos \tilde{q}_{\alpha}\rangle= 0, \,\langle\sin \tilde{q}_{\alpha}\rangle= 0$, we have $\langle J_{\vect{q+k}}\rangle_{\vect{q}}= 0$. Thus Eq.~\ref{eq:BSave} gives:
\begin{align}
A_0&=S(K_s-1)(J_{\vect{k}_1}+J_{\vect{k}_2})-S(K_s-1)\frac{1}{N}\sum_{\vect{q'}}\frac{\Gamma_{\vect{q'}}(J_{\vect{k_1+q'}}+J_{\vect{k_2-q'}})}{S(\omega_{\vect{k_1+q'}}+\omega_{\vect{k_2-q'}})}\non\\
&=S(K_s-1)\Big((J_{\vect{k}_1}+J_{\vect{k}_2})-\Big\langle\frac{\Gamma_{\vect{q'}}(J_{\vect{k_1+q'}}+J_{\vect{k_2-q'}}-J_{\vect{k_1}}-J_{\vect{k_2}}+J_{\vect{k_1}}+J_{\vect{k_2}})}{S(\omega_{\vect{k_1+q'}}+\omega_{\vect{k_2-q'}})}\Big\rangle_{\vect{q'}}\Big)\non\\
&=S(K_s-1)\Big((J_{\vect{k}_1}+J_{\vect{k}_2})(1-\Big\langle\frac{\Gamma_{\vect{q'}}}{S(\omega_{\vect{k_1+q'}}+\omega_{\vect{k_2-q'}})}\Big\rangle_{\vect{q'}})-A_0/S\Big)
\label{eq:aveBS}
\end{align}
The first consistency equation gives:
\begin{align}
\frac{K_s}{K_s-1}A_0&=S(J_{\vect{k}_1}+J_{\vect{k}_2})\Big(1-\Big\langle\frac{\Gamma_{\vect{q'}}}{S(\omega_{\vect{k_1+q'}}+\omega_{\vect{k_2-q'}})}\Big\rangle_{\vect{q'}}\Big)
\label{eq:BS1}
\end{align}
Secondly, plug the expression of $V_{\vect{q}}(\vect{k_1},\vect{k_2})$ into Eq.~\ref{eq:BS}
\begin{align}
&\Gamma_{\vect{q}}(\vect{k_1},\vect{k_2})=\non\\
&\quad\frac{1}{2}(J_{\vect{q}}+J_{\kv_2-\kv_1-\vect{q}})+S(K_s-1)(J_{\vect{k}_1}+J_{\vect{k_1+q}}+J_{\vect{k}_2}+J_{\vect{k_2-q}})-\frac{1}{N}\times\non\\
&\quad\sum_{\vect{q'}}\frac{\Gamma_{\vect{q'}}\big(\frac{1}{2}(J_{\vect{q-q'}}+J_{\kv_2-\kv_1-\qv-\qv'})+S(K_s-1)(J_{\vect{k_1+q'}}+J_{\vect{k_1+q}}+J_{\vect{k_2-q'}}+J_{\vect{k_2-q}})\big)}{S(\omega_{\vect{k_1+q'}}+\omega_{\vect{k_2-q'}})}\non\\
&=\frac{1}{2}(J_{\vect{q}}+J_{\kv_2-\kv_1-\vect{q}})-\frac{1}{N}\sum_{\vect{q'}}\frac{\Gamma_{\vect{q'}}\frac{1}{2}(J_{\vect{q-q'}}+J_{\kv_2-\kv_1-\qv-\qv'})}{S(\omega_{\vect{k_1+q'}}+\omega_{\vect{k_2-q'}})}\non\\
&\quad+S(K_s-1)\big((J_{\vect{k_1}}+J_{\vect{k_2}})-\frac{1}{N}\sum_{\vect{q'}}\frac{\Gamma_{\vect{q'}}(J_{\vect{k_1+q'}}+J_{\vect{k_2-q'}})}{S(\omega_{\vect{k_1+q'}}+\omega_{\vect{k_2-q'}})})\non\\
&\quad+S(K_s-1)(J_{\vect{k_1+q}}+J_{\vect{k_2-q}})(1-\frac{1}{N}\sum_{\vect{q'}}\frac{\Gamma_{\vect{q'}}}{S(\omega_{\vect{k_1+q'}}+\omega_{\vect{k_2-q'}})})
\end{align}
The second to last line is exactly $A_0$ from the first line of Eq.~\ref{eq:aveBS}; the last line is $K_s\frac{(J_{\vect{k_1+q}}+J_{\vect{k_2-q}})}{(J_{\vect{k_1}}+J_{\vect{k_2}})}A_0$ according to Eq.~\ref{eq:BS1}, so we have:
\begin{align}
&\Gamma_{\vect{q}}=\\
&\frac{1}{2}(J_{\vect{q}}+J_{\kv_2-\kv_1-\vect{q}})-\frac{1}{N}\sum_{\vect{q'}}\frac{\Gamma_{\vect{q'}}\frac{1}{2}(J_{\vect{q-q'}}+J_{\kv_2-\kv_1-\qv-\qv'})}{S(\omega_{\vect{k_1+q'}}+\omega_{\vect{k_2-q'}})}+A_0\big(1+K_s\frac{(J_{\vect{k_1+q}}+J_{\vect{k_2-q}})}{(J_{\vect{k_1}}+J_{\vect{k_2}})}\big)
\label{eq:BS2}
\end{align}
Define a integral matrix $\tau$ as:
\begin{equation}
\tau_{\vect{k}_1,\vect{k}_2}=\frac{1}{N}\sum_{\vect{q'}}\frac{\Gamma_{\text{b}\vect{q'}}\Gamma_{\text{b}\vect{q'}}^T}{S(\omega_{\vect{k_1+q'}}+\omega_{\vect{k_2-q'}})+2\mu}
\end{equation}
The consistency equations Eq.~\ref{eq:BS1}, Eq.~\ref{eq:BS2} in the matrix form are:
\begin{align}
\label{eq:matrixBS10}\text{id}^T\Gamma_{\text{b}\vect{q}}&=\mathcal{A}^T(\frac{K_s}{K_s-1}\frac{1}{S(J_{\vect{k_1}}+J_{\kv_2})}+\tau_{\vect{k}_1,\vect{k}_2})\,\text{id}\cdot\text{id}^T\Gamma_{\text{b}\vect{q}}\quad \text{id}=(1,0,0,0,0)^T \\
\label{eq:matrixBS11}\mathcal{J}^T\Gamma_{\text{b}\vect{q}}&=\mathcal{A}^T(\mathcal{I}+\tau_{\vect{k}_1,\vect{k }_2}\mathcal{M}-\mathcal{K})\Gamma_{\text{b}\vect{q}}
\end{align}
$\mathcal{I}$ is the identity matrix. $\mathcal{J}$, $\mathcal{M}$ and $\mathcal{K}$ depend on the two incoming momenta $(\kv_1,\kv_2)$, and they are defined as:
\begin{align}
&\frac{1}{2}(J_{\vect{q}}+J_{\kv_2-\kv_1-\vect{q}})=\mathcal{J}^T \Gamma_{\text{b}\vect{q}}\\
&\mathcal{J}=\{0,1+\cos(k_{2\alpha}-k_{1\alpha}),\sin(k_{2\alpha}-k_{1\alpha}),J_2(1+\cos(\tilde{k}_{2\alpha}-\tilde{k}_{1\alpha})),J_2\sin(\tilde{k}_{2\alpha}-\tilde{k}_{1\alpha})\}^T\non
\end{align}
\begin{align}
&\frac{1}{2}(J_{\vect{q-q'}}+J_{\kv_2-\kv_1-\qv-\qv'})=\Gamma_{\text{b}\vect{q'}}^T\mathcal{M}\Gamma_{\text{b}\vect{q}}\\
&\mathcal{M}=
\begin{pmatrix}
0 & 0 & 0 & 0 & 0\\
0 & 1+\cos k_\alpha & \sin k_{\alpha} & 0 & 0\\
0 &  \sin k_{\alpha} & 1-\cos k_\alpha & 0 & 0\\
0 & 0 & 0 & J_2(1+\cos \tilde{k}_\alpha) & J_2 \sin \tilde{k}_{\alpha}\\
0 & 0 & 0& J_2 \sin \tilde{k}_{\alpha} & J_2(1-\cos \tilde{k}_\alpha)
\end{pmatrix}\non
\end{align}
$k_{\alpha},\,\tilde{k}_{\alpha}$ are defined as $\kv\cdot\vect{\delta}_{\alpha},\,\kv\cdot\vect{l}_{\alpha}$ respectively, $\kv=\kv_2-\kv_1$.
\begin{align}
&A_0\big(1+K_s\frac{(J_{\vect{k_1+q}}+J_{\vect{k_2-q}})}{(J_{\vect{k_1}}+J_{\vect{k_2}})}\big)=\mathcal{A}^T\mathcal{K}\Gamma_{\text{b}\vect{q}}\\
\text{The first row of $\mc{K}$ is:}\non\\
&\mathcal{K}_1=\big\{1 , \frac{2K_s}{J_{\vect{k_1}}+J_{\kv_2}} (\cos k_{1\alpha}+\cos k_{2\alpha}) , \frac{2K_s}{J_{\vect{k_1}}+J_{\kv_2}} (-\sin k_{1\alpha}+\sin k_{2\alpha}) , \non\\&\qquad\qquad\frac{2J_2K_s}{J_{\vect{k_1}}+J_{\kv_2}} (\cos \tilde{k}_{1\alpha}+\cos \tilde{k}_{2\alpha}) , \frac{2J_2K_s}{J_{\vect{k_1}}+J_{\kv_2}} (-\sin \tilde{k}_{1\alpha}+\sin \tilde{k}_{2\alpha})\big\}\non
\end{align}
Other rows of $\mc{K}$ are zero.

Note that the matrix on the RHS of Eq.~\ref{eq:matrixBS10} is nonzero only in the first column, and the matrix on the RHS of Eq.~\ref{eq:matrixBS11} is zero in the first column, i.e. combing the two equations by adding them doesn't lose any information, and we have:
\begin{align}\label{eq:matrixBS2}
\mathcal{A}^T\mathcal{O}_{\vect{k_1},\vect{k_2}}\Gamma_{\text{b}\vect{q}}&=\bar{\mathcal{J}}^T\Gamma_{\text{b}\vect{q}}\\
\text{where\quad}\mathcal{O}_{\vect{k_1},\vect{k_2}}&=
\mathcal{I}+\tau_{\kk}(\mathcal{M}+\text{id}\cdot\text{id}^T)-(\mathcal{K}_{\kv_1,\kv_2}-\frac{K_s}{K_s-1}\frac{1}{S(J_{\vect{k_1}}+J_{\kv_2})}\text{id}\cdot\text{id}^T)\non\\
\bar{\mathcal{J}}&=\mathcal{J}+\text{id}\non
\end{align}
As matrix $\mc{M}$ doesn't depend on $K_s$, $\tau_{\kv_1,\,\kv_2}$ in $\mc{O}_{\kv_1,\,\kv_2}$ couple to $K_s$. To solve for $\mathcal{A}^T$
\begin{align}
\mathcal{A}^T=\bar{\mathcal{J}}^T\mathcal{O}_{\vect{k_1},\vect{k_2}}^{-1}
\label{eq:solveA}
\end{align}
Near \hsat, at $h=\mu+h_{\text{sat}}$ when $\mu\rightarrow0^+$, the matrix elements of $\tau_{\vect{k_1},\vect{k_2}}$ is $\log$ divergent. $\tau_{\kk}$ is expressed as the sum of the $\log$ divergent part and the finite part.
\begin{equation}
\tau_{\kk}=\frac{|\log{\mu}|}{S}\tau^{(0)}_{\kk}+\frac{1}{S}\tau^{(1)}_{\kk}
\label{eq:tau}
\end{equation}
We keep the spin $S$ explicit, and $\tau^{(0)},\,\tau^{(1)}$ are independent of $S$. So we can solve for $\mathcal{A}$ order by order in $\InvLog$. It is straight forward to find $\tau^{(0)}_{\kk}$ using:
\begin{align}
&\frac{1}{N}\sum_{\vect{q}'}\rightarrow\frac{1}{\mc{A}_{B.Z.}}\int_{B.Z.}\text{d}\vect{q}'\non\\
&\frac{1}{\mc{A}_{B.Z.}}\int_{\Lambda}\text{d}\vect{q}'\frac{q_x^{\alpha}q_y^{\beta}}{a q_x^2+b q_y^2+\mu}=\frac{\pi}{\mc{A}_{B.Z.}}\frac{1}{\sqrt{a}^{\alpha+1}}\frac{1}{\sqrt{b}^{\beta+1}}|\log{\frac{\mu}{\Lambda}}|\delta_{\alpha,0}\delta_{\beta,0}
\end{align}
We find that the leading order of $\Gamma_{\qv}=\mc{A}^T\Gamma_{\text{b}\qv}$ only depend on the $\log$ divergent part of $\mathcal{O}_{\kk}$, so it doesn't depend on $K_s=\sqrt{1-\frac{1}{2S}}$. The leading order of $\Gamma_{\qv}$ should be $\sim\frac{S}{|\log{\mu}|}$, we get it by
\begin{equation}
\Gamma^{(0)}_{\qv}=(\mathcal{A}^{(0)})^T\Gamma_{\text{b}\qv}=\frac{S}{|\log{\mu}|}\lim_{\mu\rightarrow0}\frac{|\log{\mu}|}{S}\bar{\mathcal{J}}^T\big( \mathcal{I}+\frac{|\log{\mu}|}{S}\tau^{(0)}_{\vect{k}_1,\vect{k}_2}(\mathcal{M}+\text{id}\cdot\text{id}^T)\big)^{-1}\Gamma_{\text{b}\qv}
\end{equation}
 To solve $\Gamma_{\qv}$ at higher orders, we solve Eq.~\ref{eq:solveA} exactly. The quartic couplings can be obtained following Eq.~\ref{eq:vertexdef}.
\section{Spectrum calculation in high field}
\label{app:HighFieldSW}
In this section, we show details of calculating the magnon spectrum of the $V$ phase right below \hsat. We expand the H-P Hamiltonian in powers of the tilt angle $\theta_1$, $\theta_2$, and then relate them with the magnetic field $h\lesssim \hsa$. To the leading order in the tilt angle $\theta_1$, as $\theta_2=-2\theta_1+\mc{O}(\theta_1^3)$, it is enough to replace $\theta_2$ with $-2\theta_1$ in expression of the quadratic and cubic terms. And we define $\theta\equiv\theta_1$. The Hamiltonian up to the quartic order in terms of the magnons $a,\,b,\,c$ defined in the local coordinates of $\vect{S}_a,\,\vect{S}_b,\,\vect{S}_c$ are:
\begin{align}
\mathcal{H}^{(1)} & =\frac{\iu S\sqrt{S}}{\sqrt{2}} \sqrt{N}\sum_{\kv}\, [(a_{\vect{k}}+b_{\vect{k}})\delta_{\vect{k},0}(h\sin{\theta_1}-3\sin{(\theta_1-\theta_2)})\non\\
&\quad +c_{\vect{k}}\delta_{\vect{k},0}(h\sin{\theta_2}+6\sin{(\theta_1-\theta_2)})]+h.c.
\end{align}
Applying the first classical constraint $\sin\theta_2=-2\sin\theta_1$,
\begin{equation}
\mathcal{H}^{(1)} =\sqrt{N}\frac{ \iu S\sqrt{S}}{\sqrt{2}}\sin\theta_1 \sum_{\kv}\, (a_{\vect{k}}+b_{\vect{k}}-2c_{\kv})\delta_{\vect{k},0}(h-3(\cos\theta_2+2\cos\theta_1))+h.c.
\end{equation}
\begin{align}
\mathcal{H}^{(2)}= &S\sum_{\kv}\big(-\gamma_0(2-9\theta^2/2)+h(1-\theta^2/2)\big)(\ad_{\kv}a_{\kv}+\bd_{\kv}b_{\kv})\\
&+\big(-\gamma_0(2-9\theta^2)+h(1-2\theta^2)\big)\cd_{\kv}c_{\kv} +\Big(\gamma_{\kv}\big(\bd_{\kv}a_{\kv}+(1-9\theta^2/4)(\cd_{\kv}b_{\kv}+\ad_{\kv}c_{\kv})\big)+h.c.\Big)\non\\
& +\Big(\gamma_{\kv} \,9\theta^2/4 (\cd_{\kv}\bd_{-\kv}+\ad_{\kv}\cd_{-\kv})+h.c.\Big)\non
\end{align}
\begin{align}
\mathcal{H}^{(3)}=-\frac{ 3\iu\theta\sqrt{S}}{\sqrt{2N}} &\sum_{1,2}\Big(\big(\gamma_1(\ad_1\cd_2c_{1+2}-\cd_1\bd_2b_{1+2})-\gamma_{-1}(\cd_1\ad_2a_{1+2}-\bd_1\cd_2c_{1+2})\big)\\
&+\frac{1}{3}(K_s-1)(h/S-9(1-\theta^2))(\ad_1\ad_{2}a_{1+2}+\bd_1\bd_{2}b_{1+2}-2\cd_1\cd_{2}c_{1+2})\Big)+h.c.\non
\end{align}
\begin{align}
\mathcal{H}^{(4)}_{\bot}=\frac{1}{N} \sum_{1,2,3}&\{ S(K_s-1)\big(\gamma_1(\bd_1\ad_2a_3a_{1+2-3} +\cd_1\bd_2b_3b_{1+2-3} +\ad_1\cd_2c_3c_{1+2-3})\\
&+\gamma_{-1}(\ad_1\bd_2b_3b_{1+2-3} +\bd_1\cd_2c_3c_{1+2-3} +\cd_1\ad_2a_3a_{1+2-3})\big)+h.c.\}\non
\end{align}
\begin{equation}
\mathcal{H}^{(4)}_{\parallel}=\frac{1}{N} \sum_{1,2,3}\gamma_{1-2}(\bd_{1}b_2\ad_{3}a_{1-2+3}+\cd_{1}c_2\bd_{3}b_{1-2+3}+\ad_{1}a_2\cd_{3}c_{1-2+3})
\end{equation}
\begin{align}
\mathcal{H}^{(4)}_{\text{n.n.n}}=\frac{J_2}{N} \sum_{1,2,3}&\big(S(K_s-1)(\mu_1+\mu_2+\mu_3+\mu_{1+2-3})+\frac{1}{2}(\mu_{1-3}+\mu_{2-3})\big)\times\\
&(\ad_{1}\ad_2 a_{3} a_{1+2-3}+a\rightarrow b+ a\rightarrow c)\non
\end{align}
$K_s$ is defined as $K_s=\sqrt{1-1/2S}$. $\gamma_{\kv},\,\mu_{\kv}$ are defined as $\gamma_{\kv}=\sum_{\vect{\delta}_i}\mathrm{e}^{ i \kv\cdot\vect{\delta}_i}=\mathrm{e}^{i k_x}+2\mathrm{e}^{-i k_x/2} \cos \frac{\sqrt{3}k_y}{2}$, $\mu_{\kv}=\frac{1}{2}\sum_{\pm\vect{l}_i}\mathrm{e}^{ \pm i\vect{k}\cdot\vect{l}_i}=\cos \sqrt{3}k_y+2\cos\frac{\sqrt{3}k_y}{2}\cos \frac{3k_x}{2}$.
\subsection{Calculate quantum corrections}
We now show how the quantum corrections at the order of $\theta^2$ to the normal and anomalous self-energy are obtained for generic spin diagrammatically. We write the bare $\mathcal{H}^{(n)}$, $n=1,\,2,\,3,\,4$ in terms of the eigenmodes $\phi_{\mu,\kv}=\{A_{\kv},\,B_{\kv},\,C_{\kv}\}$ defined in Eq.~\ref{eq:FerroBasis}. The Bogoliubov basis is defined as $\Phi_{\mu,\kv}=\{\phi_{\mu,\kv},\phi^{\dagger}_{\mu,-\kv}\}$.
\begin{align}
\mathcal{H}^{(1)} &=\sqrt{N}\frac{ \iu S\sqrt{S}}{\sqrt{6}} \sum_{\kv}\Gamma^{(0)}_{\mu,\kv} \phi_{\mu,\kv}+h.c.\\
\Gamma^{(0)}_{\mu,\kv}&=
\begin{cases}
\delta_{\kv,0}\, h\big(2  \sin\theta_1+ \sin \theta_2\big) & \mu=A\\
\delta_{\kv,0}\big( h \sin\theta_2-h \sin \theta_1+9\sin (\theta_1-\theta_2)\big) & \mu=B,\,C
\end{cases}
\label{eq:VHF1}
\end{align}
\begin{align}
\mathcal{H}^{(2)}=&\frac{S}{2}\sum_{\kv,\mu,\nu} \Phi^{\dagger}_{\mu,\kv}\Gamma^{(0)}_{\mu\nu,\kv}\Phi_{\nu,\kv}\non\\
=&S\Big{\{}\sum_{\mu,\kv}(\omega^{(0)}_{\mu,\kv}+\theta^2 E^{(0)}_{\mu,\kv}) \phid_{\mu,\kv}\phi_{\mu,\kv}+ \theta^2\Delta^{(0)}_{\mu,\kv}\opemu{\phid}\phid_{\mu,-\kv}\non\\
&\quad+\frac{1}{2} \theta^2\sum_{\mu\neq\nu,\kv}\{\phid_{\mu,\kv}E^{(0)}_{\mu\nu,\kv}\openu{\phi}+\phi^{\dagger}_{\mu,\kv}\Delta^{(0)}_{\mu\nu,\kv}\phi^{\dagger}_{\nu,-\kv}+h.c.\}\Big{\}}
\end{align}
The relevant quadratic vertex in the order of $\theta^2$ can be found easily. They are:
\begin{align}
&E^{(0)}_{A,\kv}=-3\Re[\gk] &
&E^{(0)}_{B(C),\kv}=3/2\big(\Re[\gk]\mp\sqrt{3}\Im[\gk]\big)\non\\
&E^{(0)}_{AB(AC),\kv}=-3/4\big(\Re[\gk]\pm\sqrt{3}\Im[\gk]\big) &
&E^{(0)}_{BC,\kv}=3/2\Re[\gk]\non\\
&\Delta^{(0)}_{A,\kv}=3/2\Re[\gk] &
&\Delta^{(0)}_{B(C),\kv}=-3/4\Re[\gk]\non\\
&\Delta^{(0)}_{AB(AC),\kv}=3/4\big(\Re[\gk]\mp\sqrt{3}\Im[\gk]\big) &
&\Delta^{(0)}_{BC,\kv}=-3/2\big(\Re[\gk]-\sqrt{3}\Im[\gk]\big)
\end{align}
\begin{align}
\mathcal{H}^{(3)}=-\iu\sqrt{\frac{1}{6}}\sqrt{\frac{S}{N}}\theta\sum_{1,2,\mu,\nu,\rho}\Gamma^{(0)}_{\mu\nu\rho}(1,2,1+2)\phi^{\dagger}_{\mu,1}\phi^{\dagger}_{\nu,2}\phi_{\rho,1+2}+h.c.
\end{align}
The three-point vertex functions relevant to the calculations are:
\begin{align}
\Gamma^{(0)}_{BBB}(1,2,1+2)&=-\Gamma^{(0)}_{AAB}(1,2,1+2)=-\frac{1}{2}[(1-\bar{\jmath})(\gamma_1+\gamma_2)+h.c].\non\\
\Gamma^{(0)}_{CBB}(1,2,1+2)&=-\Gamma^{(0)}_{ACB}(1,2,1+2)=-[(1-\jmath)\gamma_1+(\jmath-\bar{\jmath})\gamma_2+h.c.]\non\\
\Gamma^{(0)}_{CCC}(1,2,1+2)&=-\Gamma^{(0)}_{AAC}(1,2,1+2)=-\frac{1}{2}[(1-\jmath)(\gamma_1+\gamma_2)+h.c].\non\\
\Gamma^{(0)}_{BCC}(1,2,1+2)&=-\Gamma^{(0)}_{ABC}(1,2,1+2)=-[(1-\bar{\jmath})\gamma_1+(\bar{\jmath}-\jmath)\gamma_2+h.c.]
\end{align}
\begin{align}
\mathcal{H}^{(4)}=\frac{1}{3N}\sum{\vphantom{\sum}}'_{1,2,3,\mu,\nu,\rho,\sigma}\{\Gamma^{(0)}_{\mu\nu\rho\sigma}(1,2,3,1+2-3)\phi^{\dagger}_{\mu,1}\phi^{\dagger}_{\nu,2}\phi_{\rho,3}\phi_{\sigma,1+2-3}+h.c.\}
\end{align}
$\sum{\vphantom{\sum}}'$ means the double counting of $\mu,\nu,\rho,\sigma$ is avoided. Thus the four-point vertex functions are found as:
\begin{align}\non
\Gamma^{(0)}_{BBBB}(1,2,3,4)&=\big(S(K_s-1)\Re[\bar{\jmath}\gamma_1+\barj\gamma_2+\jmath\bar{\gamma_3}+\jmath\bar{\gamma_4}]+\frac{1}{2}\Re[\gamma_{1-3}+\gamma_{2-3}]+\Gamma_{n.n.n.}\big)\\\non
\Gamma^{(0)}_{BCBC}(1,2,3,4)&=4\big(S(K_s-1)\Re[\bar{\jmath}\gamma_1+\jmath\gamma_2+\jmath\bar{\gamma_3}+\barj\bar{\gamma_4}]+\frac{1}{2}\Re[\gamma_{1-3}+\barj\gamma_{2-3}]+\Gamma_{n.n.n}\big)\\\non
\Gamma^{(0)}_{ACBB}(1,2,3,4)&=2\big(S(K_s-1)\Re[\gamma_1+\jmath\gamma_2+\jmath\bar{\gamma_3}+\jmath\bar{\gamma_4}]+\frac{1}{2}\Re[\jmath\gamma_{1-3}+\barj\gamma_{2-3}]+\Gamma_{n.n.n}\big)\\
\Gamma^{(0)}_{AABC}(1,2,3,4)&=2\big(S(K_s-1)\Re[\gamma_1+\gamma_2+\jmath\bar{\gamma_3}+\barj\bar{\gamma_4}]+\frac{1}{2}\Re[\jmath\gamma_{1-3}+\jmath\gamma_{2-3}]+\Gamma_{n.n.n}\big)
\label{eq:highHquartic}
\end{align}
where $\Gamma_{n.n.n}=J_2\big((S(K_s-1)(\mu_1+\mu_2+\mu_3+\mu_{4})+\frac{1}{2}(\mu_{1-3}+\mu_{2-3})\big)$. The first term in Eq.~\ref{eq:highHquartic} comes from $\mathcal{H}^{(4)}_{\bot}$, and the second term comes form $\mathcal{H}^{(4)}_{\parallel}$.

\begin{figure}[tbp]
\centering
\subfigure[]{\includegraphics[scale=0.6]{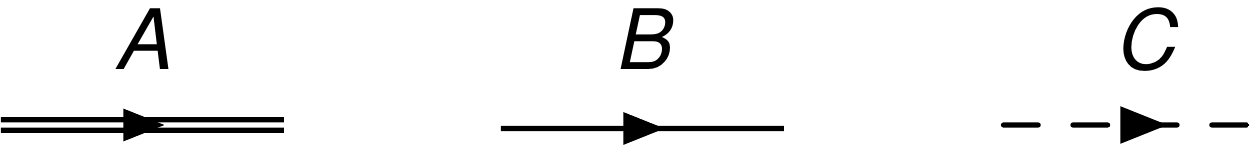}}\\
\subfigure[]{\includegraphics[scale=0.6]{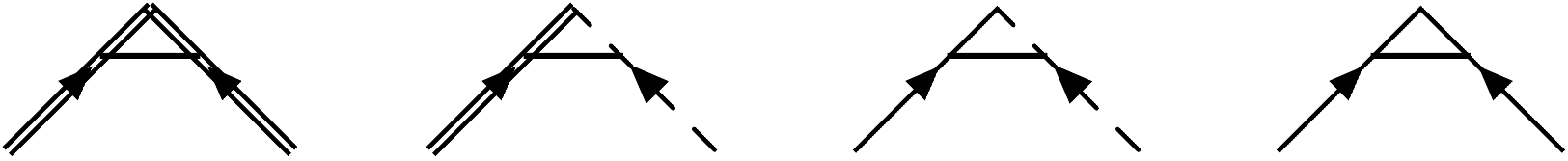}}
\caption{(a) Bare propagators of the magnon modes A, B, C. (b) Bare anomalous Green's functions $\Delta^{(0)}_{AA},\,\Delta^{(0)}_{AC},\,\Delta^{(0)}_{BC},\,\Delta^{(0)}_{BB}$.
\label{fig:GreenFunc}}
\end{figure}

We use double line for the heavy mode A; solid line for one of the light modes B; dashed line for another light mode C. The bare Green function can be written as $G^0_{\mu,\kv}(\omega)^{-1}=\iu\omega-\epsilon_{\mu,\kv}$, $\epsilon_{\mu,\kv}=S(\omega^{(0)}_{\mu,\kv}+\theta^2 E^{(0)}_{\mu,\kv})$. The renormalized Green function is $G_{\mu,\kv}(\omega)^{-1}=\iu\omega-\epsilon_{\mu,\kv}-\Sigma_{\mu,\kv}(\omega)$. $\delta_1\theta^2,\,\delta_2\theta^2$ defined in Eq.~\ref{eq:coplanarLE} relate to the normal and anomalous self-energy as: $S9\delta_1\theta^2=\Sigma_{B,\vect{K}_1}(\omega=0)=\Sigma_{C,-\vect{K}_1}(\omega=0)$ and $S(-3+9\delta_2)\theta^2=\Delta_{BC,\vect{K}_1}(\omega=0)$. Here, vertex functions (e.g. $\Delta$) without superscript $(0)$ are the renormalized vertex functions. The propagators for the bare normal and anomalous Green's functions are shown as in Fig.~\ref{fig:GreenFunc}.
\begin{figure}[tbp]
\centering
\subfigure[]{\includegraphics[scale=0.7]{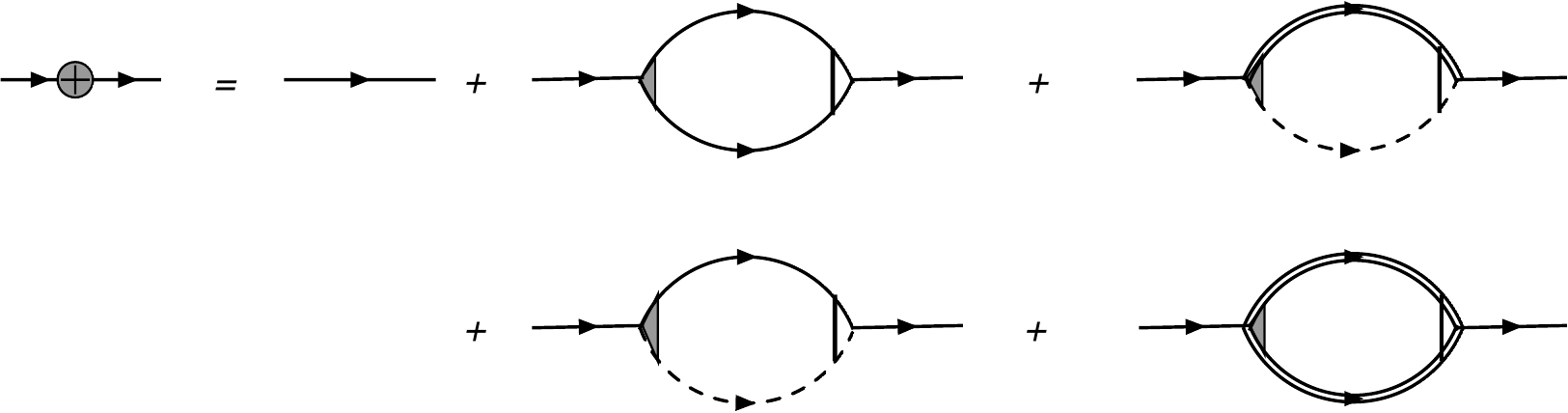}\label{fig:selfE}}\\
\subfigure[]{\includegraphics[scale=0.7]{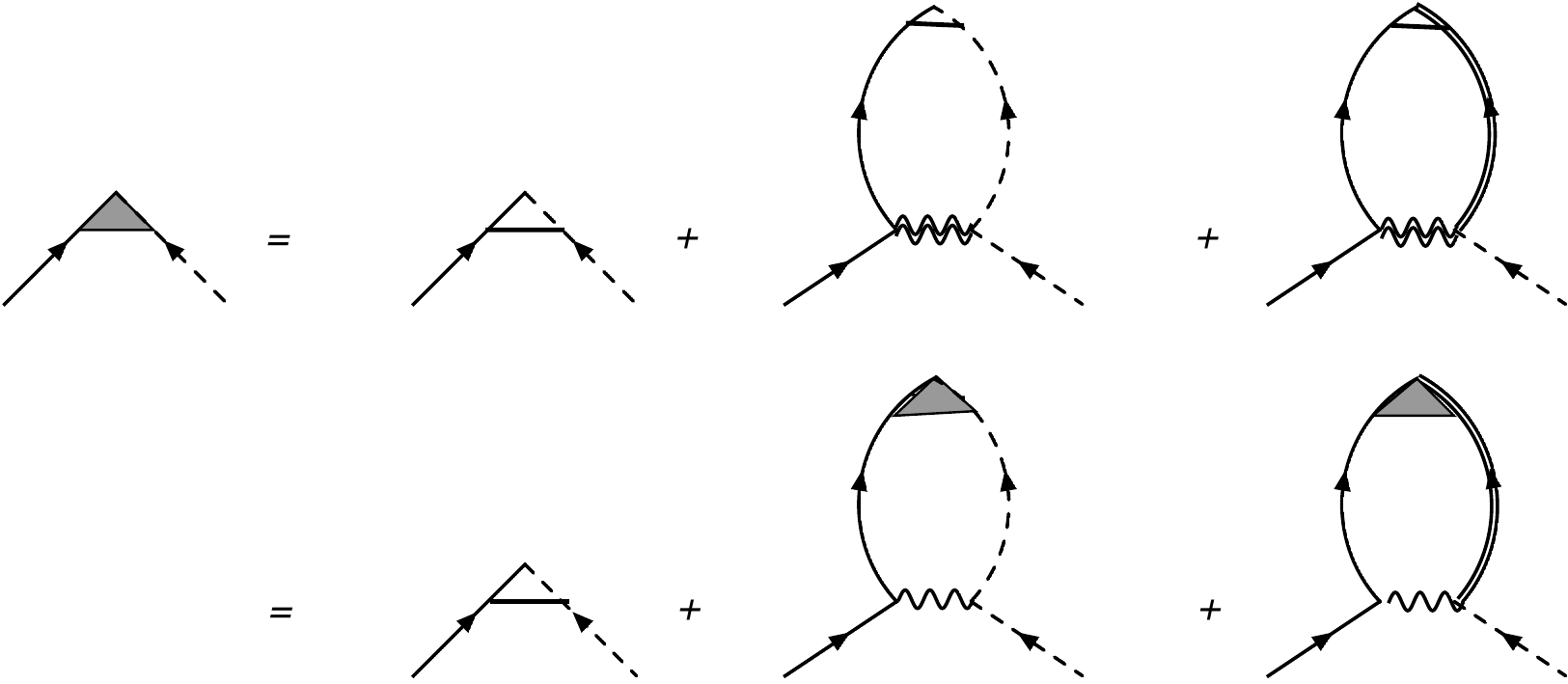}\label{fig:pairing}}
\caption{Renormalized (a) normal and (b) anomalous Green's functions with quantum corrections up at order $\theta^2$. The three-point vertex with shaded triangle and four-point vertex with double wavy line are the fully renormalized vertices. The first propagator on the RHS of panel (a) includes the quantum correction to the relation between the magnetic field $h$ and $\theta^2$.}
\end{figure}
\begin{figure}[tbp]
\centering
\includegraphics[scale=0.7]{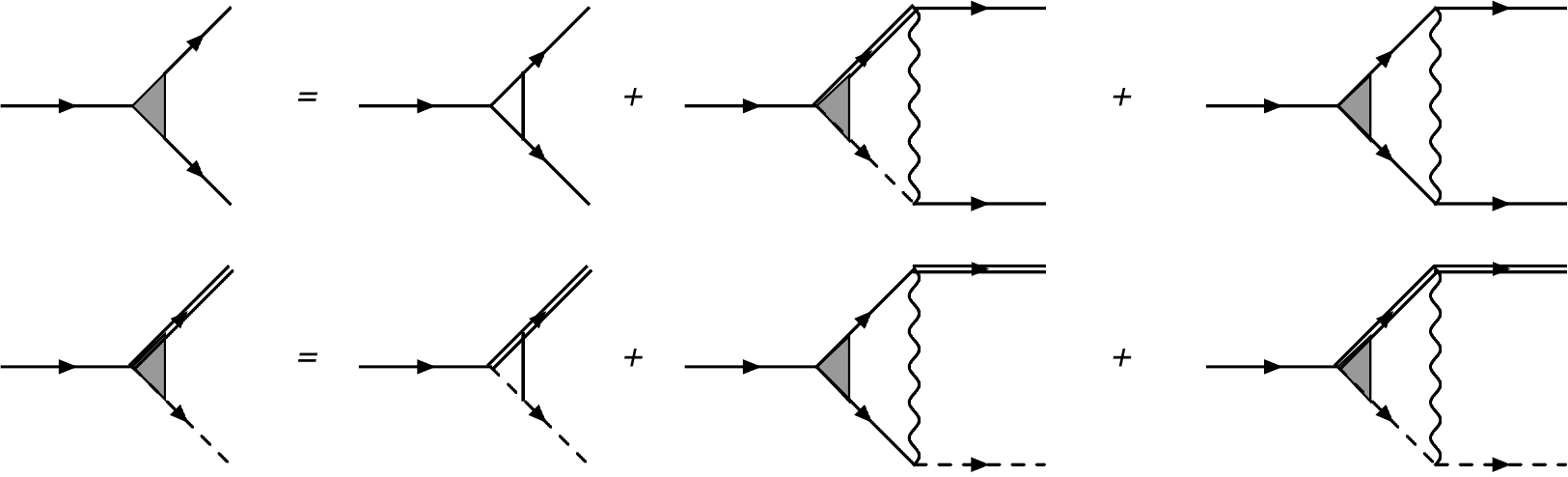}
\caption{Self-consistency equation for the fully renormalized vertices $\Gamma_{BBB}$ and $\Gamma_{ACB}$. Vertices with shaded triangle are the fully renormalized ones, those with hollow triangle are bare vertices. Similarly, we find the self-consistency equation of $\Gamma_{BCB}$ and $\Gamma_{AAB}$.
\label{fig:3ptvertex}}
\end{figure}
\begin{figure}[tbp]
\centering
\includegraphics[scale=0.7]{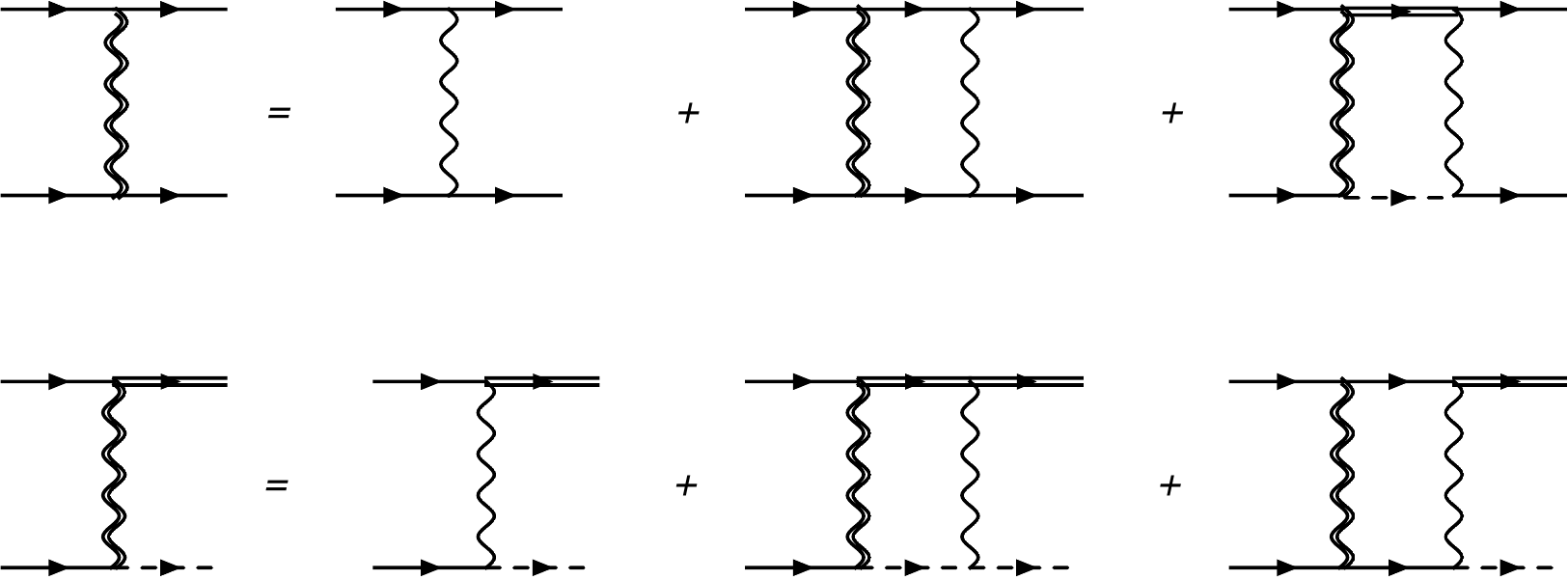}
\caption{Self-consistency equation for the fully renormalized vertices $\Gamma_{BBBB}$ and $\Gamma_{ACBB}$. Four-point vertices with double wavy lines are the fully renormalized ones, those with single wavy line are bare ones. Similarly, we find the self-consistency equation of $\Gamma_{BCBC}$ and $\Gamma_{AABC}$.
\label{fig:4ptvertex}}
\end{figure}
\begin{figure}[tbp]
\centering
\includegraphics[scale=0.7]{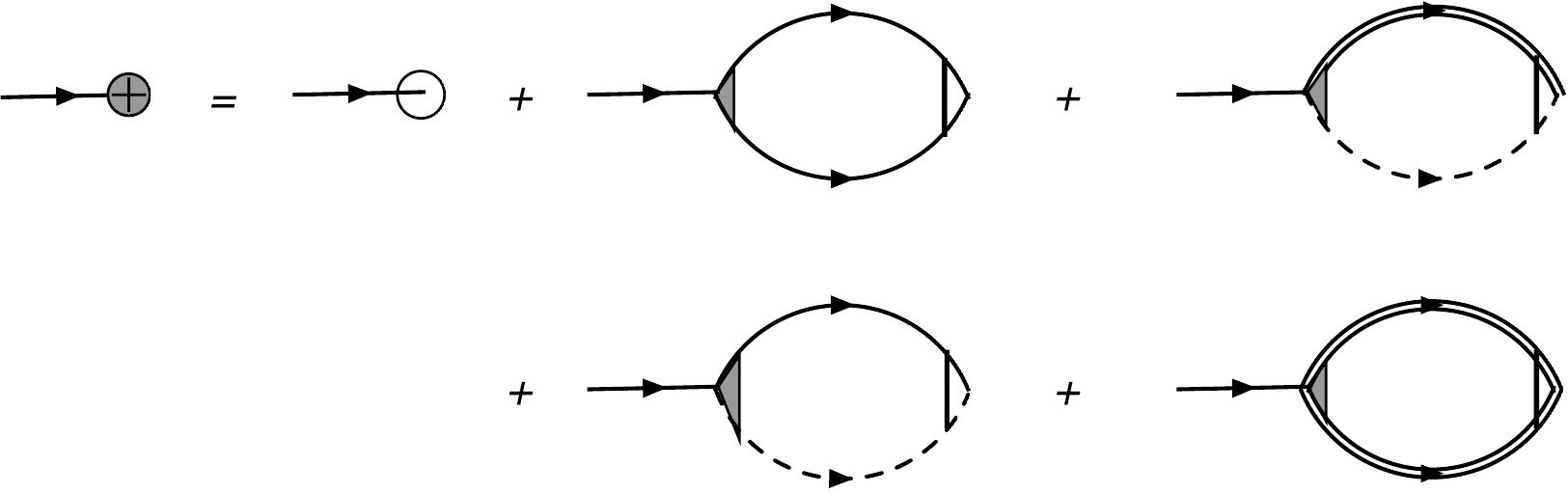}
\caption{Fully renormalized linear term. The first term on the RHS is the bare linear term. The stability of the ordered phase ensures the condition that the prefactor of the linear term is zero.
\label{fig:dH}}
\end{figure}
\begin{figure}[tbp]
\includegraphics[scale=0.7]{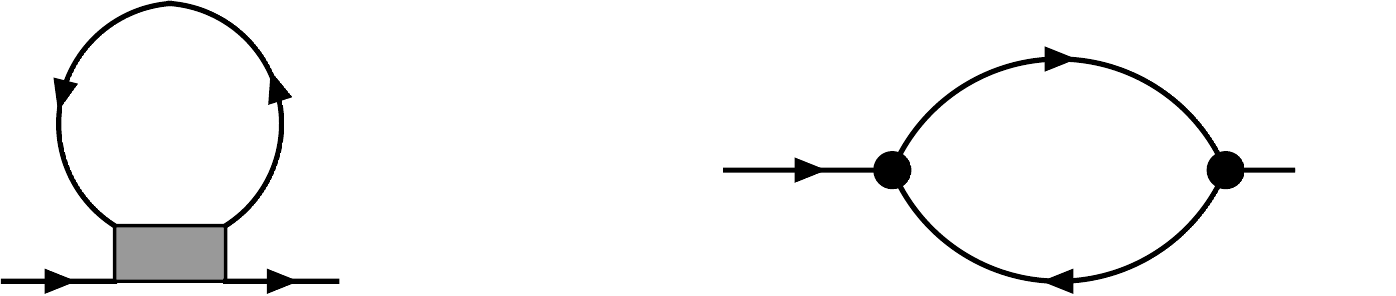}
\caption{Diagrams not considered. Due to the renormalization of the vertex, the above two diagrams are at the order of $\mc{O}(\theta^2/\loghsat)$.
\label{fig:IrreDiags}}
\end{figure}

The self-energy corrections to find $\delta_1\theta^2,\,\delta_2\theta^2$ can be found diagrammatically as shown in Fig.~\ref{fig:selfE} and Fig.~\ref{fig:pairing}. We outline the key steps and results in obtaining $\Sigma_{\alpha,\kv}(\omega=0)$ and $\Delta_{\alpha\beta,\kv}(\omega=0)$:
\begin{itemize}
\item To obtain the self-energy and estimate the contributions from different diagrams, it is essential to obtain the fully renormalized (FR) four-point and three-point vertex functions by solving the consistency equations shown diagrammatically in Fig.~\ref{fig:3ptvertex} and Fig.~\ref{fig:4ptvertex}. Same as finding the FR vertex function Eq.~\ref{eq:BS} for the ferromagnet, we convert the integral equation to matrix algebraic equation. Due to the lattice symmetry of the problem, the vertex function ansatz is determined by only a few free parameters, thus the matrix size is greatly reduced and the inverse matrix problem can be solved exactly. The FR four-point vertex functions are at the order of $\mc{O}(1/\loghsat)$ as long as the total incoming momentum $\kv_{\text{in}}$ satisfies $\omega_{\qv}+\omega_{\kv_{\text{in}}-\qv}=0$ at certain $\qv$. The FR three-point vertex functions scale as $\mc{O}(1)$ in general. But when the momenta of $\Gamma_{\qv,\kv_{\text{in}}-\qv,\kv_{\text{in}}}$ satisfy $\omega_{\qv}=0$ and $\omega_{\kv_{\text{in}}-\qv}=0$, $\Gamma\sim\mc{O}(1/\loghsat)$.
\item The correction to the magnetic field contributes to the normal self-energy in addition to the bubble diagrams in Fig.~\ref{fig:selfE}. The quantum corrections to the magnetic field is determined from the condition that the renormalized prefactor in the linear term $\Gamma_{\mu,\kv}$ is zero. One can show that the quantum corrections to the prefactor of the heavy mode $A_{\kv}$, $\Gamma_{A,\kv=0}$, is suppressed at least logarithmically, $\delta h$ to the leading order $\theta^2$ is determined by the quantum corrections to the prefactor of mode $B$ or $C$, and it can be obtained from diagrams in Fig.~\ref{fig:dH}. $h$ and $\theta$ are related as:
\begin{equation}\label{eq:thetah}
(\hsa-h)\sim \frac{\theta^2}{\loghsat}
 \end{equation}
\item Diagrams like the ones shown in Fig.~\ref{fig:IrreDiags} are suppressed logarithmically and scale as $\theta^2/\loghsat$. The contributions from the particle-hole bubble are at order $\theta^2$ and it comes from the light mode with momentum $|\qv|\lesssim\theta$. On the other hand, both the FR four-point and three-point vertex functions at momentum $\qv\sim 0$ are at the order of $1/\loghsat$. Thus, diagrams like Fig.~\ref{fig:IrreDiags} don't contribute at the order of $\theta^2$.
\item We argue that $\Delta_{BC,\vect{K}_1}(\omega=0)=(-3+9\delta_2)\theta^2=\mc{O}(\frac{\theta^2}{\loghsat})$ for all spins when $\frac{S}{\loghsat}\ll 1$. As the FR four-point vertex function is of $\mc{O}(1/\loghsat)$, the heavy modes don't contribute to the anomalous self-energy at the leading order $\mc{O}(1)$, thus the last diagram with heavy modes in the bubble in the second line of Fig.~\ref{fig:pairing} can be ignored. As a result, the paring vertex can be obtained by solving a single consistency equation, one can easily find that the prefactor of the renormalized anomalous self-energy $(-3+9\delta_2)$ is 0 at order $\mc{O}(1)$.
\item To find the normal self-energy at the leading order, i.e. $(3+9\delta_1)$ at $\mc{O}(1)$, one should include the heavy modes' contributions and find the exact FR three-point vertex functions $\Gamma_{B,B,B}\text{ and }\Gamma_{B,C,B}$ up to the order of $1/\loghsat$.
The renormalized normal self-energy depends on spin S non-trivially, and that of a small spin $S=1/2$ and $S=1$ as well as in the large $S$ limit are obtained. When $S=1/2$, one can unambiguously show that $1/3+\delta_1<0$ and $1/3+\delta_1=-0.1$. When $S=1$, $1/3+\delta_1= -0.02$. In the limit $1\ll S \ll \loghsat$, $1/3+\delta_1=0.03/S$.
 \end{itemize}

In the following, we show explicitly how the $1\ll S \ll \loghsat$ calculation is formulated. The calculations for a generic spin follow the same idea, but is solved exactly at the order of $\mc{O}(1)$. Right below \hsat, there are two scales, the tilt angle (density of condensate field) $\theta$ and $1/S$. We discuss the limit $\frac{S}{\loghsat}\ll 1,\,\frac{1}{S}\ll 1$, where $(\hsa-h)\sim\theta^2$ (mod logarithms). One can show in this limit, $\theta\ll \frac{1}{\loghsat} \ll 1/S$. Thus the leading order correction beyond $\theta^2$ should be $\theta^2/S$. To find the normal self-energy at the order of $\theta^2/S$, the FR three-point vertex functions $\Gamma_{BBB}\text{ and }\Gamma_{BCB}$, which involve soft virtual scatterings, should be exact at the order of $\mc{O}(1+\frac{S}{\loghsat}+\frac{1}{\loghsat})$; the FR three-point vertex functions $\Gamma_{ACB}$ and $\Gamma_{AAB}$, which involve high energy virtual scatterings only, should be exact at the order of $\mc{O}(1)$. The consistency equations for three-point vertex function (Fig.~\ref{fig:3ptvertex}) can be written in the matrix form as:
\begin{align}\label{eq:SC3ptvertex}
\mc{A}_1^T\cdot \Gamma_{b\qv}&=\big(\mc{A}^{(0)\,T}_{1}-\mc{A}_1^T\tau_1V_1-\mc{A}_2^T\tau_2V_2\big)\cdot\Gamma_{b\qv}\non\\
\mc{A}_2^T\cdot \Gamma_{b\qv}&=\big(\mc{A}^{(0)\,T}_{2}-\mc{A}_1^T\tau_1V_2^T-\mc{A}_2^T\tau_2\tilde{V}_1\big)\cdot\Gamma_{b\qv}
\end{align}
The matrices are defined as:
\begin{align}
\label{eq:CoplanarMatrix}
&\Gamma_{BBB}(\qv,\kv-\qv,\kv)=\mc{A}_1^T\cdot \Gamma_{b\qv}\non\\
&\Gamma_{ACB}(\qv,\kv-\qv,\kv)=\mc{A}_2^T\cdot \Gamma_{b\qv}\non\\
&\Gamma^{(0)}_{BBB}(\qv,\kv-\qv,\kv)=\mc{A}_1^{(0)\,T}\cdot \Gamma_{b\qv}\non\\
&\Gamma^{(0)}_{ACB}(\qv,\kv-\qv,\kv)=\mc{A}_2^{(0)\,T}\cdot \Gamma_{b\qv}\non\\
&\Gamma^{(0)}_{BBBB}(\qv',\kv-\qv',\qv,\kv-\qv)=\Gamma_{b\qv'}^T\cdot V_1\cdot \Gamma_{b\qv}\non\\
&\Gamma^{(0)}_{ACBB}(\qv',\kv-\qv',\qv,\kv-\qv)=\Gamma_{b\qv'}^T\cdot V_2\cdot \Gamma_{b\qv}\non\\
&\Gamma^{(0)}_{ACAC}(\qv',\kv-\qv',\qv,\kv-\qv)=\Gamma_{b\qv'}^T\cdot \tilde{V}_1\cdot \Gamma_{b\qv}\non\\
&\tau_{1}=\frac{1}{N}\sum_{\vect{q'}}\frac{\Gamma_{\text{b}\vect{q'}}\Gamma_{\text{b}\vect{q'}}^T}{S(\omega_{B,\qv'}+\omega_{B,\kv-\qv'})}\sim\frac{\loghsat}{S}+\frac{1}{S}\non\\
&\tau_{2}=\frac{1}{N}\sum_{\vect{q'}}\frac{\Gamma_{\text{b}\vect{q'}}\Gamma_{\text{b}\vect{q'}}^T}{S(\omega_{A,\qv'}+\omega_{C,\kv-\qv'})}\sim \frac{1}{S}
\end{align}
We follow the convention in the main text that vertex functions without superscript $(0)$ are the fully renormalized ones; those with superscript $(0)$ are the bare ones. We define $\kv$ as the external incoming momentum.
Eq.~\ref{eq:SC3ptvertex} can be solved as:
\begin{align}\label{eq:Slv3ptvertex}
\mc{A}_1^T&=\Big(\mc{A}^{(0)\,T}_{1}-\mc{A}^{(0)\,T}_{2}(\mc{I}+\tau_2\tilde{V}_1)^{-1}\tau_2V_2\Big)\Big(\mc{I}+\tau_1V_1-\tau_1V_2^T(\mc{I}+\tau_2\tilde{V}_1)^{-1}\tau_2V_2\Big)^{-1}\non\\
\mc{A}_2^T&=\Big(\mc{A}^{(0)\,T}_{2}-\mc{A}^{(0)\,T}_{1}(\mc{I}+\tau_1V_1)^{-1}\tau_1V_2^T\Big)\Big(\mc{I}+\tau_2V_3-\tau_2V_2(\mc{I}+\tau_1V_1)^{-1}\tau_1V_2^T\Big)^{-1}
\end{align}
To keep the accuracy as explained above Eq.~\ref{eq:SC3ptvertex}, Eq.~\ref{eq:Slv3ptvertex} can be simplified as:
\begin{align}\label{eq:Slv3ptvertex2}
\mc{A}_1^T&=\Big(\mc{A}^{(0)\,T}_{1}-\mc{A}^{(0)\,T}_{2}\tau_2V_2\Big)\Big(\mc{I}+\tau_1V_1-\tau_1V_2^T\tau_2V_2\Big)^{-1} ~\text{with the accuracy of } \non\\
&\quad\Big(\mc{A}^{(0)\,T}_{1}-\mc{A}^{(0)\,T}_{2}\tau_2V_2\Big)\sim 1+\frac{1}{S} \non\\
&\quad\Big(\mc{I}+\tau_1V_1-\tau_1V_2^T\tau_2V_2\Big)^{-1}\sim 1+\frac{S}{\loghsat}+\frac{1}{\loghsat}\\
\mc{A}_2^T&=\Big(\mc{A}^{(0)\,T}_{2}-\mc{A}^{(0)\,T}_{1}(\mc{I}+\tau_1V_1)^{-1}\tau_1V_2^T\Big) ~~\text{with the accuracy of } \non\\
&\quad\Big(\mc{A}^{(0)\,T}_{2}-\mc{A}^{(0)\,T}_{1}(\mc{I}+\tau_1V_1)^{-1}\tau_1V_2^T\Big)\sim 1
\end{align}
In the end, we find that $1/3+\delta_1=\frac{0.03}{S}+\mc{O}(\frac{1}{S^2})$, i.e. when $1\ll S \ll \loghsat$, the classical zero modes at the $\vect{ \overline M}$ points are lifted by quantum fluctuations. The mass of the spectrum at momentum $\vect{ \overline K}$ for the stripe phase, $\sim(1/3+\tilde{\delta}_1)$, is obtained for a generic spin in a similar way. The results are summarized in Table~\ref{tab:HFmass}.
\begin{table}
\begin{center}
  \begin{tabular}{|c|c|c|}
    \hline
  \quad & \quad\quad$ (1/3+\delta_1)$\quad\quad & \quad\quad$ (1/3+\tilde{\delta}_1)$\quad\quad \\
    \hline
\quad$S=1/2$\quad   &  \quad -$0.1$ \quad          &   \quad 0 \quad  \\\hline
\quad$S=1$\quad & \quad-$0.02$ \quad  &   \quad0 \quad       \\\hline
\quad$S\gg 1$\quad & \quad$ 0.03/S$  \quad &   \quad 0 \quad \\
  \hline
  \end{tabular}
\end{center}
\caption{Relevant prefactors of the quantum corrections to the spectrum at the relevant classical zero mode the $V$ phase ($\vect{\overline M}$) and the stripe phase ($\vect{\widetilde K}$) near \hsat.
\label{tab:HFmass}}
\end{table}
\bibliography{bibJ1J2MY}

\end{document}